\RequirePackage{lineno}
\def\mytwocolumn{1}

\documentclass[aps,prl,twocolumn,showpacs,superscriptaddress,groupedaddress]{revtex4} \def\mytwocolumn{1} 

\usepackage{graphicx}  
\usepackage{dcolumn}   
\usepackage{bm}        
\usepackage{amssymb}   
\usepackage{amsmath}
\usepackage{footnote}
\usepackage{setspace}
\usepackage{enumerate}
\usepackage{lineno}
\usepackage{color}

\def \emu   {\ensuremath{  e\mu                    }}
\def \ttbar {\ensuremath{ t\bar{t}                  }}

\def \ifb    {\ensuremath{ \rm fb^{-1}                          }}
\def \bfifb  {\bf fb$^{\mathbf{-1}}$}

\def \pt     {\ensuremath{ p_T                                  }}

\def \pte    {\ensuremath{ p_{T}^{e}                            }}
\def \ptm    {\ensuremath{ p_{T}^{\mu}                          }}
\def \ptl    {\ensuremath{ p_{T}^{\ell}                        }}
\def \ptone  {\ensuremath{ p_{T}^{\ell1}                        }}
\def \pttwo  {\ensuremath{ p_{T}^{\ell2}                        }}

\def \egev   {\ensuremath{\mathrm{GeV}                      }}

\def \mtmin {\ensuremath{M_T^{\rm min}}}
\def \mt    {\ensuremath{M_T}}
\def \etmisssc {\mbox{\ensuremath{E\kern-0.6em\slash_T^{\rm\kern+0.1em Sc}}}}
\def \etmiss {\mbox{\ensuremath{E\kern-0.6em\slash_T}}}

\def \ee {\ensuremath{ee}}
\def \em {\ensuremath{e\mu}}
\def \mm {\ensuremath{\mu\mu}}

\newcommand{\Eslash}{\mbox{$E \kern-0.6em\slash$                }}

\newcommand{\runII} {Run II}

\def\ie{{\it i.e.}}
\def\dzero{D0}

\def \pt     {\ensuremath{ p_{{T}}                       }}
\def \ptl   {\ensuremath{ p_{{T}}^{\ell}                 }}

\def \pte    {\ensuremath{ p_{{T}}^{e}                   }}

\def \elmu {\em}
\def \elel {\ee}
\def \mumu {\mm}

\usepackage{dcolumn}
\newcolumntype{.}{D{.}{.}{-1}}
\newcolumntype{,}{D{,}{\relax}{-1}}
\begin{document}




\hspace{5.2in} \mbox{FERMILAB-PUB-13-004-E}
\title{Search for Higgs boson production in oppositely charged dilepton and missing energy 
final states in 9.7~\bfifb\ of $\boldsymbol{p\bar{p}}$ collisions at $\boldsymbol{\sqrt{s} =}$
1.96~TeV}
\affiliation{LAFEX, Centro Brasileiro de Pesquisas F\'{i}sicas, Rio de Janeiro, Brazil}
\affiliation{Universidade do Estado do Rio de Janeiro, Rio de Janeiro, Brazil}
\affiliation{Universidade Federal do ABC, Santo Andr\'e, Brazil}
\affiliation{University of Science and Technology of China, Hefei, People's Republic of China}
\affiliation{Universidad de los Andes, Bogot\'a, Colombia}
\affiliation{Charles University, Faculty of Mathematics and Physics, Center for Particle Physics, Prague, Czech Republic}
\affiliation{Czech Technical University in Prague, Prague, Czech Republic}
\affiliation{Center for Particle Physics, Institute of Physics, Academy of Sciences of the Czech Republic, Prague, Czech Republic}
\affiliation{Universidad San Francisco de Quito, Quito, Ecuador}
\affiliation{LPC, Universit\'e Blaise Pascal, CNRS/IN2P3, Clermont, France}
\affiliation{LPSC, Universit\'e Joseph Fourier Grenoble 1, CNRS/IN2P3, Institut National Polytechnique de Grenoble, Grenoble, France}
\affiliation{CPPM, Aix-Marseille Universit\'e, CNRS/IN2P3, Marseille, France}
\affiliation{LAL, Universit\'e Paris-Sud, CNRS/IN2P3, Orsay, France}
\affiliation{LPNHE, Universit\'es Paris VI and VII, CNRS/IN2P3, Paris, France}
\affiliation{CEA, Irfu, SPP, Saclay, France}
\affiliation{IPHC, Universit\'e de Strasbourg, CNRS/IN2P3, Strasbourg, France}
\affiliation{IPNL, Universit\'e Lyon 1, CNRS/IN2P3, Villeurbanne, France and Universit\'e de Lyon, Lyon, France}
\affiliation{III. Physikalisches Institut A, RWTH Aachen University, Aachen, Germany}
\affiliation{Physikalisches Institut, Universit\"at Freiburg, Freiburg, Germany}
\affiliation{II. Physikalisches Institut, Georg-August-Universit\"at G\"ottingen, G\"ottingen, Germany}
\affiliation{Institut f\"ur Physik, Universit\"at Mainz, Mainz, Germany}
\affiliation{Ludwig-Maximilians-Universit\"at M\"unchen, M\"unchen, Germany}
\affiliation{Fachbereich Physik, Bergische Universit\"at Wuppertal, Wuppertal, Germany}
\affiliation{Panjab University, Chandigarh, India}
\affiliation{Delhi University, Delhi, India}
\affiliation{Tata Institute of Fundamental Research, Mumbai, India}
\affiliation{University College Dublin, Dublin, Ireland}
\affiliation{Korea Detector Laboratory, Korea University, Seoul, Korea}
\affiliation{CINVESTAV, Mexico City, Mexico}
\affiliation{Nikhef, Science Park, Amsterdam, the Netherlands}
\affiliation{Radboud University Nijmegen, Nijmegen, the Netherlands}
\affiliation{Joint Institute for Nuclear Research, Dubna, Russia}
\affiliation{Institute for Theoretical and Experimental Physics, Moscow, Russia}
\affiliation{Moscow State University, Moscow, Russia}
\affiliation{Institute for High Energy Physics, Protvino, Russia}
\affiliation{Petersburg Nuclear Physics Institute, St. Petersburg, Russia}
\affiliation{Instituci\'{o} Catalana de Recerca i Estudis Avan\c{c}ats (ICREA) and Institut de F\'{i}sica d'Altes Energies (IFAE), Barcelona, Spain}
\affiliation{Uppsala University, Uppsala, Sweden}
\affiliation{Lancaster University, Lancaster LA1 4YB, United Kingdom}
\affiliation{Imperial College London, London SW7 2AZ, United Kingdom}
\affiliation{The University of Manchester, Manchester M13 9PL, United Kingdom}
\affiliation{University of Arizona, Tucson, Arizona 85721, USA}
\affiliation{University of California Riverside, Riverside, California 92521, USA}
\affiliation{Florida State University, Tallahassee, Florida 32306, USA}
\affiliation{Fermi National Accelerator Laboratory, Batavia, Illinois 60510, USA}
\affiliation{University of Illinois at Chicago, Chicago, Illinois 60607, USA}
\affiliation{Northern Illinois University, DeKalb, Illinois 60115, USA}
\affiliation{Northwestern University, Evanston, Illinois 60208, USA}
\affiliation{Indiana University, Bloomington, Indiana 47405, USA}
\affiliation{Purdue University Calumet, Hammond, Indiana 46323, USA}
\affiliation{University of Notre Dame, Notre Dame, Indiana 46556, USA}
\affiliation{Iowa State University, Ames, Iowa 50011, USA}
\affiliation{University of Kansas, Lawrence, Kansas 66045, USA}
\affiliation{Kansas State University, Manhattan, Kansas 66506, USA}
\affiliation{Louisiana Tech University, Ruston, Louisiana 71272, USA}
\affiliation{Northeastern University, Boston, Massachusetts 02115, USA}
\affiliation{University of Michigan, Ann Arbor, Michigan 48109, USA}
\affiliation{Michigan State University, East Lansing, Michigan 48824, USA}
\affiliation{University of Mississippi, University, Mississippi 38677, USA}
\affiliation{University of Nebraska, Lincoln, Nebraska 68588, USA}
\affiliation{Rutgers University, Piscataway, New Jersey 08855, USA}
\affiliation{Princeton University, Princeton, New Jersey 08544, USA}
\affiliation{State University of New York, Buffalo, New York 14260, USA}
\affiliation{University of Rochester, Rochester, New York 14627, USA}
\affiliation{State University of New York, Stony Brook, New York 11794, USA}
\affiliation{Brookhaven National Laboratory, Upton, New York 11973, USA}
\affiliation{Langston University, Langston, Oklahoma 73050, USA}
\affiliation{University of Oklahoma, Norman, Oklahoma 73019, USA}
\affiliation{Oklahoma State University, Stillwater, Oklahoma 74078, USA}
\affiliation{Brown University, Providence, Rhode Island 02912, USA}
\affiliation{University of Texas, Arlington, Texas 76019, USA}
\affiliation{Southern Methodist University, Dallas, Texas 75275, USA}
\affiliation{Rice University, Houston, Texas 77005, USA}
\affiliation{University of Virginia, Charlottesville, Virginia 22904, USA}
\affiliation{University of Washington, Seattle, Washington 98195, USA}
\author{V.M.~Abazov} \affiliation{Joint Institute for Nuclear Research, Dubna, Russia}
\author{B.~Abbott} \affiliation{University of Oklahoma, Norman, Oklahoma 73019, USA}
\author{B.S.~Acharya} \affiliation{Tata Institute of Fundamental Research, Mumbai, India}
\author{M.~Adams} \affiliation{University of Illinois at Chicago, Chicago, Illinois 60607, USA}
\author{T.~Adams} \affiliation{Florida State University, Tallahassee, Florida 32306, USA}
\author{G.D.~Alexeev} \affiliation{Joint Institute for Nuclear Research, Dubna, Russia}
\author{G.~Alkhazov} \affiliation{Petersburg Nuclear Physics Institute, St. Petersburg, Russia}
\author{A.~Alton$^{a}$} \affiliation{University of Michigan, Ann Arbor, Michigan 48109, USA}
\author{A.~Askew} \affiliation{Florida State University, Tallahassee, Florida 32306, USA}
\author{S.~Atkins} \affiliation{Louisiana Tech University, Ruston, Louisiana 71272, USA}
% ...
%
%
\collaboration{The D0 Collaboration\footnote{with visitors from
$^{a}$Augustana College, Sioux Falls, SD, USA,
$^{b}$The University of Liverpool, Liverpool, UK,
$^{c}$UPIITA-IPN, Mexico City, Mexico,
$^{d}$DESY, Hamburg, Germany,
$^{e}$SLAC, Menlo Park, CA, USA,
$^{f}$University College London, London, UK,
$^{g}$Centro de Investigacion en Computacion - IPN, Mexico City, Mexico,
$^{h}$ECFM, Universidad Autonoma de Sinaloa, Culiac\'an, Mexico
and
$^{i}$Universidade Estadual Paulista, S\~ao Paulo, Brazil.
}} \noaffiliation
\vskip 0.25cm
\date{January 7, 2013}


\begin{abstract}

We present a search for Higgs boson in final states
with two oppositely charged leptons and large missing transverse
energy as expected in $H\rightarrow WW \rightarrow \ell \nu \ell' \nu'$ decays.
The events are selected from the full Run II data sample of
 9.7~\ifb\ of  $p\bar{p}$ collisions
collected with the D0 detector at the Fermilab
Tevatron Collider at $\sqrt s=1.96$ TeV.
To validate our search methodology, we measure
the non-resonant $WW$ production cross section and find
 $\sigma_{WW}=11.6\pm 0.7$~pb, in agreement with the standard model prediction.
In the Higgs boson search,
no significant excess above the  background expectation  is observed.
Upper limits  at the 95\% confidence level on the Higgs boson production cross
section are therefore derived. Within the standard model, the Higgs boson mass range
$159 < M_{H} < 176$~\egev\ is excluded
while the expected exclusion sensitivity is $156 < M_{H} < 172$~\egev.
For a mass hypothesis of $M_H=125$~\egev, we exclude Higgs boson production cross sections 4.1 times larger than the
standard model expectation, which is compatible with the presence of a Higgs boson at this mass.
Within a theoretical framework with a fourth generation of fermions,
the mass range  $125 < M_{H} < 218$~\egev\ is excluded.
The search results are also interpreted in the context of fermiophobic Higgs boson couplings,
which yields an exclusion of fermiophobic Higgs boson production cross sections 3.1 times larger than the expectation for $M_H=125$~\egev. 

\end{abstract}

\pacs{ 14.80.Bn, 13.85.Qk, 13.85.Rm, 14.65.Jk, 14.80.Ec, 14.70.Fm }
\maketitle


\section{\label{sec:intro}INTRODUCTION}

Spontaneous breaking of the $SU(2) \times U(1)$ electroweak symmetry  explains why the $W$ and $Z$ 
weak vector bosons are massive particles.
However the details of the symmetry breaking mechanism are yet to be fully explored.
In the standard model (SM), it results from
the existence of a single elementary scalar field doublet
that acquires a non-zero vacuum expectation value.
After accounting 
for the mass of the weak vector bosons,
one degree of freedom remains, manifesting itself as
a single scalar particle, the Higgs boson. 
Its mass, $M_{H}$, is a free parameter of the model.
A lower limit of 114.4~\egev\ was set on $M_H$ by 
the CERN LEP experiments~\cite{bib:lephiggs}. This
experimental constraint was extended by the combined results from the CDF and \dzero\ experiments
that excluded the Higgs boson mass range
from 156~\egev\ to 177~\egev~\cite{bib:tevcomb-hww,CDFandD0:2011aa}.
Upper (lower) limits of  131 (122)~\egev~\cite{bib:atlas-hdiscovery} and 128 (121.5)~\egev~\cite{bib:cms-hdiscovery} have then been
established by the ATLAS and CMS Collaborations, respectively.
These exclusion limits and those reported hereafter are all defined at the 95\% C.L.
In both Ref.~\cite{bib:atlas-hdiscovery} and Ref.~\cite{bib:cms-hdiscovery}, excesses above  
background expectations at the five standard deviation (s.d.) level have been reported,
consistent with the observation of a Higgs boson of $M_H \approx 125$~\egev.
The CDF and D0 Collaborations have reported excesses
above background expectations in the $H\rightarrow b\bar b$
search channels~\cite{bib:CDFhbb,bib:D0hbb}. 
Their combination yields an excess at the three s.d.\ level,
consistent with the production of a Higgs boson of mass $M_H \approx 125$~\egev~\cite{bib:TeVhbb}.

In this Letter, we present a search for the SM Higgs boson in final
states containing two oppositely charged leptons ($\ell\ell'=e\mu$,
$ee$, or $\mu\mu$, where small contributions from leptonic
$\tau$~decays are also included) and missing transverse energy
($\etmiss$). The search relies on the full Run~II data set of
$9.7$~\ifb\ of $p\bar{p}$ collisions collected with
the D0 detector at the Fermilab Tevatron Collider at $\sqrt s= 1.96$\, TeV.
This analysis supersedes our previously published results in the same final states,
obtained after analyzing 5.4~\ifb~\cite{bib:hww} and 8.6~\ifb~\cite{bib:hww_prd}
of integrated luminosity. 
A similar search has been conducted by the CDF Collaboration 
using 4.8~\ifb\ of integrated luminosity~\cite{bib:cdf-hww}.
%
The results from Refs.~\cite{bib:cdf-hww,bib:hww} have been combined in Ref.~\cite{bib:tevcomb-hww}.
More recently,
searches in dilepton plus missing transverse energy final states
have been conducted by the ATLAS~\cite{bib:atlas-hww} and
CMS~\cite{bib:cms-hww} Collaborations 
using 4.7~\ifb\ and 4.6~\ifb\ of integrated luminosity, respectively.

The main Higgs boson production and decay channel resulting in opposite charge 
    dilepton plus \etmiss\ final states at the Tevatron is the 
       gluon fusion production,
$gg\rightarrow H$, with subsequent decay $H \rightarrow W^+ W^-
\rightarrow \ell^+\nu\ell^{\prime -}\bar{\nu}'$,
where one of the $W$ bosons is virtual for $M_H <160$~\egev.
 This final state receives additional contributions
from Higgs boson production via vector boson fusion (VBF),
     $q\bar{q}'\rightarrow q\bar{q}'VV\rightarrow q\bar{q}'H$,
and from production in association with a vector boson, $q\bar{q}'\rightarrow VH$ ($V=W,Z$).
The primary source of background is due to diboson production,
in particular the non-resonant $p\bar{p} \rightarrow WW$ process.
Other background sources are
the  Drell-Yan (DY) process, $p\bar{p}\rightarrow Z/\gamma^*\rightarrow \ell\ell$,
with a mismeasured \etmiss, the leptonic decays of top-quark pairs ($\ttbar$),
$W$+jets/$\gamma$ and multijet events in which jets (photons) are misidentified as leptons
(electrons).

The initial selection of Higgs boson candidate events is based on the reconstruction of two high transverse momentum (\pt) leptons.
This selection is followed by additional
requirements, involving \etmiss, and the usage
of multivariate techniques based on boosted decision trees (BDT)~\cite{bib:bdt}, 
to suppress the large DY background.
To increase the sensitivity, the events are separately analyzed
according to the lepton flavors (\ee, \em, and \mm) and jet multiplicity, and they are also categorized into
$WW$-enriched and $WW$-depleted sub-samples. 
Additional BDTs are trained to separate the signal from the remaining background events.
To demonstrate the validity of the techniques used in this search, we use similar BDTs to measure the cross section
for the SM non-resonant $WW$ production cross section.
For the Higgs boson searches, the outputs of the BDTs are the final discriminants used for the statistical interpretation of the data,
within the SM framework, but also in the contexts of a fourth generation of fermions and a fermiophobic Higgs boson.
These models are described in more detail in later sections.

\section{Detector and object reconstruction}
\label{sec:detector}

The \dzero\ detector used for \runII~(2002 -- 2011) is described in detail in Ref.~\cite{bib:d0det}.
The innermost part of the detector is composed of
a central tracking system with a silicon microstrip tracker (SMT) and
a central fiber tracker embedded within a 2~T solenoidal
magnet.  The tracking system is surrounded by a central preshower
detector and a liquid-argon/uranium calorimeter with
electromagnetic, fine, and coarse hadronic sections. The 
central calorimeter (CC) covers pseudorapidity~\cite{footnote:eta} $|\eta|$ $\lesssim 1.1$.
Two end calorimeters (EC) extend the coverage to $1.4\lesssim |\eta| \lesssim 4.2$.
The pseudorapidity gap between the ECs and CC is covered by scintillating tiles.
A muon
spectrometer, with pseudorapidity coverage of $|\eta|\lesssim 2$, resides outside the calorimetry and is comprised of drift
tubes, scintillation counters, and toroidal magnets.
Trigger decisions are based
on information from the tracking detectors,
calorimeters, and muon spectrometer.

Electrons are reconstructed as isolated clusters in the electromagnetic calorimeter, and required to spatially match a track
in the central tracking system.  
They have to pass a BDT (\ee\ channel) or likelihood (\emu\ channel)
criterion (collectively called electron quality later in the text) that accounts for calorimeter shower shape observables,
a spatial track match probability estimate, and the ratio of the electron cluster energy to track momentum ($E/p$). 
Electrons are required to be in the acceptance
of the calorimeter ($|\eta|<1.1$ or $1.5<|\eta|<2.5$).
In the
dielectron channel, events with one electron in the 
EC are treated separately from events with both
electrons in the CC.
Events where both electrons are in the EC
are not considered due to a large background and a small
signal contribution.

Muons are identified by the presence of at least one track segment
reconstructed in the acceptance ($|\eta| < 2.0$) of the muon spectrometer, that is spatially consistent
with a track in the central tracking detector. The momentum and charge are
measured by the curvature in the central tracking system.
To select isolated muons, criteria based
on the momenta of central tracks emitted in the approximately same direction as the muon
and criteria based on the energy deposited around the muon trajectory in the calorimeter are employed.
The number of hits in the wire chambers and in the scintillators
are combined to define a muon quality variable 
used in the final stage of the analysis.

Jets are reconstructed from energy deposits in the calorimeter using
an iterative midpoint cone algorithm~\cite{bib:jet} with a cone radius  ${\cal R}= 0.5$~\cite{footnote:DR}.
The jet energies are
calibrated using transverse momentum balance in $\gamma+$jet
events~\cite{bib:jetCalib}. 
Jets are considered in this analysis only if they have $p_{T} >20$~\egev and $|\eta| < 2.4$.
Each jet is also required to be matched to at least two
tracks associated to the $p\bar{p}$ interaction vertex.

The $\etmiss$ and its direction are obtained from the vector sum of the transverse
components of energy deposits in the calorimeter, corrected for
the differences in detector response of the reconstructed muons, electrons, and jets.

\section{DATA AND SIMULATED SAMPLES }
\label{sec:data}\label{sec:MC}



Signal and SM background processes except multijet are simulated with
\textsc{pythia}~\cite{bib:pythia} or \textsc{alpgen}~\cite{bib:alpgen} generators, with \textsc{pythia}
providing showering and hadronization in the latter case, using the
CTEQ6L1~\cite{bib:cteq6} parton distribution functions (PDFs), followed by
a detailed {\sc geant3}-based~\cite{bib:geant} simulation of the \dzero\
detector. In order to model the effects of multiple $p\bar{p}$
interactions, the Monte-Carlo (MC) samples are overlaid with events from random
$p\bar{p}$ collisions with the same luminosity distribution as data. Then, these events are
reconstructed with the same software as used for the data.
Jet energy calibration and calorimeter response to unclustered objects are adjusted in simulated events to match those measured in data.
Corrections for residual differences between
data and simulation are applied to electrons, muons, and
jets for both  identification efficiencies and energy resolutions.


Higgs boson signal samples are simulated using {\sc
pythia} for $100 \le M_{H} \le 200$~GeV in increments of $5$~GeV, and for
 $200 \le M_{H} \le 300$~GeV in increments of $10$~GeV.
For $gg \to H$
production, the cross section is calculated at next-to-next-to-leading order with resummed next-to-next-to-leading logarithm
(NNLO+NNLL)~\cite{bib:gluon_fusion_xsec},
for VBF at NNLO~\cite{bib:vbf_xsec},
and for $VH$ at NNLO~\cite{bib:vh-xs}.
All signal cross sections are computed using the MSTW 2008 NNLO PDF
set~\cite{bib:mstw08}.
The distribution of the Higgs boson \pt\ for $gg\to H$ process
is weighted to match the calculation of
the \textsc{hqt} generator, which has NNLO and NNLL accuracy~\cite{bib:Higgs_pT}.
The Higgs boson branching ratio predictions are taken from {\sc
hdecay}~\cite{bib:hdecay}.

The $W+$jets and $Z+$jets backgrounds are modeled using
{\sc alpgen}.  The $W+$jets and $Z+$jets
processes are normalized using the NNLO cross section calculations of
Ref.~\cite{bib:v-xs}. 
The $p_{T}$ distribution of $Z$ bosons is weighted to match the distribution observed in
data~\cite{bib:zbosonpT},
taking into account its dependence on the number of reconstructed jets. 
The $p_{T}$ distribution of $W$ bosons is
weighted to match the  measured $Z$ boson $p_{T}$
spectrum,
corrected for the differences between the 
$W$ and $ Z$ $\pt$ spectra predicted in NNLO QCD~\cite{bib:w_z_bosonpT_ratio}.
In the \ee\ and \emu\ channels, the $W$+jets
simulation includes contributions from events where a misidentified electron
originates from a jet or a photon.  The size of each of these
contributions is corrected such that the distribution of the number of
hits in the innermost silicon layer, associated to the electron track,
matches that observed in a
$W$+jets enriched control sample.

The \ttbar\ process is modeled using {\sc alpgen} with showering and hadronization
        provided by {\sc pythia}, and normalized to the approximate  NNLO cross section~\cite{bib:tt-xs}.

Diboson production processes ($WW$, $WZ$,
and $ZZ$) are simulated using {\sc pythia}, normalized to NLO cross sections~\cite{bib:dibo-xs}.  An additional correction,
determined using the \textsc{powheg} generator~\cite{bib:powheg},
accounts for $Z/\gamma^\star$ interference in $WZ$ production.
For the irreducible background arising from $WW$ production, the $p_{T}$ of the diboson system is modeled using
the {\sc mc@nlo} simulation~\cite{bib:mcatnlo},
and the distribution of the
opening angle of the two leptons is corrected for the
contribution of the non-resonant $gg\rightarrow WW$ process~\cite{bib:ggWW}.  

The background due to multijet production, where jets are
misidentified as leptons, is determined from data by inverting
some of the lepton selection criteria. All other event selection criteria
are applied in order to model the kinematic distributions of the
multijet background in the signal region. 
In the \mm\ channel, the
opposite-charge requirement for muons is reversed, and a correction for
the presence of non-multijet events in the like-sign sample,
estimated from simulation, is applied.  For the \ee\ and \emu\ channels,
the electron quality requirement is reversed, and the normalization
is determined from control samples in which the leptons have the same charge.

\section{EVENT PRESELECTION}
\label{sec:selection}

\begin{table*}[!]
\caption{\label{tab:presel_cutflow}
Observed and expected number of events after preselection in
   the \ee, \em, and \mm\ final states.  The signal is for a Higgs
   boson mass of 125~GeV. The uncertainty quoted on the background combines both statistical and systematic uncertainties,
after the normalization procedure described in the main text.
}
\begin{ruledtabular}

\begin{tabular}{c,rcl c ,, ,,,,,,,,,,,,,cc}
 &   \multicolumn{1}{c}{Data} & \multicolumn{3}{c}{Total background} & Signal &  \multicolumn{1}{c}{$Z/\gamma^\star \to ee$  }
 & \multicolumn{1}{c}{$Z/\gamma^\star \to \mu\mu$  }&  \multicolumn{1}{c} {$Z/\gamma^\star \to \tau\tau$} & \multicolumn{1}{c}{ $\ttbar$ }&  \multicolumn{1}{c}{$W$+jets} &  \multicolumn{1}{c}{Diboson} &  \multicolumn{1}{c}{Multijet} \\

\elel{}:    &659,570 &664460 &$\pm$& 13290      &16.1 &653,263         & ,\mbox{--} &5,494   &21,0    &7,95  &9,45   &37,52 \\
\elmu{}:    &14,936 &15142   &$\pm$&303         &16.6 & ,408          &  1,211      &8,671   &53,7    &12,25 &9,06  &21,84 \\
\mumu{}:    &811,549 &818269 & $\pm$&16370      &18.7  & ,\mbox{--}   & 807,642     &6,459   &35,6    &4,38 &13,14  &20,60   \\
\end{tabular}

\end{ruledtabular}
\end{table*}

A first selection is applied on the data by requiring
two high-$\pt$ leptons, that have opposite charge and that
originate from the same
location, within 2~cm, along the beamline.
In the \ee~and \mm~channels, the  highest-$p_{T}$ and next highest $\pt$ leptons are
required to satisfy $\ptone>15$~GeV and $\pttwo>10$~GeV,
whereas in the \em\ channel,~$\pte>15$~GeV and
$\ptm>10$~GeV are required.
Additionally, in the \ee\ and \mm\ final states, the
dilepton invariant mass $M_{\ell\ell}$ is required to be
greater than 15~GeV. A veto against additional leptons is applied to prevent overlap with dedicated Higgs searches in trilepton
final states~\cite{bib:trilepton}. 
These criteria define the ``preselection'' stage
of the analysis, and they select samples dominated by DY production.
Most events selected at this level pass single-lepton 
trigger conditions. 
But, as no specific trigger requirement is made,
the trigger acceptance with respect
to off-line selections is enhanced to $\approx 92\%$ for the $\mu\mu$
channel and $\approx 100\%$ for the $ee$ and $e\mu$ channels,
due to additional events passing lepton+jets or dilepton triggers. The
remaining trigger inefficiency is modeled
in the simulation by corrections derived from $Z\to\ell\ell$ samples selected with different trigger requirements.
The preselected samples are further
subdivided according to the number of jets in the event.
Namely 0-, 1-, and 
$(\ge 2)$-jet multiplicity bins are considered.
Dividing the
analysis into different jet multiplicity bins significantly increases
the sensitivity of this search as the signal and background
compositions are different in each sample.

To correct for any possible mismodeling of the lepton reconstruction
and trigger efficiencies, and to reduce the impact of the luminosity
uncertainty, scale factors are applied to the MC samples at the
preselection stage to match the data.
The $Z$ boson mass peak regions in the preselected samples are used to 
determine normalization factors. Their differences from unity are found to be consistent with the luminosity uncertainty of 6.1\,\%~\cite{bib:lumi}.
This procedure is repeated for each jet multiplicity to derive
jet-bin-dependent DY background normalizations  to correct for possible
            mismodeling of the DY jet multiplicity.

The number of events
after the preselection is presented in Table~\ref{tab:presel_cutflow}~\cite{bib:aux}; all sub-samples are dominated by
DY production.
Figure~\ref{fig:presel_all}(a) shows
the dilepton invariant mass distributions~\cite{bib:aux} for data and the background prediction
            for the combined sub-samples.

\begin{figure*}[!] 
\center
\includegraphics[ height=0.237\textheight]{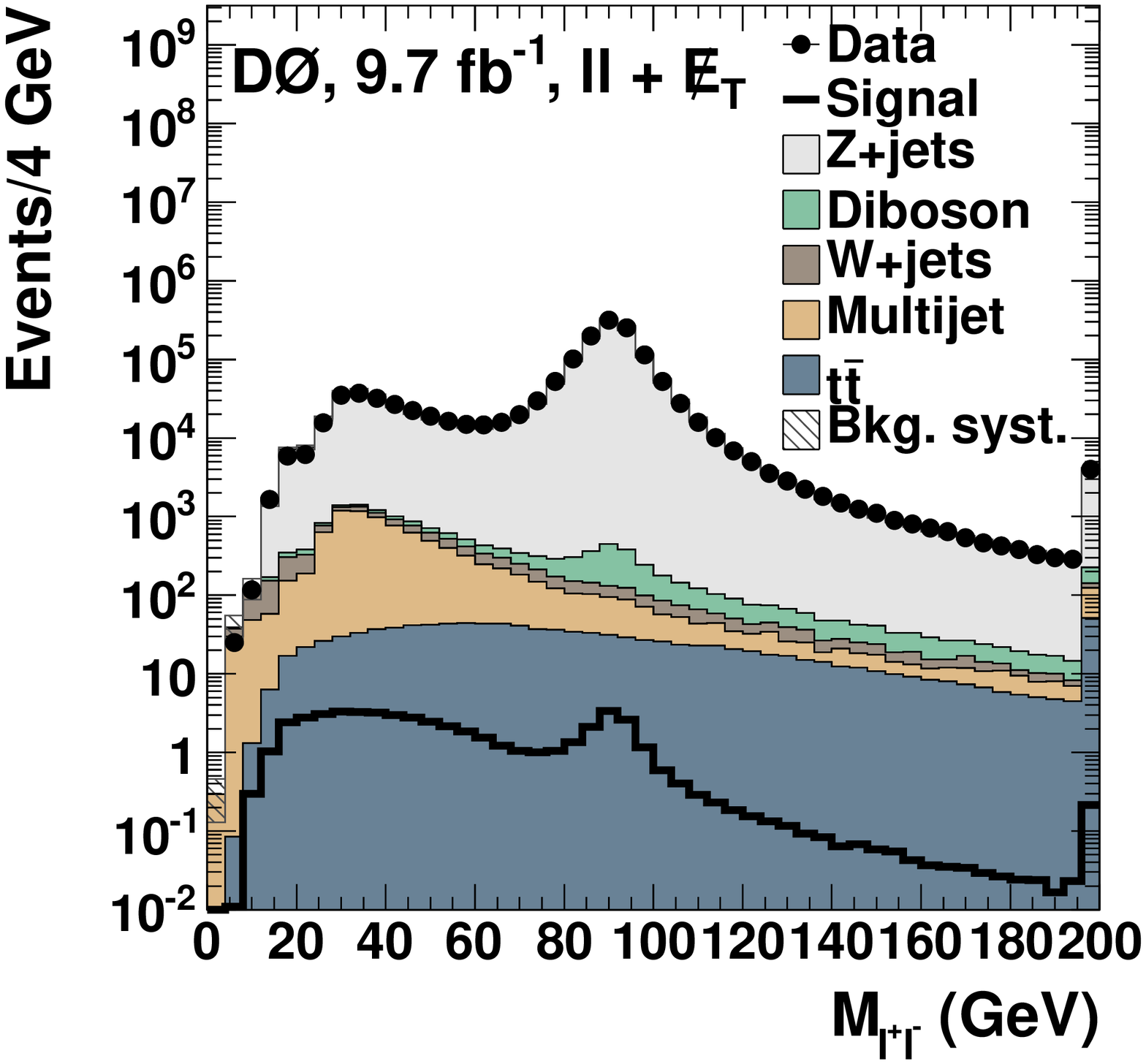} 
\includegraphics[height=0.237\textheight]{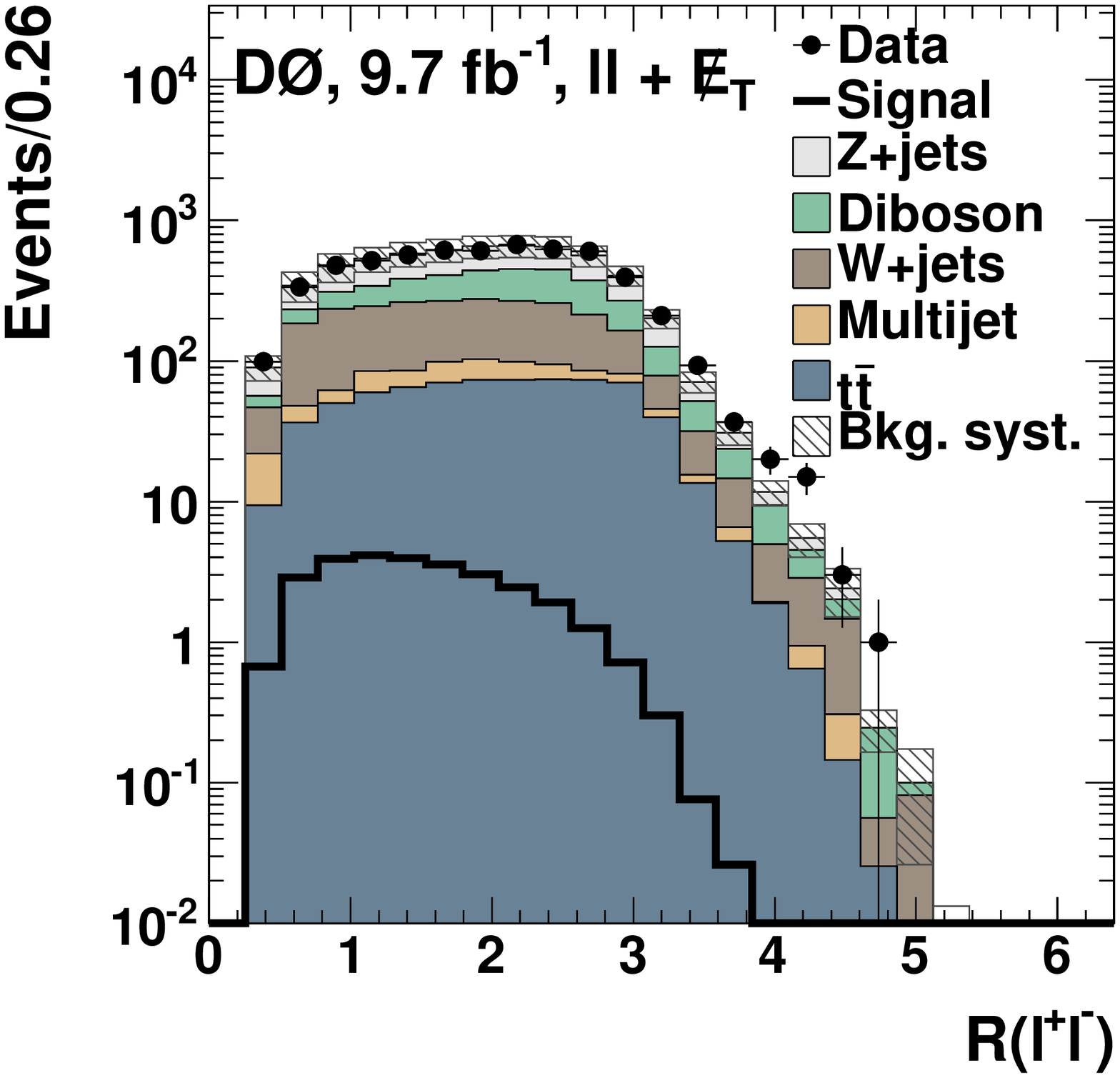} 
\includegraphics[ height=0.237\textheight]{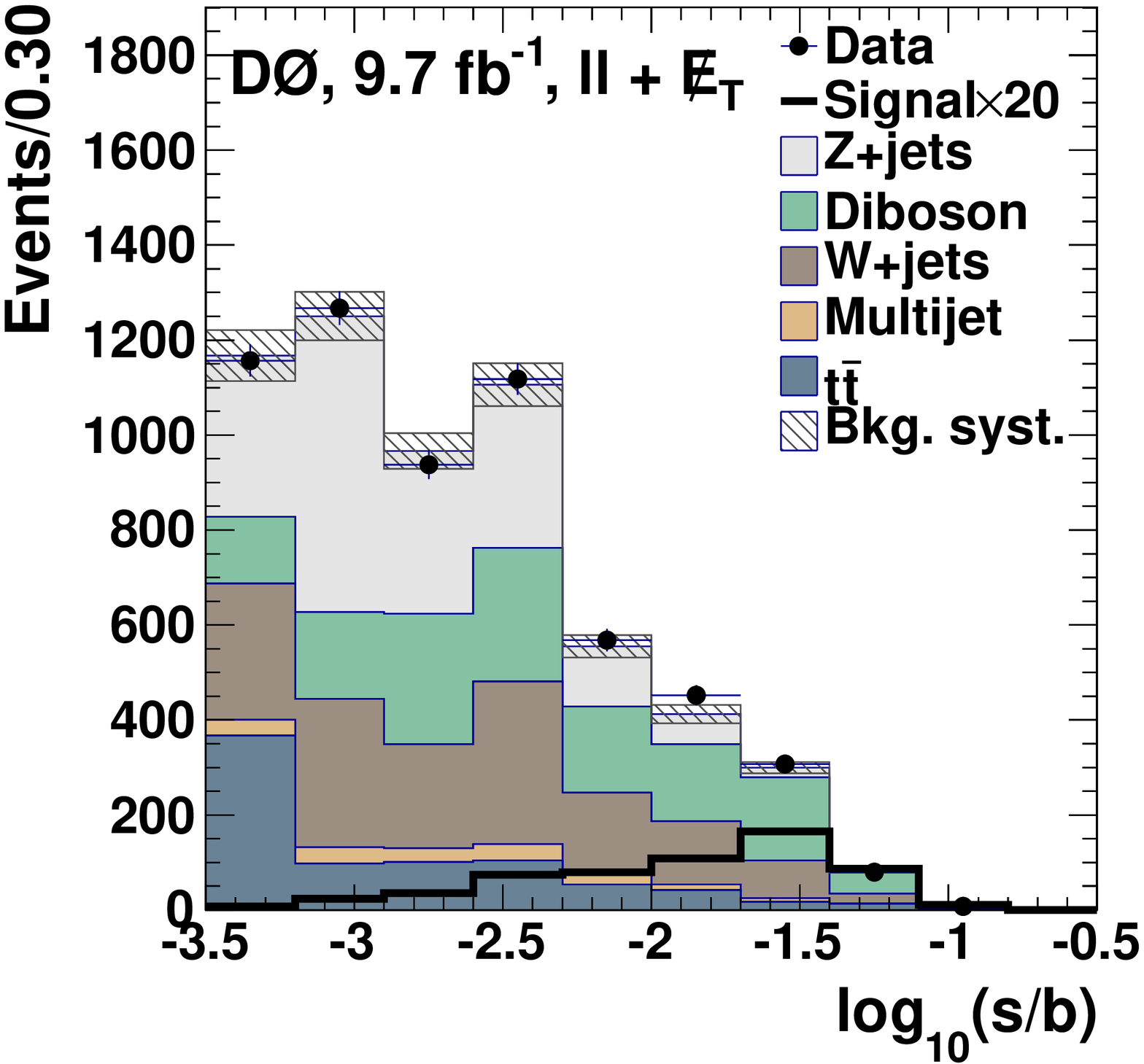} 

\unitlength=1mm
\begin{picture}(00,00)

\if \mytwocolumn 1
\put(-77,50){\text{\bf (a)}} 
\put(-17,50){\text{\bf (b)}} 
\put(43,50){\text{\bf (c)}}
\else
\put(-47,106){\text{\bf (a)}} 
\put(-17,50){\text{\bf (c)}}
\put(13,106){\text{\bf (b)}}
\fi

\end{picture}

  \caption{ [color online]
    The (a) dilepton invariant mass at preselection level, (b)
   the angular separation $\cal R(\ell,\ell)$ between the leptons at preselection level, 
    and (c) output of the final BDT discriminants after the final selection,
sorted as a function of signal over background ratio. In (a), the last bin includes all events
above the upper bound of the histogram.
In these plots,
the hatched bands show the total  systematic uncertainty on the background predictions, and
the signal distributions are those expected from a  Higgs boson of mass $M_H=125$~GeV. It is scaled by a factor 20 in (c).
\label{fig:presel_all}
\label{fig:sel_all}
}

\end{figure*}

\section{Analysis using decision trees}
In the \ee\ and \mm\ channels, BDTs are trained for each Higgs boson mass value and each jet multiplicity bin
to discriminate between the signal and the dominant DY background.
The input variables to these ``anti-DY BDTs'' are kinematic quantities,
such as the lepton momenta, the azimuthal opening angle between the two leptons,
$\etmiss$, 
variables
that take into account both $\etmiss$ and its direction relative to a lepton or a jet,
and
observables   that differentiate between real and misreconstructed $\etmiss$.
This multivariate technique follows the method defined in the previous publication~\cite{bib:hww_prd} where more details
on the BDTs' input variables are given.
The final selection stage for the \ee\ and \mm\ channels is obtained
by applying cuts on the anti-DY BDT discriminants~\cite{bib:aux}.
The thresholds are chosen to obtain similar background rejection
as the cut-based rejection employed in Ref.~\cite{bib:hww}. 

In the \emu\ channel the final selection stage requires
$\mtmin>20$~GeV and $M_{T2}>15$~GeV~\cite{bib:aux},
where
$\mtmin$  is the  minimum value, over the two possible lepton choices, of
the transverse mass, $\mt(\ell,\etmiss) =
\sqrt{2\cdot\ptl\cdot\etmiss\cdot[1 - \cos\Delta\phi(\ell,\etmiss )]}$
,
and $M_{T2}$ is an extension of the
transverse mass suitable for final states with two visible and two invisible
particles~\cite{bib:mt2}.

The number of events at the final selection
stage can be found in Table~\ref{tab:final_cutflow}, and
the distribution of the angular separation between the leptons,  combined for all dilepton final states, $\cal R(\ell^+\ell^-)$,
can be seen in Fig.~\ref{fig:sel_all}(b)~\cite{bib:aux}.

\begin{table*}[!]
\caption{\label{tab:final_cutflow} 
Expected and observed number of
events after the final selection in the \ee,  \em, and \mm\ final
states. The signal is for a Higgs boson mass of 125~GeV.
 The numbers in parentheses correspond to
the efficiency of the final selection with respect to the
preselection, shown in Table~\ref{tab:presel_cutflow}, for both the
total background and signal. The uncertainty quoted on the total background combines both statistical and systematic uncertainties.
}
\begin{ruledtabular}

\begin{tabular}{ccrcllrl   ccc ccccccccccccccccccccccccccccccrr                       ccccccc}

             & Data & \multicolumn{4}{c}{Total background} & \multicolumn{2}{c}{Signal} &{ $Z\to ee$} & $Z\to \mu\mu$ & $Z\to \tau\tau$ & $\ttbar$ & $W$+jets & Diboson & Multijet \\
\hline \hline 
\elel{}:   & 1882 & 1859 &  $\pm$   &  205  & (0.3\,\%) &  7.5 & (46.8\,\%)  & 746   &  \phantom{000}-- & 55  & 151 & 518 &  371 & \phantom{0}18 \\
0 jet        & 1289 & 1317&  $\pm$  &  145  & (0.2\,\%) &  4.6  & (64.8\,\%) & 528   & \phantom{000}-- & 32  &\phantom{0}12  & 424 & 307 &\phantom{0}13 \\
1 jet        &\phantom{0}379  & 343 &  $\pm$  &  38   & (0.4\,\%) &  1.8 & (36.6\,\%)  & 152  &  \phantom{000}-- &\phantom{0}6  &\phantom{0}47  &\phantom{0}80  &\phantom{0}53  &\phantom{00}4 \\
$\ge 2$ jets &\phantom{0}214  & 199 &  $\pm$  &  22   & (1.7\,\%) &  1.1  & (27.5\,\%) & \phantom{0}65 &  \phantom{000}-- & 16  &\phantom{0}91  &\phantom{0}13  &\phantom{0}11  &\phantom{00}1 \\
\hline 
\elmu{}:    & 1954 & 1960 &  $\pm$  &  212 & (12.9\,\%) & 12.3  & (74.1\,\%) &\phantom{0}11   &\phantom{00}71 & 11  & 332 & 871   & 628 &\phantom{0}35 \\
0 jet         & 1266 & 1340 &  $\pm$&  129 & (10.8\,\%) & 8.0  & (82.5\,\%)  &\phantom{00}7   &\phantom{00}55 &\phantom{0}8  &\phantom{0}11  & 716 & 522 &\phantom{0}22 \\
1 jet         &\phantom{0}367  & 336 &  $\pm$ &  43  & (16.5\,\%) & 3.1  & (67.4\,\%)  &\phantom{00}3  &\phantom{00}13 &\phantom{0}3  &\phantom{0}97  & 116  &\phantom{0}94  &\phantom{0}11 \\
$\ge 2$ jets  &\phantom{0}321  & 283 &  $\pm$ &  40  & (38.1\,\%) & 1.2  & (52.2\,\%)  &\phantom{00}1  &\phantom{000}3 &\phantom{0}1   & 225 &\phantom{0}39  &\phantom{0}12  &\phantom{00}2 \\
\hline 
\mumu{}:      & 2057  & 2109 &  $\pm$ & 325 &  (0.3\,\%) & 9.1 &  (48.6\,\%) &  \phantom{00}--            & 1055           & 45          & 235           & 231           & 378   & 165   \\
0 jet         &\phantom{0}767   & 785 &  $\pm$ & 100  &  (0.1\,\%) & 5.1 &  (57.0\,\%) &  \phantom{00}--  &\phantom{0}210  &\phantom{0}3 &\phantom{00}4  & 178           & 275   & 115    \\
1 jet         &\phantom{0}485   & 464 &  $\pm$ & 72   &  (0.4\,\%) & 2.3  &  (43.4\,\%) &  \phantom{00}-- &\phantom{0}238  & 23          &\phantom{0}53  &\phantom{0}42  &\phantom{0}73    &\phantom{0}34    \\
$\ge 2$ jets  &\phantom{0}805   & 860 &  $\pm$ & 153  &  (4.9\,\%) & 1.7  &  (38.2\,\%) &  \phantom{00}-- &\phantom{0}607  & 19          &178            &\phantom{0}11  &\phantom{0}30    &\phantom{0}16    \\

\end{tabular}
\end{ruledtabular}
\end{table*}

At the final selection stage,
a series of new BDTs is built:
the ``$WW$-BDTs'' are trained to separate the non-resonant $WW$ production from the other backgrounds, while
the ``final BDTs'' are trained to separate the signal from all the backgrounds. In the former case,
    the Higgs signal is not used in the training.
These BDTs rely on the same input variables as for the anti-DY BDTs,
but supplementary variables are added
characterizing the lepton reconstruction quality and the lepton
isolation, to discriminate against the instrumental backgrounds (multijet and $W$+jets backgrounds).
Outputs of jet $b$-tagging multivariate discriminants~\cite{bib:bid} are
also added as inputs to separate the signal from the \ttbar\ background.

Using the $WW$-BDT discriminants, we split the 0- and 1-jet samples into $WW$-depleted and 
$WW$-enriched regions for the \elel{} and \mumu{} analyses.
In the {\em} channel, splitting only the 0-jet sample according to the lepton reconstruction quality
achieves a sufficiently pure separation of the data sample into a $WW$-depleted and $WW$-enriched sub-samples. 
The final BDTs are then trained 
separately for each jet multiplicity bin, for each dilepton final state,
and for the  $WW$-depleted and $WW$-enriched samples, resulting in 14 BDTs for each mass hypothesis~\cite{bib:aux}.
The outputs of these BDTs are used as final discriminants. 
Figure~\ref{fig:sel_all}(c) shows the BDT
distributions of the 14 sub-samples summed in bins with similar signal to background ratios ($s/b$).

\section{Systematic uncertainties}

Systematic uncertainties are estimated for each final state,
background, and signal process. They can affect only the normalization
or both the normalization and the shape of the final discriminants.

Sources of systematic uncertainty that affect only the
normalization arise from
the overall normalization uncertainty due to theoretical inclusive cross sections of $Z$+jets ($4\%$), $W$+jets ($6\%$), diboson ($6\%$) and
\ttbar~($7\%$) processes; multijet normalization ($30\%$); the $W$+jets jet-bin-dependent
normalization ($15\%$--$30\%$); the $Z$+jets jet-bin-dependent
normalization ($2\%$--$15\%$);
and the 
modeling of the $\etmiss$ measurement for the $Z$+jets background ($5\%$--$19\%$).

The uncertainties on $\sigma(gg\to H)$ production are estimated following the
prescription described in Ref.~\cite{ref:errMatr}, \ie,
by considering  as uncorrelated the scale uncertainties of the NNLL
inclusive~\cite{ref:Anastasiou,bib:gluon_fusion_xsec}, NLO $\ge 1$
jet~\cite{bib:ggH01jetUncert}, and NLO $\ge 2$
jets~\cite{ref:ggH2jetUncert} cross sections. This prescription
results in the following covariance matrix for the
the exclusive production of  ${gg\to H}+0$ jet, $+1$ jet, and $+2$ jets or more , respectively:
\begin{align}
\left( \begin{array}{ccc} 
\phantom{-}(26.6\%) ^2             &  -(28.3\%) ^2             &   \phantom{-(0}0\phantom{.0)\%^2} \\
-(28.3\%) ^2                       & \phantom{-}(41.8\%) ^2      &   -(20.5\%)^2                          \\
 \phantom{-(0}0\phantom{.0)\%^2}      &           -(20.5\%) ^2      &  \phantom{-}( 33.0\% ) ^2                \\
\end{array}\right)
\end{align}
The PDF uncertainties
for ${gg\to H}$ production, obtained using the prescription from
Refs.~\cite{bib:ggH01jetUncert,bib:gluon_fusion_xsec}, are 
7.6\%, 13.8\%, and 29.7\% for  the exclusive production of  ${gg\to H}+0$ jet, $+ 1$ jet, and $+ 2$ jets or more, respectively.

We also consider sources of systematic uncertainty that affect the
shape of the final discriminant distribution (and we quote here the
average fractional uncertainty across bins of the final discriminant
distribution for all backgrounds): jet energy scale ($4\%$); jet
resolution ($0.5\%$); jet identification ($2\%$); jet association to the hard-scatter 
primary $p\bar{p}$ interaction vertex ($2\%$); $b$-tagging ($<2\%$);
and $W$+jets  modeling (10\%--30\%), depending on jet multiplicity bin and final state.
The systematic uncertainties due to the modeling of $\pt(WW)$ and $\Delta \phi$ between
          leptons, and the $\pt$ of the vector boson from the $V$+jets production, are at the level of
                    $<1\%$ and taken into account.

\section{Measurement of the non-resonant $\mathbf{p\bar p\to WW}$ cross section}

To validate the analysis techniques  employed to search for the Higgs boson,
a measurement of the non-resonant $WW$ production cross section is performed.
This is motivated by the fact that $WW$ production yields similar particle content
and topology as the Higgs boson signal.
The same analysis methods are employed as for the Higgs bosons searches,
and same sources of systematic uncertainty are accounted for,
but the outputs of the $WW$ discriminants, described in the
``Analysis using decision trees'' section, are considered.
The $WW$ cross section is obtained as the result of a  maximum likelihood fit to the data, with maximization
over the $WW$ signal normalization and over the systematic uncertainties treated as nuisance parameters,
as  for the SM Higgs boson search results described in the next section.
The measurement is carried out using discriminants from the three dilepton final states,
in the 0- and 1-jet multiplicity bins.
Figure~\ref{fig:bk_sub_WW} shows the combined output distribution of these discriminants~\cite{bib:aux}, rebinned according to $s/b$ and
after the expected backgrounds have been subtracted.
The measured value  $\sigma_{WW}=11.6\pm 0.4\, \mathrm{(stat)}\pm 0.6\, \mathrm{(syst)}$~pb~\cite{bib:aux}  is in agreement with the SM
prediction of $11.3\pm0.7$ pb~\cite{bib:dibo-xs}. 
The possible presence of a 
SM Higgs boson  of 125~\egev\ in the data is not accounted for,
but it is expected to bias this measurement upward
by $\sim 0.1$~pb.

\begin{figure}[h]
  \begin{center}
    \includegraphics[height=0.25\textheight]{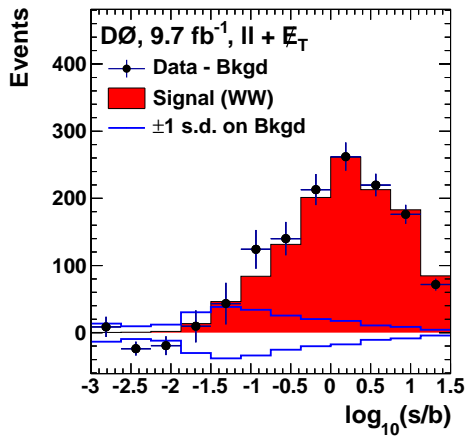}
  \end{center}
  \caption{[color online]\label{fig:bk_sub_WW} The post-fit
    background-subtracted data distribution for the final
    discriminant, summed in bins with similar signal to background
    ratios, for the $WW$ cross section measurement. The uncertainties
    shown on the background-subtracted data points are the square
    roots of the post-fit predictions for signal plus background events in each bin,
    representing the expected statistical uncertainty on the
    data points. Also shown is the $\pm 1$~s.d.\ band on
    the total background after fitting.
}
\end{figure}

\begin{figure*}[!]
\center
  \includegraphics[ height=0.236\textheight]{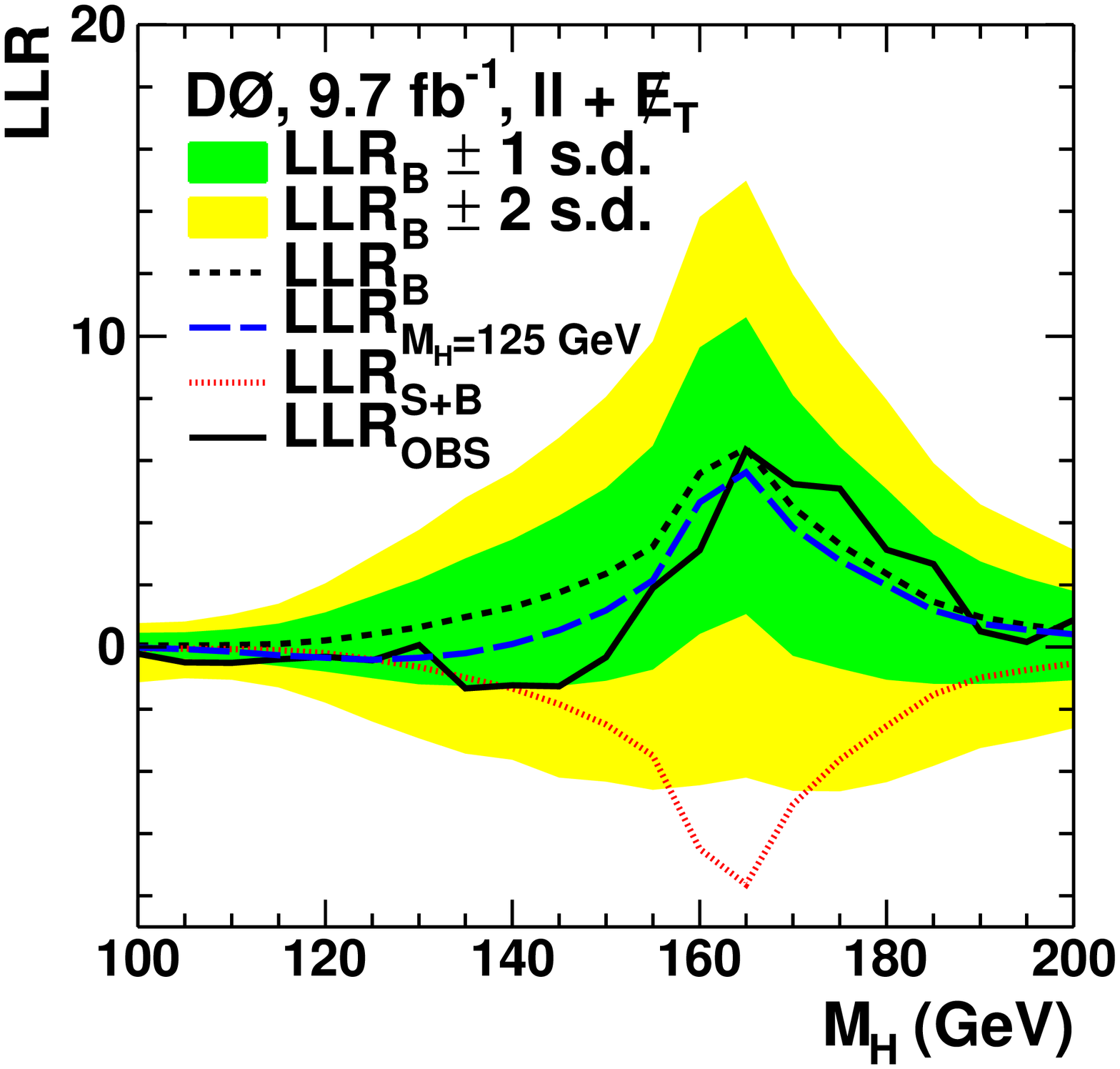} 
  \includegraphics[ height=0.236\textheight]{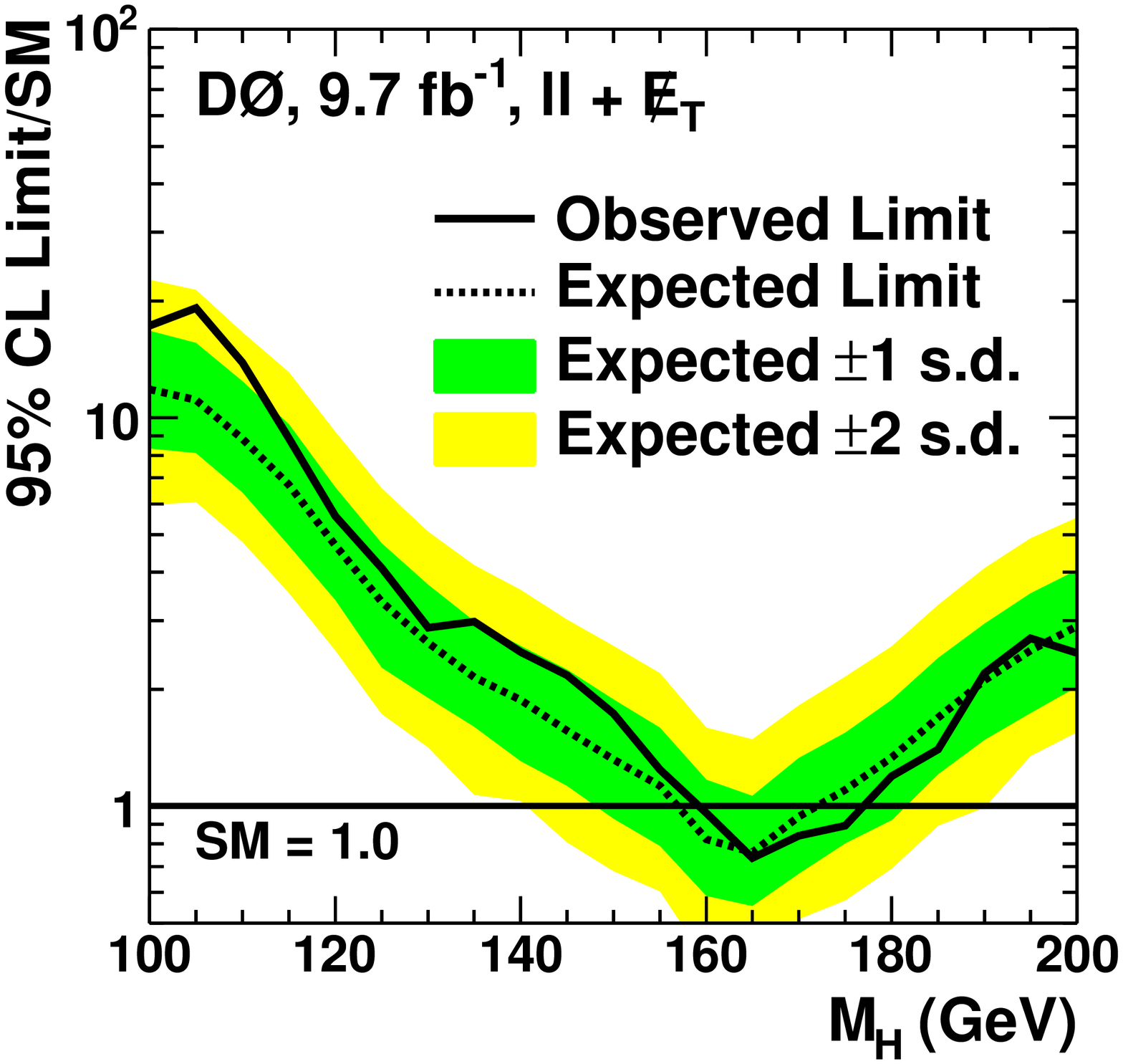} 
  \includegraphics[height=0.236\textheight]{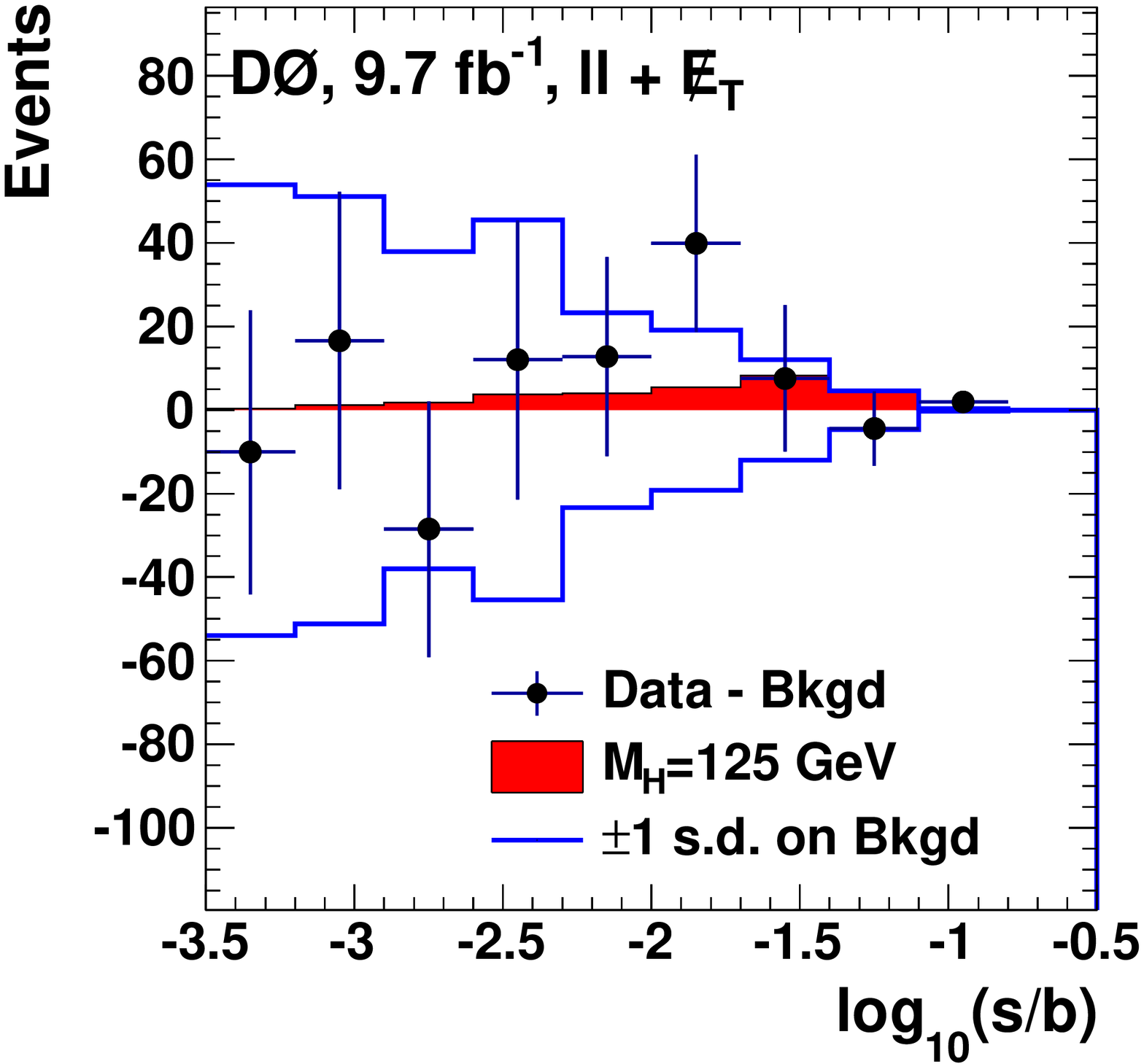}
\unitlength=1mm
\begin{picture}(00,00)

\if \mytwocolumn 1
\put(-130,48){\text{\bf (a)}} 
\put(-71,48){\text{\bf (b)}} 
\put(-11,48){\text{\bf (c)}}
\else
\put(-41,103){\text{\bf (a)}} 
\put(18,103){\text{\bf (b)}}
\put(-12,47){\text{\bf (c)}}
\fi

\end{picture}
  \caption{[color online]\label{fig:alllimit} \label{fig:allllr}
   (a) The observed LLR  as a function of $M_H$.
    Also shown are the 
    expected LLRs  for the background-only hypothesis,
    for the signal+background hypothesis,
    and the expectation in the presence of a signal of $M_H=125$~\egev.
    (b) Excluded cross section,
    $\sigma(p\bar{p}\rightarrow H+X)$, at the 95\% C.L.\ in units
    of the SM cross section as a function of $M_H$. 
    In (a) and (b),  the green and yellow shaded bands indicate
    $\pm1$ and $\pm2$ s.d.\ uncertainties of the expected observation
    for the background-only hypothesis, respectively.
    (c) Background-subtracted data distribution for the final discriminants,
    summed in bins with similar signal to background ratios, for $M_H=125$~GeV. The uncertainties shown on the
background-subtracted data points are the square roots of the post-fit
background predictions of number of events in each bin, representing the expected
statistical uncertainty on the data points. Also shown is the $\pm$1
s.d.\ band on the total background after fitting.
}
\end{figure*}

\section{\label{sec:result}SM Higgs boson search results}

Table~\ref{tab:final_cutflow} and Fig.~\ref{fig:sel_all}(c) demonstrate
good agreement between the data and the predicted background,
in both the numbers of selected events and the distributions
of final discriminants.
The modified frequentist $CL_s$ method~\cite{bib:modified_freq} is
employed to set limits on SM Higgs boson production,
where the test statistic is a log-likelihood ratio (LLR)
for the background-only and signal+background hypotheses.
The LLR is obtained by summing the LLR values of the bins of the 14 BDT outputs
from the different sub-channels.
In the LLR calculation the signal and background rates
are functions of the systematic uncertainties which are taken into
account as nuisance parameters with Gaussian priors.
Their degrading effect is reduced by
fitting signal and background contributions to the data by 
maximizing the profile likelihood function for the background-only and
signal+background hypotheses separately, appropriately
taking into account all correlations between the systematic
uncertainties~\cite{bib:collie}.

Figure~\ref{fig:allllr}(a)  shows the LLR values as a function of the tested Higgs boson
mass hypothesis. The LLR values expected in the absence of signal and in the presence of a SM Higgs boson of mass 125~\egev\ are also
displayed for comparison.
Figure~\ref{fig:alllimit}(b)~\cite{bib:aux}
presents expected and observed upper limits for $\sigma(p\bar{p} \rightarrow H + X)$ relative to SM predictions.
For  $M_{H}=165$~\egev\ (125~\egev), the expected limit is 0.76 (3.4)
times the SM prediction and the observed limit reaches
0.74 (4.1) in the same units. 
A SM Higgs boson in the mass range  $159 < M_{H} < 176$~\egev\ is
excluded at the 95\% C.L. while the expected exclusion sensitivity is  $156 < M_{H} < 172$~\egev.
In these figures, a slight excess of signal-like candidates yields a limit roughly one s.d.\ above the background expectation,
in the mass range $100<M_H<145$~\egev. Figure~\ref{fig:alllimit}(c) shows
a comparison of the BDT output distributions, sorted as a function of signal over background ratio,
expected for the signal of $M_H=125$~\egev, and observed in the data after subtracting the fitted backgrounds.

\section{{Upper limit on \lowercase{$\boldsymbol{ gg}$}$\boldsymbol{ \to H\rightarrow WW}$ and Fourth generation fermion interpretation}\label{sec:fourth_gen}}

Additional generations of fermions can occur naturally in models of
grand unification, CP violation, gauge-mediated supersymmetry breaking,
and others. Measurements of the $Z$ boson decay
width~\cite{zwidth} exclude models in which the fourth neutrino mass
eigenstate is lighter than $45$~\egev, but fourth generation models can
still be accommodated for a large fourth-generation neutrino mass.
Production of $gg\to H$ occurs via top-quark loops in the SM. 
With respect to the SM, the quarks from the fourth generation
will provide additional contributions to the quark loop diagram, enhancing production by a factor of 7 to 9, depending on their masses and the Higgs boson mass~\cite{bib:4thGenAnastasiou,bib:4thGenKribs,bib:4thGenArik}.  
A previous combined \dzero\ and CDF result using up to 5.4\,fb$^{-1}$ of data excluded
the existence of a SM-like Higgs boson  in the
mass range between 131~\egev~and 204~\egev~\cite{bib:tev_fourthGen}, assuming  the presence of
a fourth sequential generation of fermions with large masses.
Similar searches have been conducted by the ATLAS and CMS Collaborations, yielding and exclusion of $140 < m_H < 185$~\egev~\cite{ATLAS_fourthGen} and  $144 < m_H < 207$~\egev~\cite{CMS_fourthGen}, respectively.

To test such models, we derive upper limits on the $gg\to H \to WW$ production cross section.
The same analysis as described in the previous sections is performed,
but the  VBF and $VH$ contributions are excluded
from the overall signal yield when constructing the LLR.
The upper limits  are reported in Fig.~\ref{fig:alllimit_4thGen}~\cite{bib:aux},
compared to the expected yield of the $gg\to H$
production in two models of fourth generation fermions.
In the ``low-mass'' scenario, the masses of the fourth generation charged lepton
and neutrino are assumed to be respectively  $m_{\ell4}=100$~GeV and $m_{\nu4}=80$~GeV,
just beyond the experimental limits, which yields a reduction by up to 15\%
in the branching ratio for $H\to WW$.
On the contrary, in the ``high-mass''
scenario, where $m_{\ell4}=m_{\nu4}=1$~TeV, the leptons are too heavy to
contribute to the Higgs boson decay width and the branching ratio for $H\to WW$ remains basically unchanged relative to the SM branching ratio.
For both scenarios, the masses of the
fourth-generation down-type ($m_{d4}$) and up-type ($m_{u4}$) quarks
are fixed to $m_{d4}=400$~GeV and $m_{u4}=450$~GeV
\cite{bib:4thGenKribs,bib:4thGenAnastasiou}.
From this figure, we derive exclusion of  the Higgs boson mass range  $125 < M_{H} < 218$~\egev\ and  
$125 < M_{H} < 228$~\egev, in the low-mass  and high-mass scenarios, respectively. 


\begin{figure}[h]
\begin{center}
\includegraphics[height=0.25\textheight]{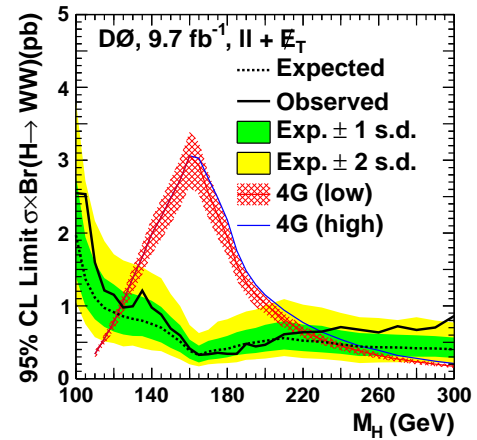}
\end{center}
\caption{\label{fig:alllimit_4thGen} [color online] Excluded cross section $\sigma(gg\to H)\times BR(H\to WW)$ 
in pb as a function of $M_H$ using all channels. The red
and blue lines correspond to the theoretical prediction for a
sequential fourth generation assumption in the ``low mass'' and ``high
mass'' scenarios, respectively (see main text). The hatched region corresponds to
the PDF and scale uncertainties on the fourth generation ``low mass''
prediction~\cite{bib:4thGenAnastasiou}.
  The green and yellow shaded bands indicate
    $\pm 1$ and $\pm 2$ s.d. uncertainties of
    the background-only hypothesis, respectively.
}
\end{figure}

\section{{ Fermiophobic Higgs boson interpretation}}

The mechanism of electroweak symmetry breaking may offer a richer
phenomenology than expected in the SM.
Several Higgs bosons may exist, or the Higgs boson(s) may have
couplings different from those predicted by the SM.
In this section, we explore the possibility
that the lightest
Higgs boson does not couple to fermions at the tree level,
but still behaves like the SM  Higgs boson for its other properties,
in particular for the coupling to vector bosons.
In this model, the VBF and $VH$ production have the same cross sections as in the SM.
The main consequences of the vanishing fermion couplings are the
suppression of production via gluon fusion, 
and the enhancement of the branching ratios to vector bosons,
$H\to WW$, $H\to ZZ$, and $H\to\gamma\gamma$,
particularly sizeable below the $WW$ threshold,  $m_H< 160$~\egev.
To provide masses to the fermions, additional degrees of freedom must
exist in the Higgs sector, as predicted in models with Higgs doublets
or triplets~\cite{fhm1}, but it is assumed that those other particles
do not have phenomenological impact in our search. In this model, the CMS and ATLAS collaborations exclude 
a Higgs boson in the mass range 
$110 < m_H < 194$~\egev~\cite{cms_fhm,atlas_fhm},
while masses below 110~\egev{} are excluded by LEP experiments~\cite{aleph_fhm,delphi_fhm,l3_fhm,opal_fhm}
and Tevatron experiments~\cite{cdf_fhm,d0_fhm}.
The same analysis steps are performed as described for the SM Higgs boson searches,
but the various BDTs are retrained,  accounting for the
fermiophobic Higgs branching ratios, computed using {\sc hdecay}, the VBF and $VH$ production at the SM rate, and the suppression of $gg\to H$ production.
The data are in good agreement with background expectation and
upper limits on the fermiophobic Higgs are derived, following the
same method as for the SM Higgs.
They are reported in  Fig.~\ref{fig:FHlimit}~\cite{bib:aux}.
We obtain a cross section upper limit of  3.1 times the fermiophobic Higgs boson production cross section for $M_H=125$~\egev, while the expected sensitivity is 2.5.

\begin{figure}[h]
\begin{center}
\includegraphics[height=0.25\textheight]{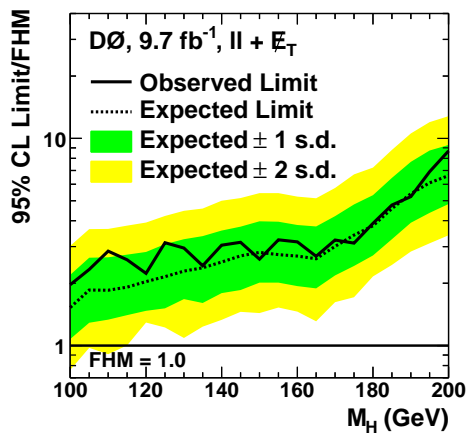}
\end{center}
\caption{\label{fig:FHlimit} [color online]  Excluded cross section,
    $\sigma(p\bar{p}\rightarrow H+X)$, as a function of $M_H$ using all channels, in units
    of the Higgs boson  production rate expected from the fermiophobic Higgs boson model (FHM) described in the main text.
}
\end{figure}

\section{\label{sec:conclusion}CONCLUSIONS}

We have performed a search for Higgs boson production using final
states with two oppositely charged leptons and large missing
transverse energy in the {\em}, \ee, and~\mm\ channels.
To validate our search methodology we have measured the non-resonant
$WW$ production cross section, which yields $\sigma_{WW}=11.6\pm 0.7$~pb,
in good agreement with the SM prediction of $11.3\pm 0.7$~pb.
For the Higgs boson searches, we observe agreement between data and the expected backgrounds.
We set upper limits on SM Higgs boson
production at the 95\% C.L.
that exclude the mass range  $159 < M_{H} < 176$~\egev,
while the expected exclusion sensitivity is $156 < M_{H} < 172$~\egev.
For a mass hypothesis of $M_H=125$~\egev, we exclude 4.1 times the
expected SM Higgs boson production cross section, while the expected sensitivity is 3.4.
This upper limit is compatible with the presence of a SM Higgs boson of  $M_H=125$~\egev.
We also interpret our search results as cross section upper limit for $gg\to H$ production,
which allows us to exclude the mass range  $125 < M_{H} < 218$~\egev\ in the context of a fourth generation of fermions.
The search results are also interpreted in the framework of a fermiophobic Higgs boson,
which yields an exclusion of 3.1 times the fermiophobic Higgs
boson production rate for $M_H=125$~\egev, while the expected sensitivity is 2.5.

\section{\label{sec:ackn}ACKNOWLEDGMENTS}

%
We thank the staffs at Fermilab and collaborating institutions,
and acknowledge support from the
DOE and NSF (USA);
CEA and CNRS/IN2P3 (France);
MON, NRC KI and RFBR (Russia);
CNPq, FAPERJ, FAPESP and FUNDUNESP (Brazil);
DAE and DST (India);
Colciencias (Colombia);
CONACyT (Mexico);
NRF (Korea);
FOM (The Netherlands);
STFC and the Royal Society (United Kingdom);
MSMT and GACR (Czech Republic);
BMBF and DFG (Germany);
SFI (Ireland);
The Swedish Research Council (Sweden);
and
CAS and CNSF (China).
%

\clearpage
\begin{figure*}[!]
{ \vspace{10cm}
\huge \bf Auxiliary Material
\vspace{2cm}}

To appear as an Electronic Physics Auxiliary Publication (EPAPS).
\end{figure*}
\onecolumngrid\clearpage

Event yields after the preselection are shown in
Table~\ref{tab:presel_cutflow_full}. The distributions of the dilepton
invariant mass (Fig.~\ref{fig:aux_presel_M}), angular separation
        between the leptons (Fig.~\ref{fig:aux_presel_dr}) and missing
transverse energy (Fig.~\ref{fig:aux_presel_met}) show the good
agreement between the data and the simulation after the preselection
in the \ee, \em\ and \mm\ channels.
Figures~\ref{fig:aux_DYDT_ee} and \ref{fig:aux_DYDT_mm} show the
distributions of the Drell-Yan (DY) BDT discriminants for the \ee\ and
\mm\ channels respectively. Figure~\ref{fig:aux_minMt_em} shows the
distributions of the variables \mtmin{} and $M_{T2}$, used by the \em\
channel to reject the DY background. The distributions of the dilepton
invariant mass (Fig.~\ref{fig:aux_finsel_M}), angular separation
        between the leptons (Fig.~\ref{fig:aux_finsel_dr}) and missing
transverse energy (Fig.~\ref{fig:aux_finsel_met}) are shown for the
\ee, \em\ and \mm\ channels after the final selection. The final
discriminant distributions for the three channels are shown in
Figs.~\ref{fig:aux_FDDT_ee}, \ref{fig:aux_FDDT_em} and \ref{fig:aux_FDDT_mm}.
Figure~\ref{fig:bk_sub_higgs} shows the background-substracted data
distributions of the final discriminants for $M_H = 125\,\egev$ and
$M_H = 165\,\egev$. Tables~\ref{tab:alllimit},
\ref{tab:alllimit_4thGen} and \ref{tab:FHlimit} give the expected and
observed upper limits at the 95\% C.L. for Higgs boson production in
the SM, in models with a fourth generation of fermions and assuming
fermiophobic couplings, respectively. Figures~\ref{fig:aux_WWDT_ee},
\ref{fig:aux_WWDT_em} and \ref{fig:aux_WWDT_mm} show the $WW$
discriminant distributions for the \ee, \em\ and \mm\ channels.
Figure~\ref{fig:WW_disc} shows the distribution of the combination of
$WW$ discriminants for the \ee, \em\ and \mm\ channels. The
results of the $p\bar{p}\to WW$ cross section measurement for the
\ee, \em\, \mm\ channels, and their combination are shown in
Fig.~\ref{fig:wwXsec}.

\begin{figure*}[!]
\section*{\Large Tables and distributions after the preselection}
\end{figure*}

\begin{table*}[tb]
\begin{center}
\caption{\label{tab:presel_cutflow_full}
Expected and observed numbers of events after the preselection in
   the  \ee, \em, and \mm\ final states in the different jet multiplicity bins.  The signal is for a Higgs
   boson mass of 125~GeV.
}
\end{center}
\begin{ruledtabular}
\begin{tabular}{ccrllcccccccc}
 & Data & \multicolumn{3}{c}{Total background} & Signal & $Z\to ee$ & $Z\to \mu\mu$ & $Z\to \tau\tau$ & $\ttbar$ & $W$+jets & Diboson & Multijet \\

\hline
\elel{}: &&&&&&&&&&\\

0 jet     & 572831 &575445   &$\pm$ &11509 &\phantom{0}7.1 & 566846 & \phantom{00000}-- & 4727 & \phantom{0}15  & 623 & 517 & 2718 \\
1 jet      &  \phantom{0}75326 & 77130   &$\pm$ &4628  &\phantom{0}5.0 & \phantom{0}75162  & \phantom{00000}-- & \phantom{0}663  & \phantom{0}66  & 143 & 243 & \phantom{0}853 \\
$\ge 2$ jets & \phantom{0}11413 & 11885 &$\pm$ &1783  &\phantom{0}4.0 & \phantom{0}11256  & \phantom{00000}-- & \phantom{0}105  & 129 & \phantom{0}29  & 185 & \phantom{0}181 \\

\hline 
\elmu{}: &&&&&&&&&&\\

0 jet          &\phantom{0}12131 &12361 &$\pm$ &247  &\phantom{0}9.7  &\phantom{000}348 &\phantom{00}1043 &7546 &\phantom{0}16  &972 &728 &1709 \\
1 jet          &\phantom{00}2039  &2040  &$\pm$ &122  &\phantom{0}4.6  &\phantom{0000}51  &\phantom{000}139  &\phantom{0}946  &155 &191 &153 &\phantom{0}406 \\
$\ge 2$ jets   &\phantom{000}766   &741   &$\pm$ &111  &\phantom{0}2.3  &\phantom{00000}9   &\phantom{0000}29   &\phantom{0}180  &366 &\phantom{0}63  &\phantom{0}25  &\phantom{00}70  \\

\hline
\mumu{}: &&&&&&&&&&\\

0 jet            &699513 &701663 &$\pm$ &14033  &\phantom{0}8.9  & \phantom{00000}--   &693390 &5663  &\phantom{00}9   &343 &673   &1585   \\
1 jet            &\phantom{0}95615  &98840  &$\pm$ &5930   &\phantom{0}5.3  &\phantom{00000}--   &\phantom{0}97278  &\phantom{0}686   &\phantom{0}87  &\phantom{0}78  &329   &\phantom{0}382   \\
$\geq 2$ jets    &\phantom{0}16421 &17766   &$\pm$ &2665   &\phantom{0}4.5  &\phantom{00000}--   &\phantom{0}16974  &\phantom{0}110   &260 &\phantom{0}16  &313   &\phantom{00}94   \\

\end{tabular}
\end{ruledtabular}
\end{table*}

\begin{figure*}[!]

\includegraphics[ height=0.237\textheight]{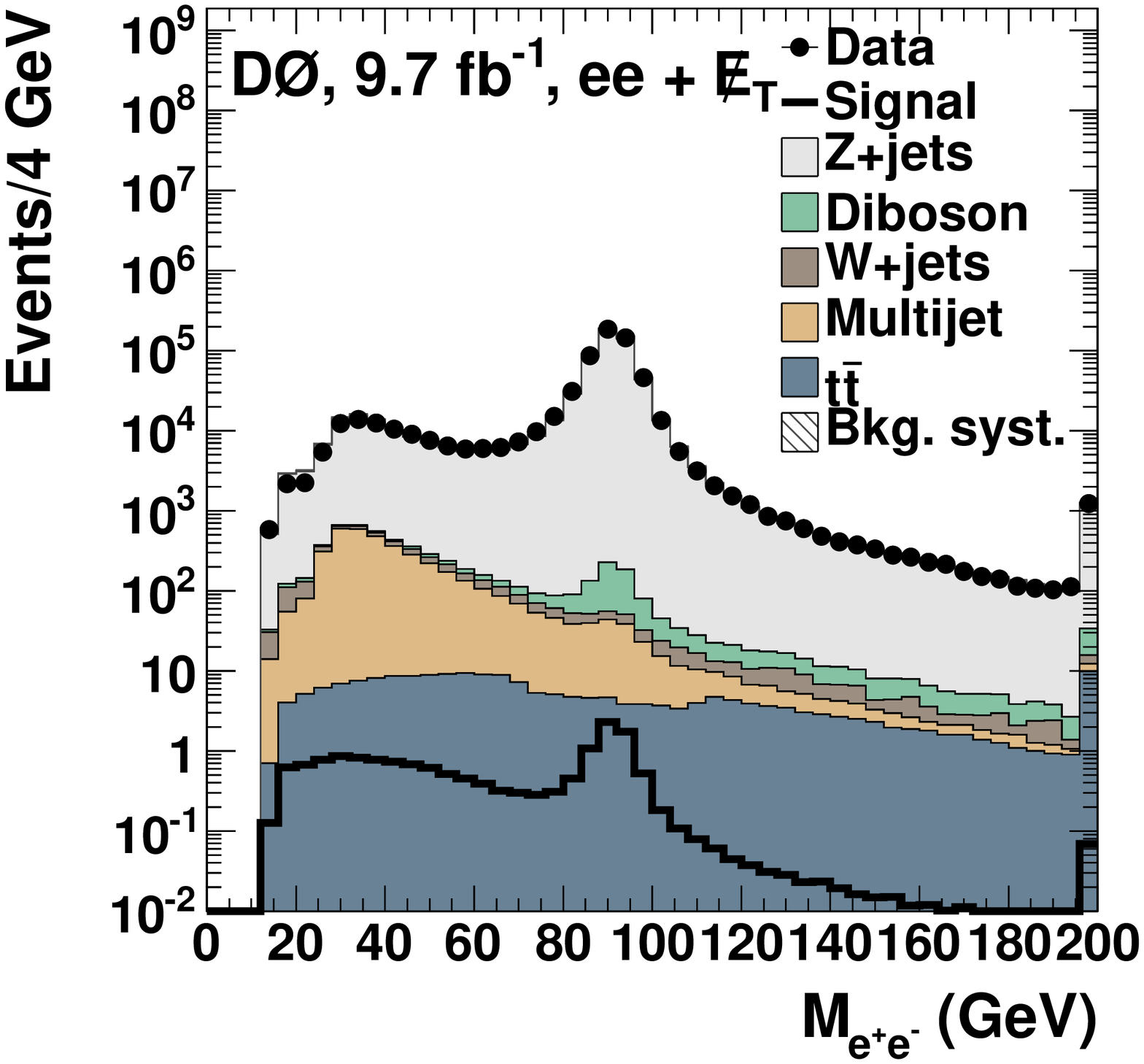}
\includegraphics[height=0.237\textheight]{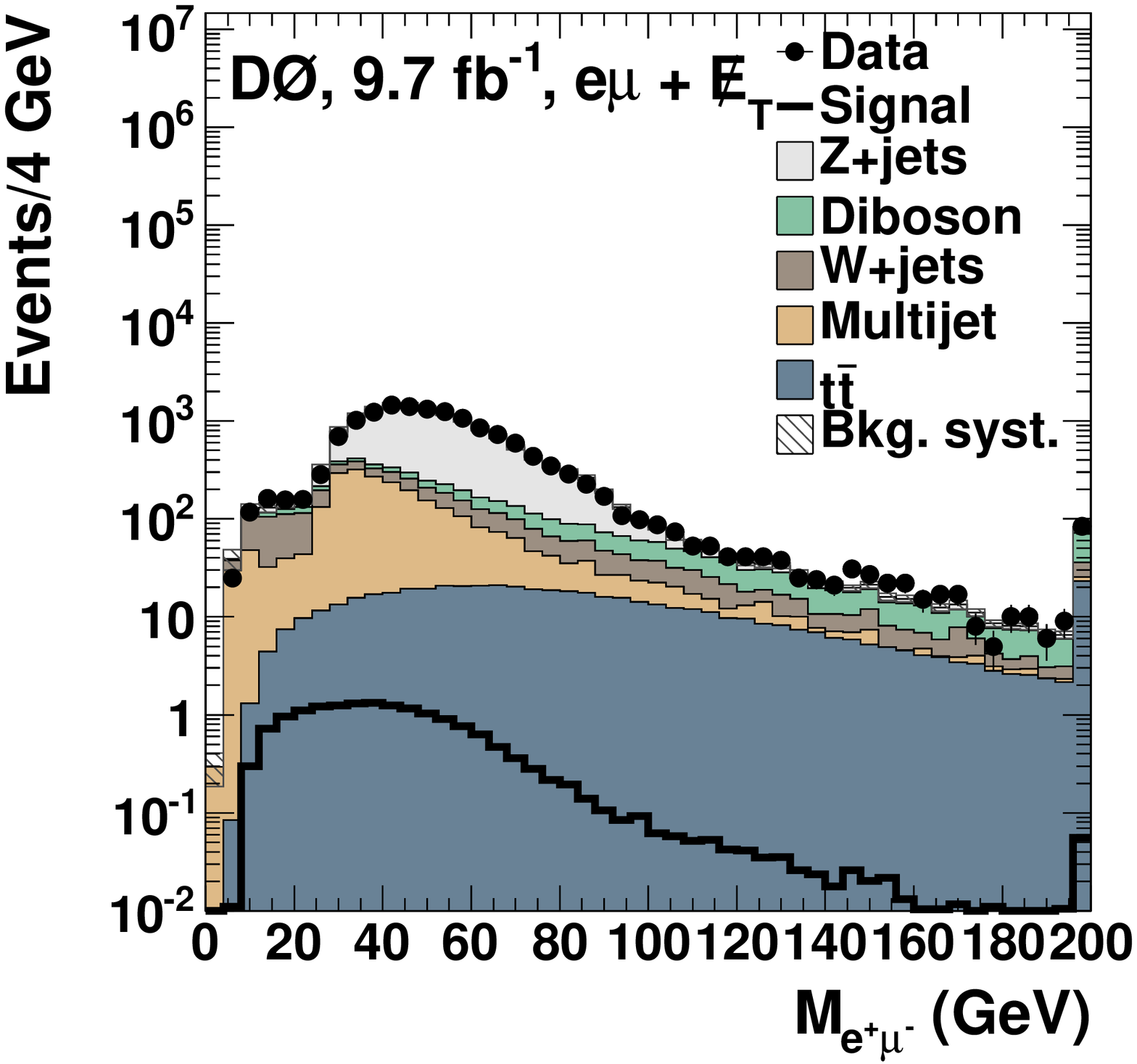}
\includegraphics[ height=0.237\textheight]{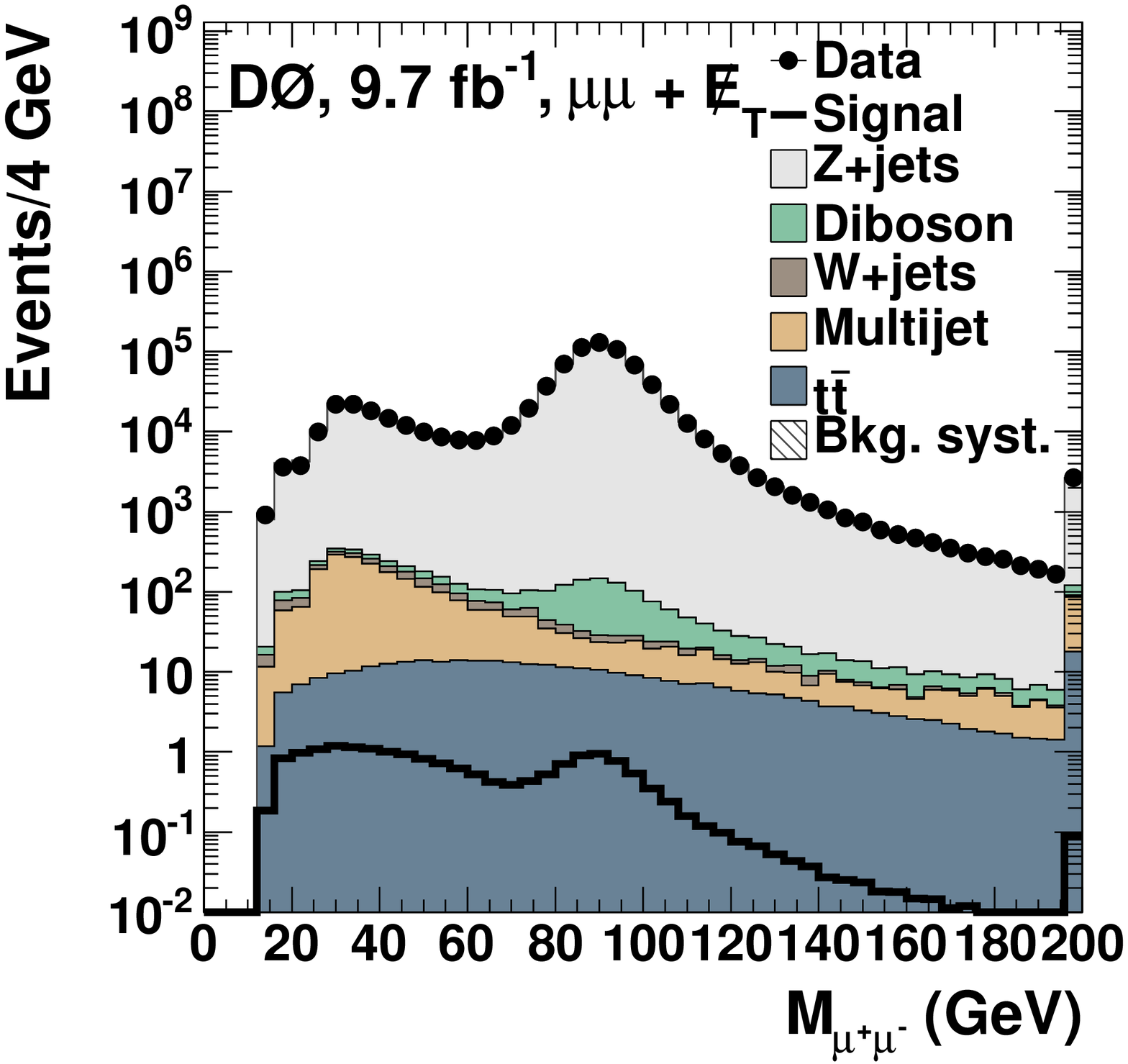}
\unitlength=1mm

\begin{picture}(00,00)

\if \mytwocolumn 1
\put(-77,50){\text{\bf (a)}} 
\put(-17,50){\text{\bf (b)}} 
\put(43,50){\text{\bf (c)}}
\else
\put(-47,106){\text{\bf (a)}} 
\put(-17,50){\text{\bf (c)}}
\put(13,106){\text{\bf (b)}}
\fi

\end{picture}

\caption{
  Distributions of the  dilepton invariant mass for the (a) \ee\ channel,  (b) \em\ channel,
  and (c) \mm\ channel  after the preselection.
  In (a), (b) and (c) the last bin includes all events above the upper bound of the histogram.
  In these plots, the hatched bands show the total  systematic uncertainty on the background predictions, and
  the signal distributions are those expected from a  Higgs boson of mass $M_H=125$ GeV.
}
  \label{fig:aux_presel_M}
\end{figure*}

\begin{figure*}[!] 
 
\includegraphics[ height=0.237\textheight]{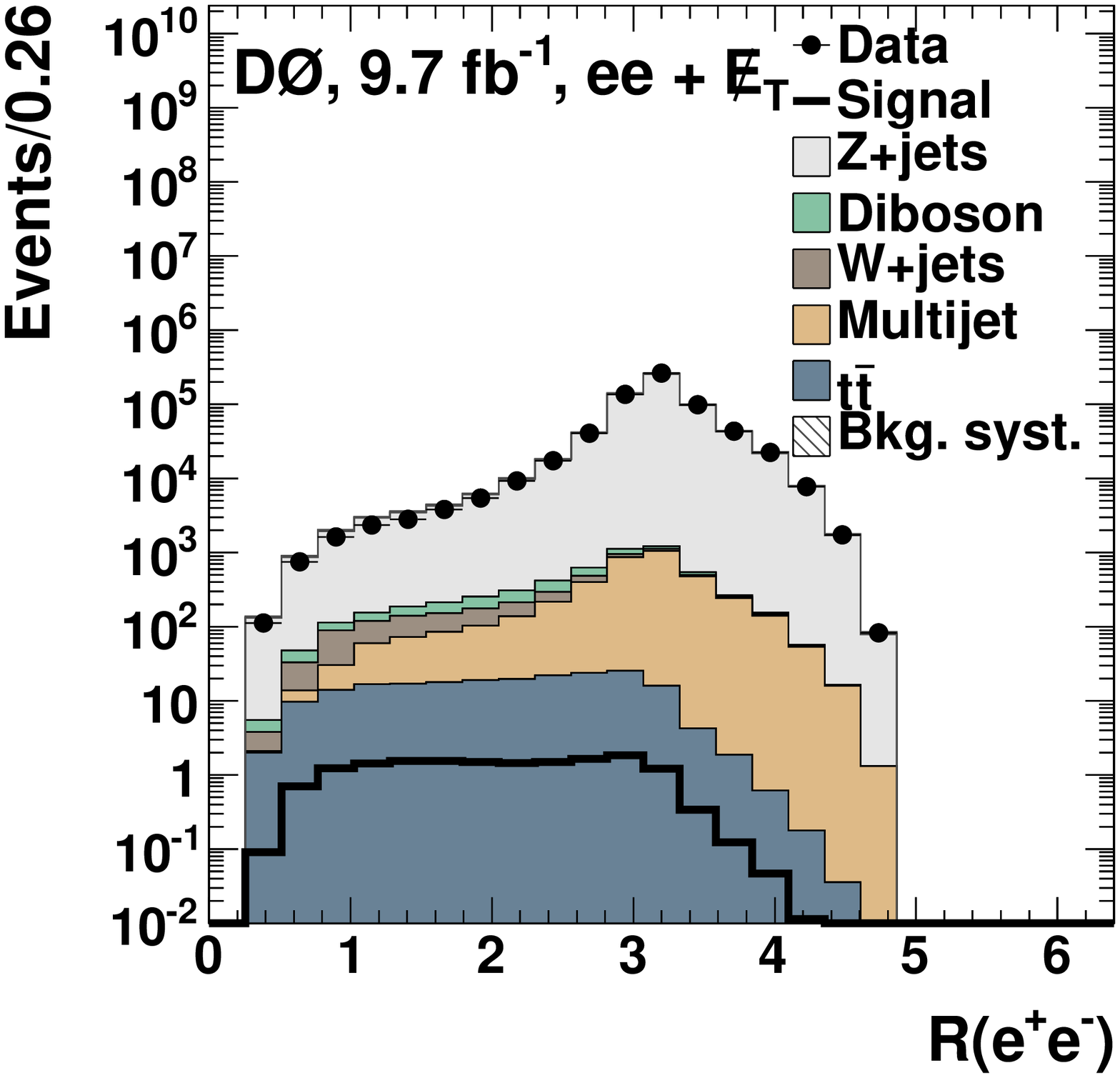}
\includegraphics[height=0.237\textheight]{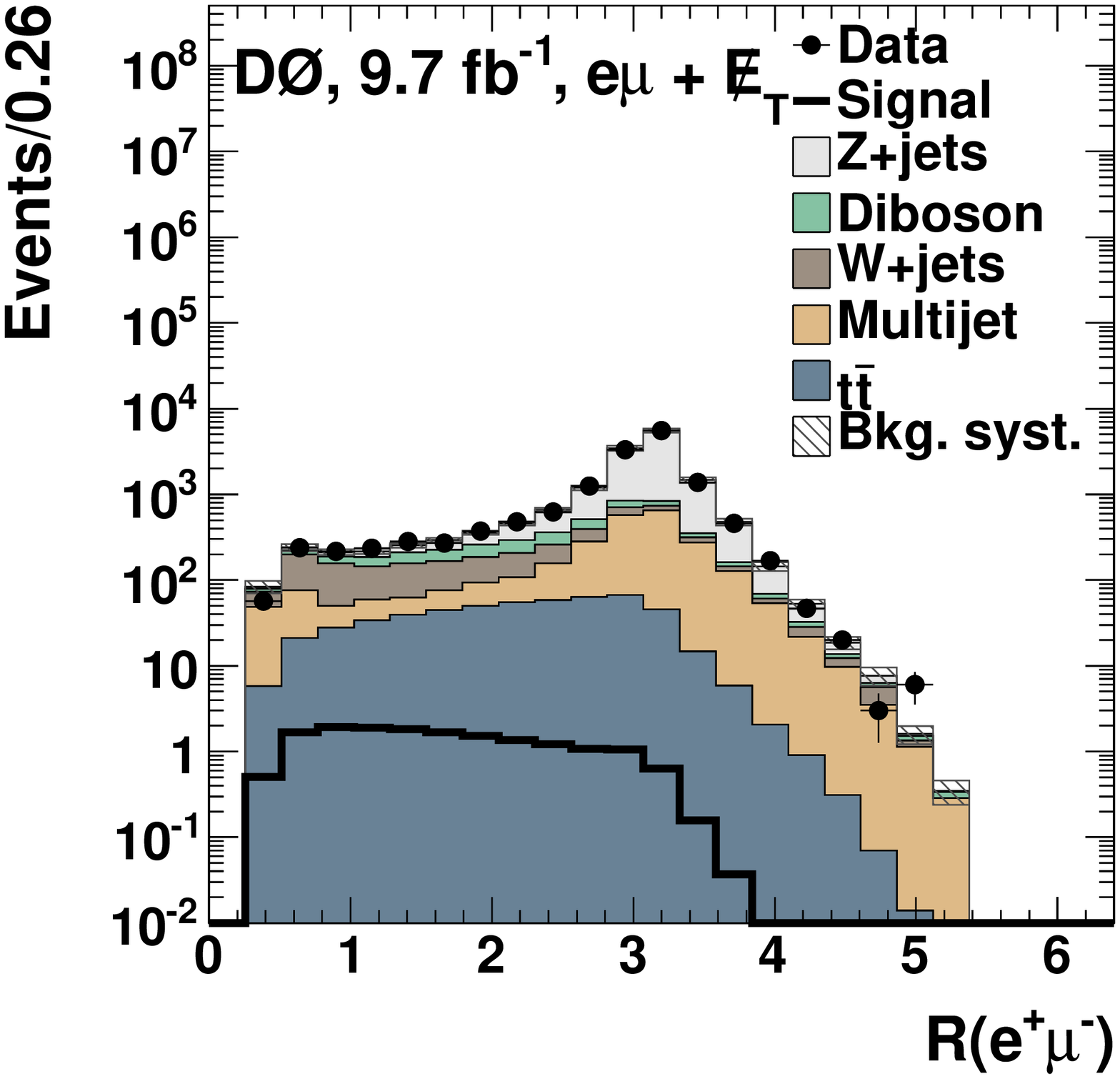}
\includegraphics[ height=0.237\textheight]{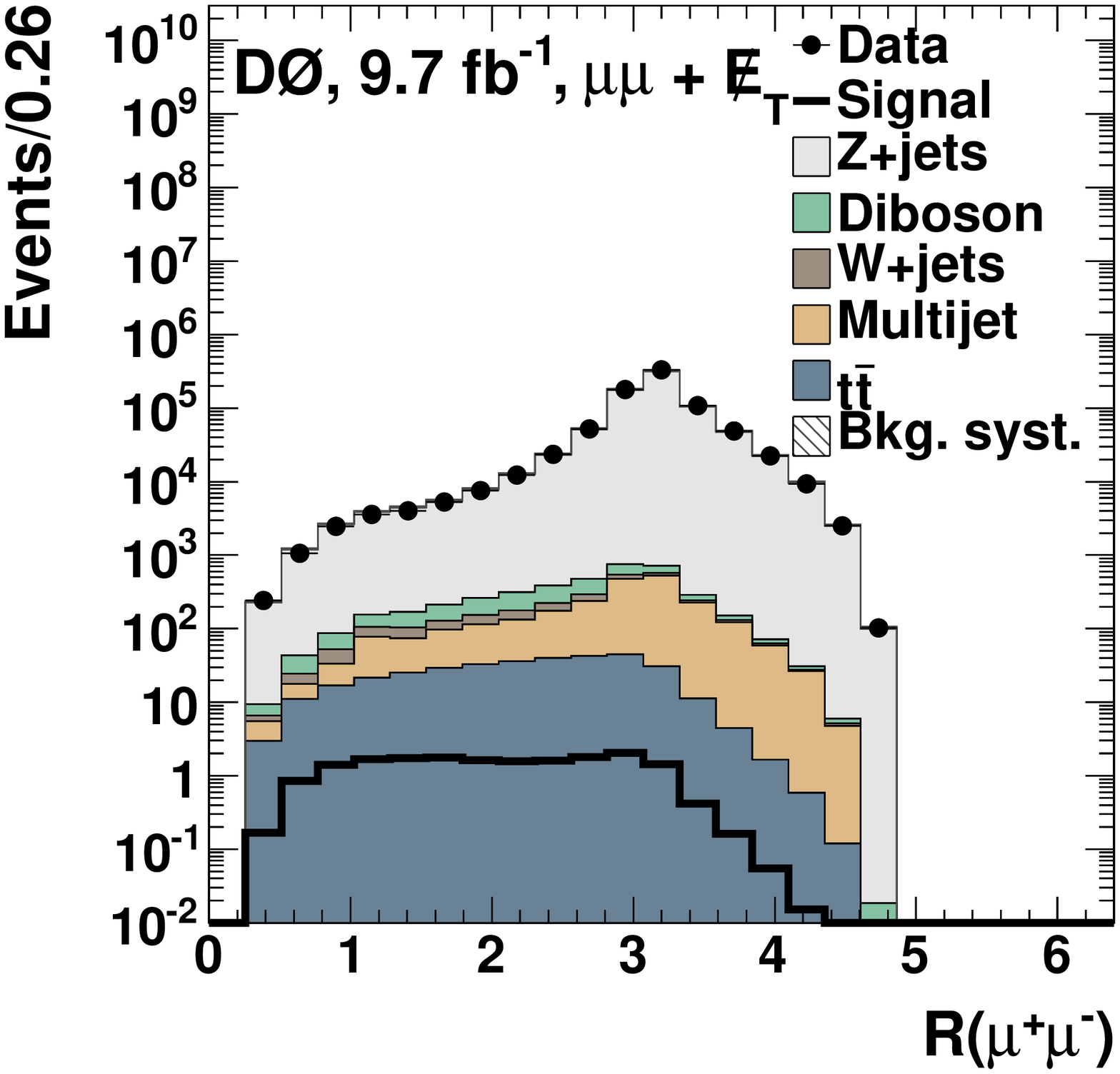}
\unitlength=1mm

\begin{picture}(00,00)
\if \mytwocolumn 1
\put(-77,50){\text{\bf (a)}} 
\put(-17,50){\text{\bf (b)}} 
\put(43,50){\text{\bf (c)}}
\else
\put(-47,106){\text{\bf (a)}} 
\put(-17,50){\text{\bf (c)}}
\put(13,106){\text{\bf (b)}}
\fi

\end{picture}

\caption{
  Distributions of the angular separation $\cal R(\ell,\ell)$ between the leptons for the (a) \ee\ channel, 
   (b) \em\ channel,
  and (c) \mm\ channel  after the preselection.
  In these plots, the hatched bands show the total  systematic uncertainty on the background predictions, and
  the signal distributions are those expected from a  Higgs boson of mass $M_H=125$ GeV.
}
  \label{fig:aux_presel_dr}
\end{figure*}

\begin{figure*}[!]
 
\includegraphics[ height=0.237\textheight]{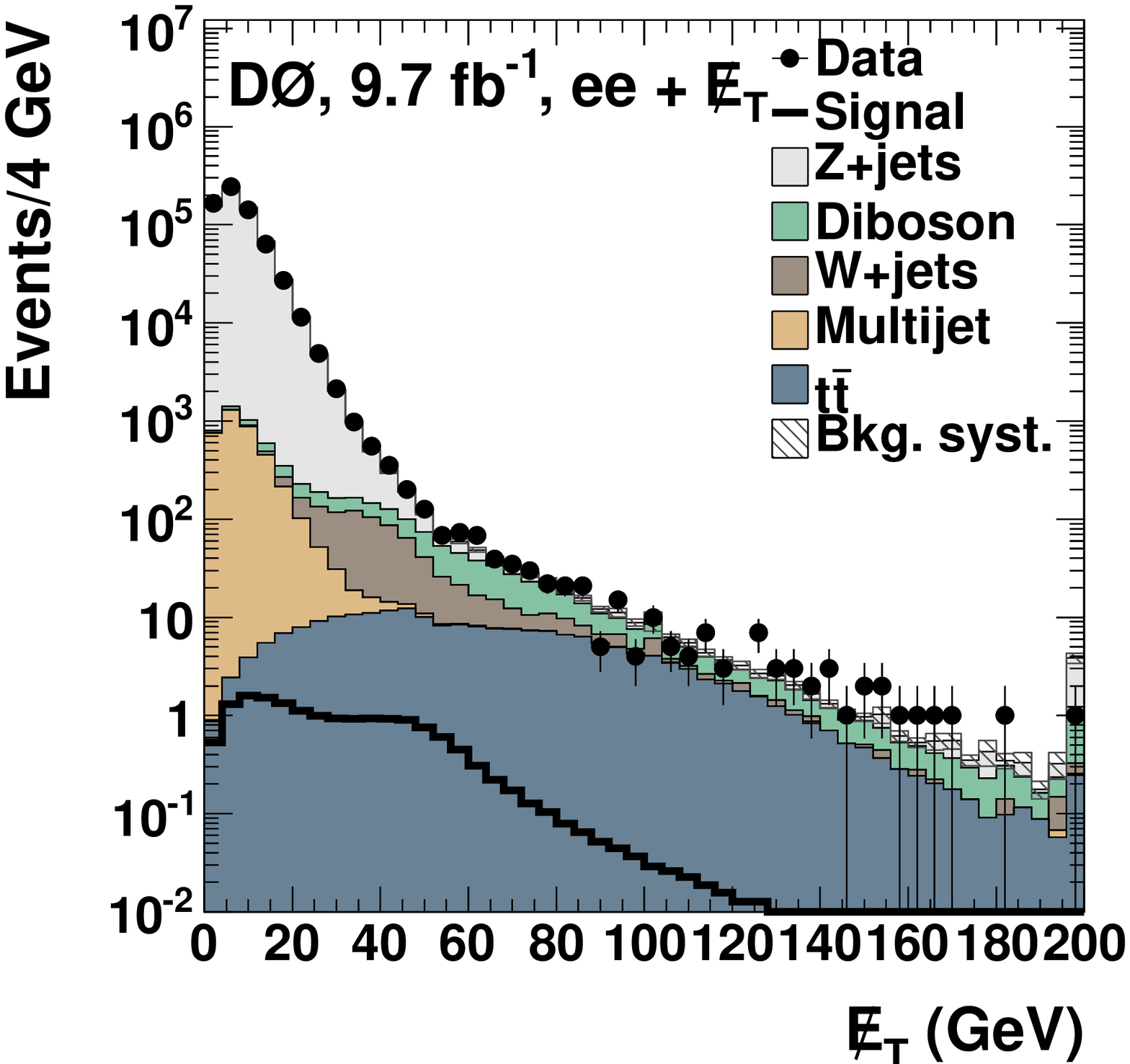}
\includegraphics[height=0.237\textheight]{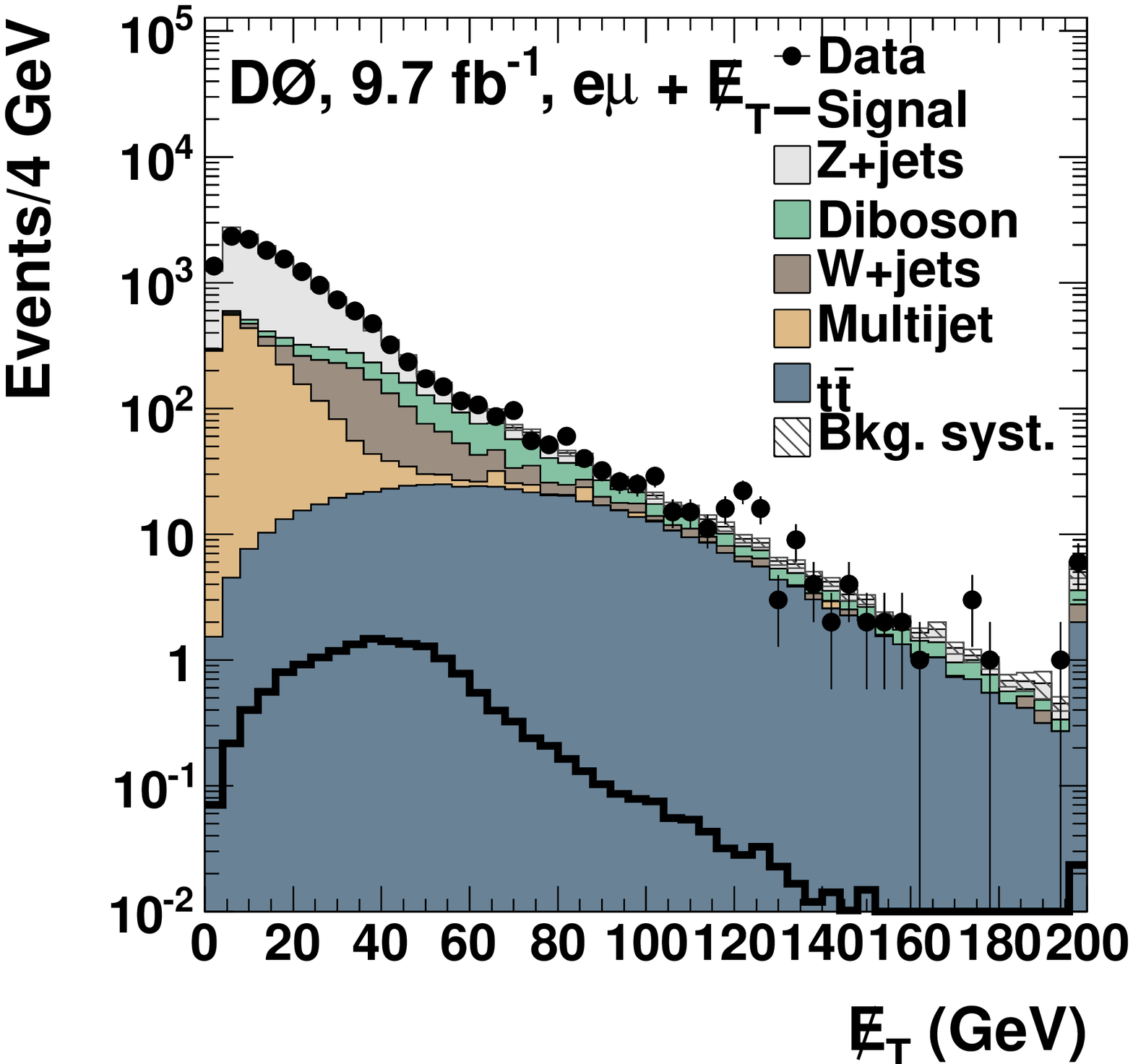}
\includegraphics[ height=0.237\textheight]{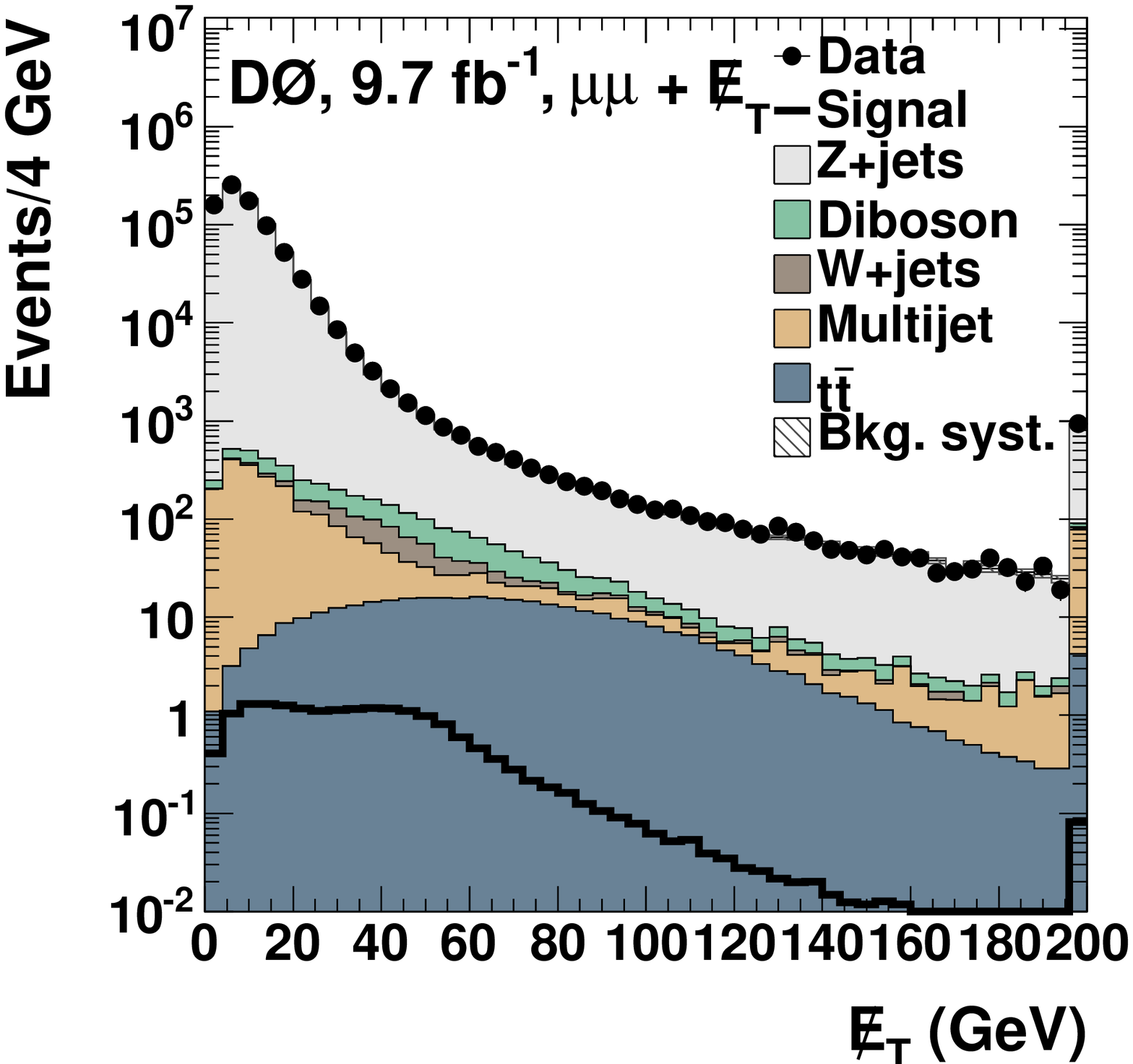}
\unitlength=1mm

\begin{picture}(00,00)
\if \mytwocolumn 1
\put(-77,50){\text{\bf (a)}} 
\put(-17,50){\text{\bf (b)}} 
\put(43,50){\text{\bf (c)}}
\else
\put(-47,106){\text{\bf (a)}} 
\put(-17,50){\text{\bf (c)}}
\put(13,106){\text{\bf (b)}}
\fi

\end{picture}

\caption{
  Distributions the missing transverse energy for the (a) \ee\ channel, 
  (b) \em\ channel,
  and (c) \mm\ channel  after the preselection.
  In (a), (b) and (c) the last bin includes all events above the upper bound of the histogram.
  In these plots, the hatched bands show the total  systematic uncertainty on the background predictions, and
  the signal distributions are those expected from a  Higgs boson of mass $M_H=125$ GeV.
}
  \label{fig:aux_presel_met}
\end{figure*}


\begin{figure*}[!] 
 
\includegraphics[height=0.237\textheight]{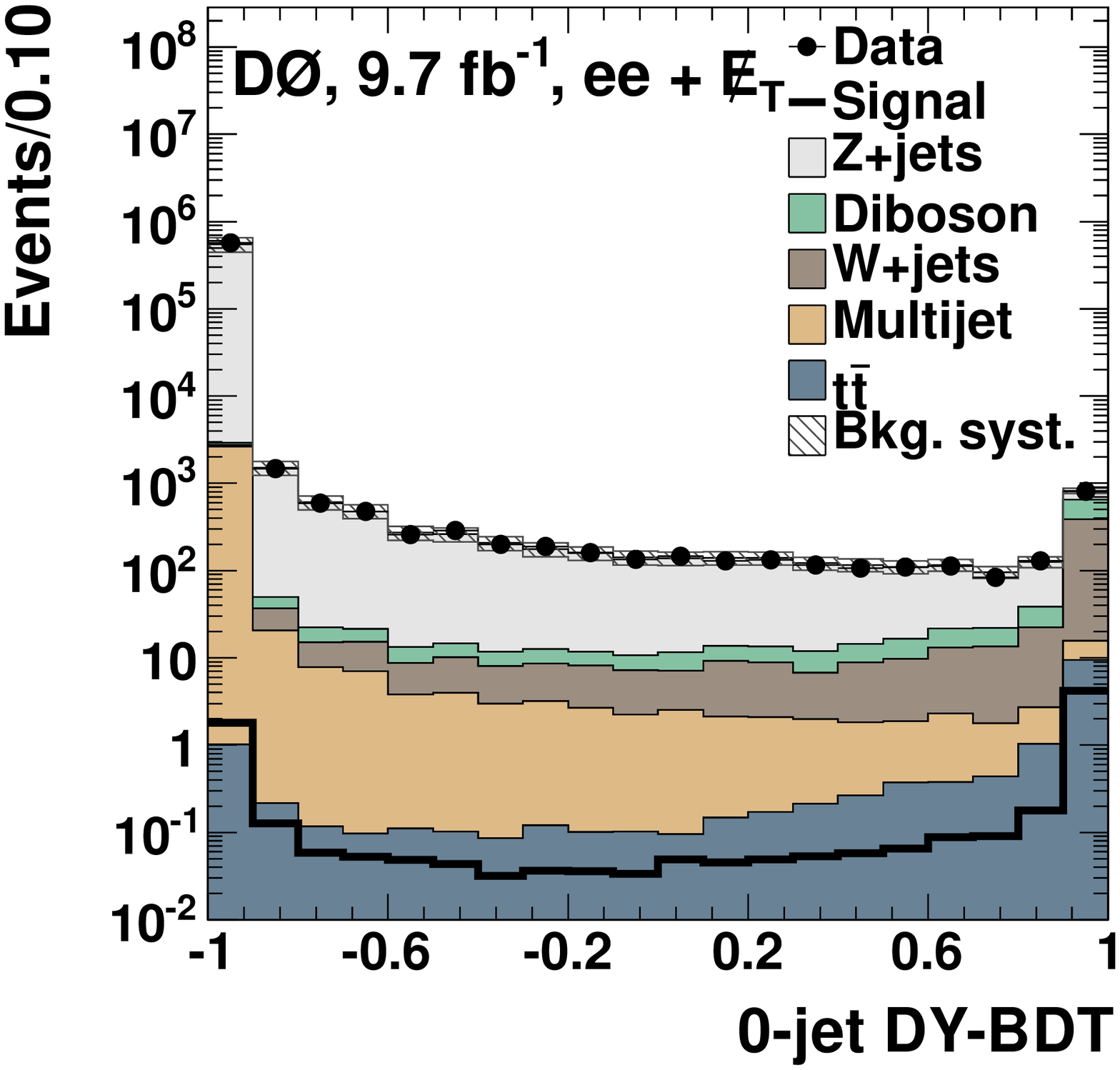}
\includegraphics[height=0.237\textheight]{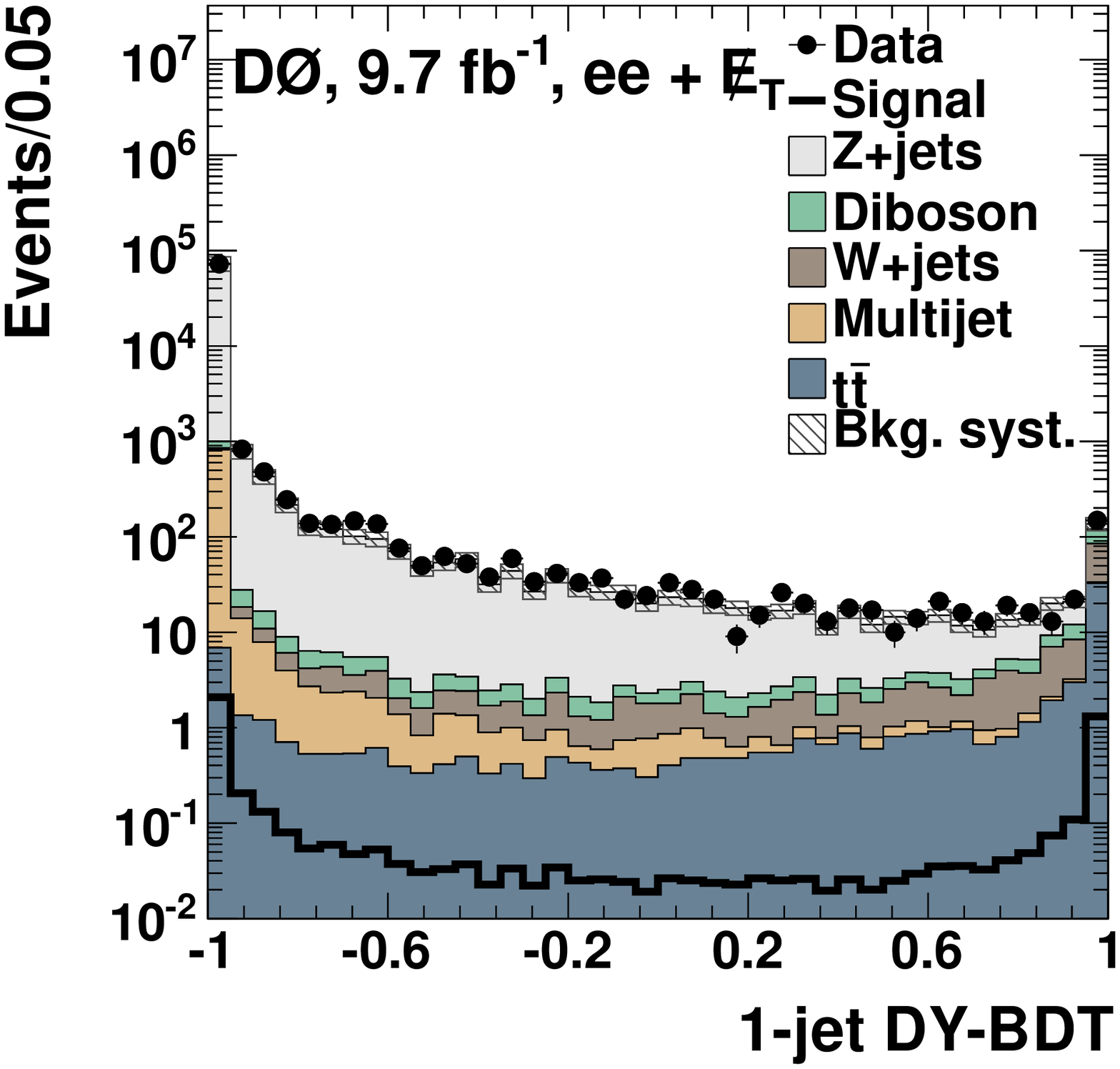}
\includegraphics[height=0.237\textheight]{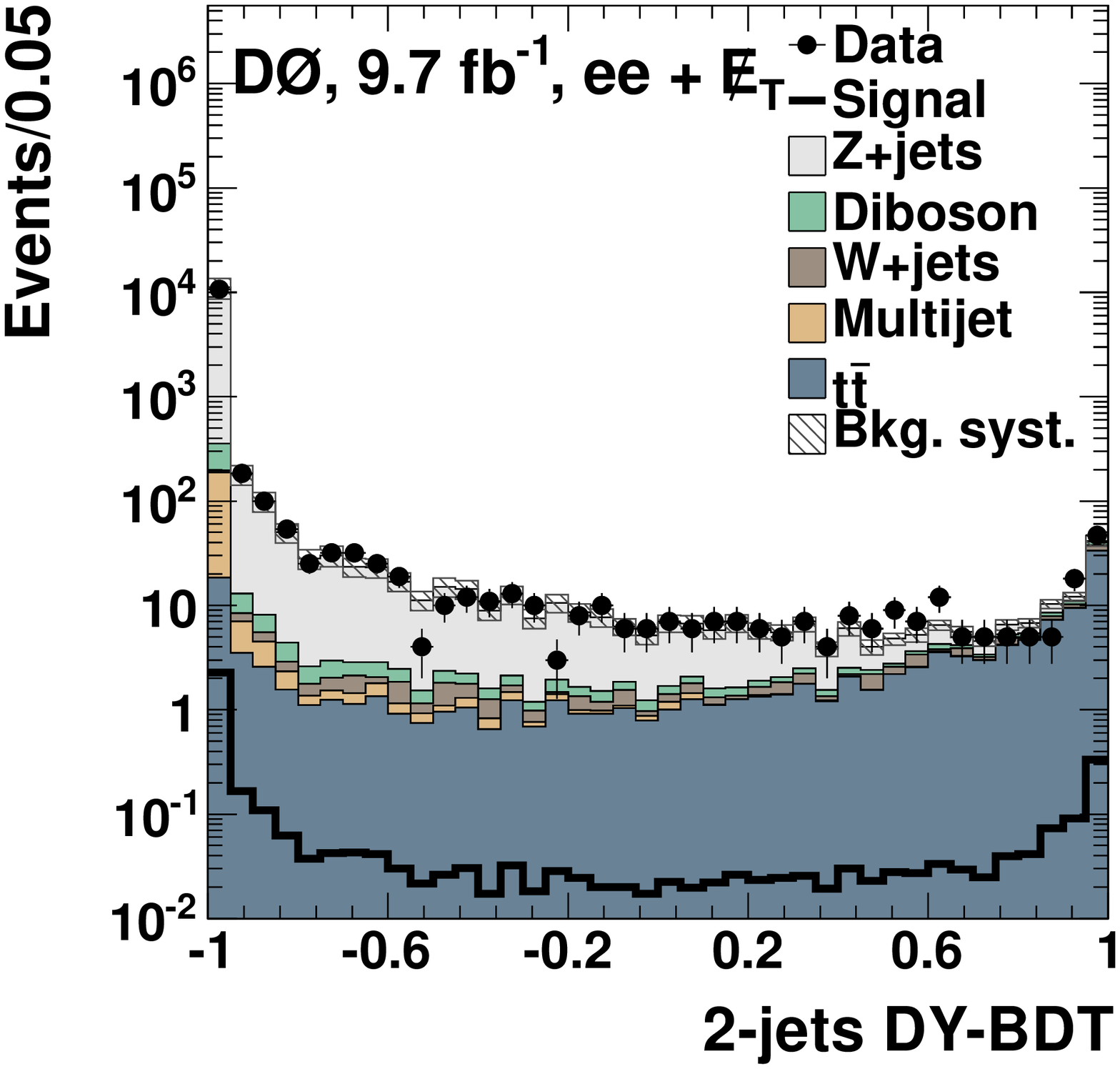}
\unitlength=1mm

\begin{picture}(00,00)
\if \mytwocolumn 1
\put(-77,50){\text{\bf (a)}} 
\put(-17,50){\text{\bf (b)}} 
\put(43,50){\text{\bf (c)}}
\else
\put(-47,106){\text{\bf (a)}} 
\put(-17,50){\text{\bf (c)}}
\put(13,106){\text{\bf (b)}}
\fi

\end{picture}

\caption{
  Distributions of the Drell-Yan (DY) BDT discriminant for \ee\ channel in the (a) 0-jet bin, (b) 1-jet bin, and (c) $\geq$ 2-jets bin.
The BDTs are trained for a Higgs mass of 125~GeV. The signal distributions are those expected from a  Higgs boson of mass $M_H=125$~GeV.
}
  \label{fig:aux_DYDT_ee}
\end{figure*}

\begin{figure*}[!] 
 
\includegraphics[height=0.237\textheight]{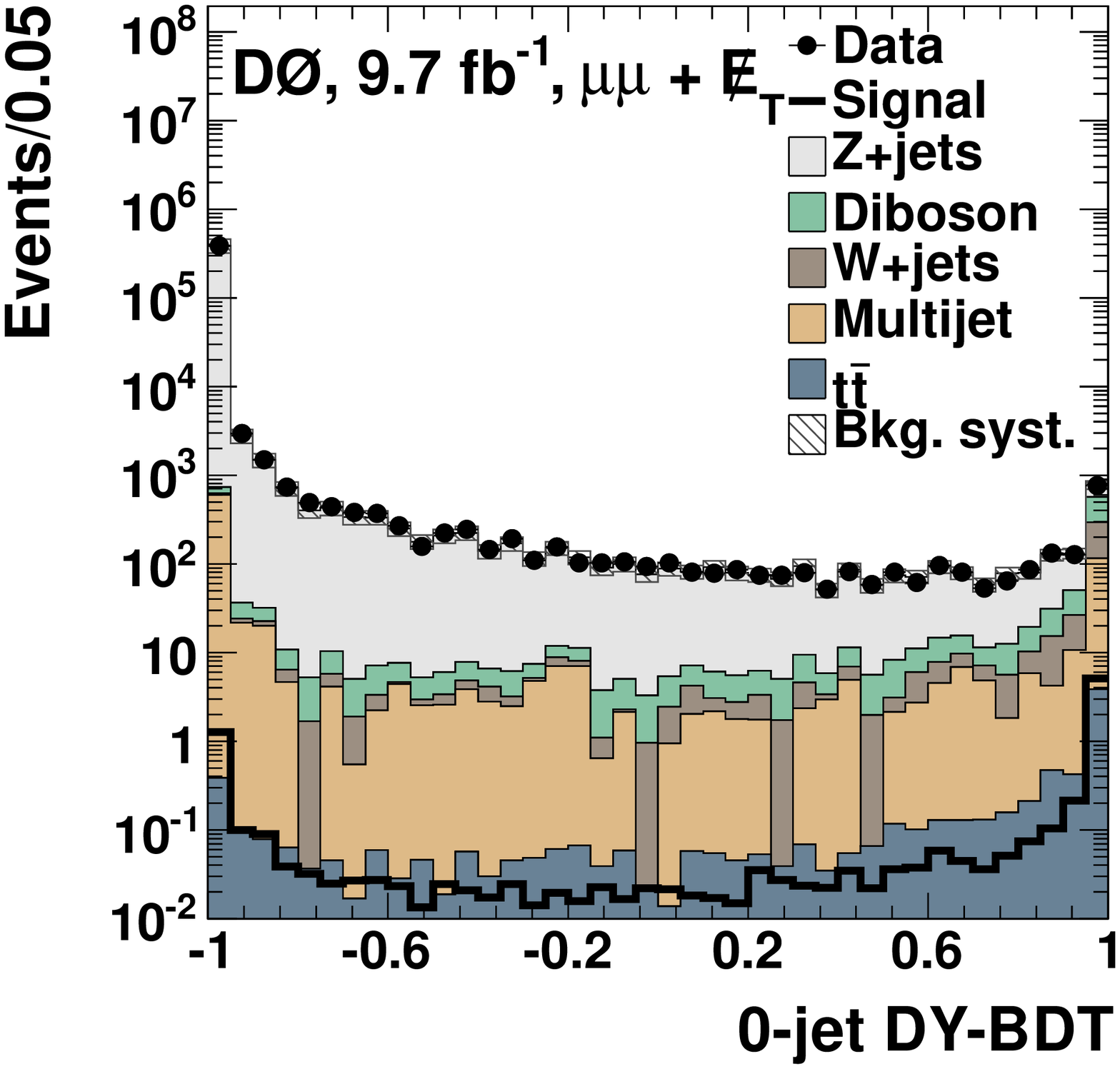}
\includegraphics[height=0.237\textheight]{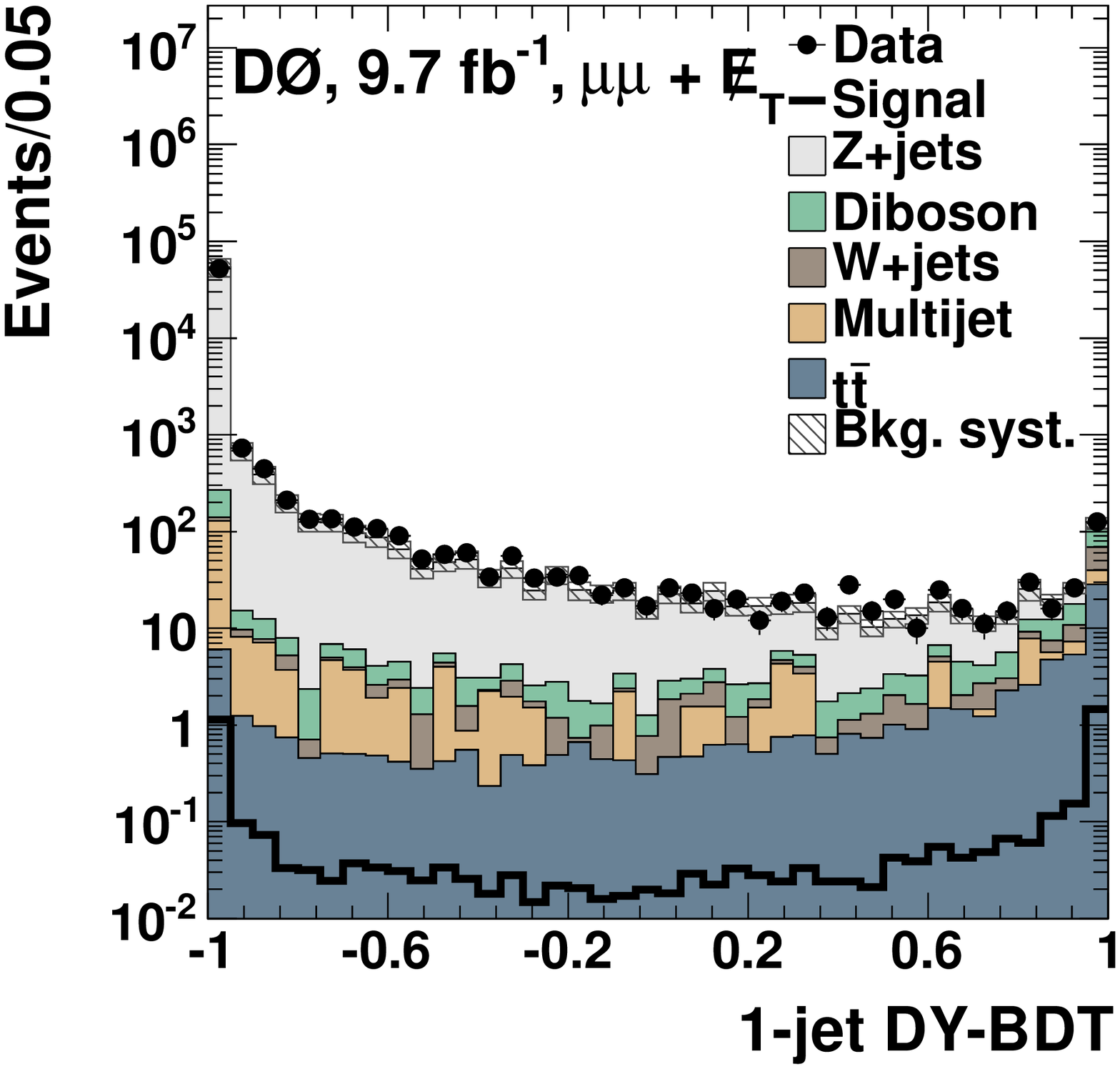}
\includegraphics[height=0.237\textheight]{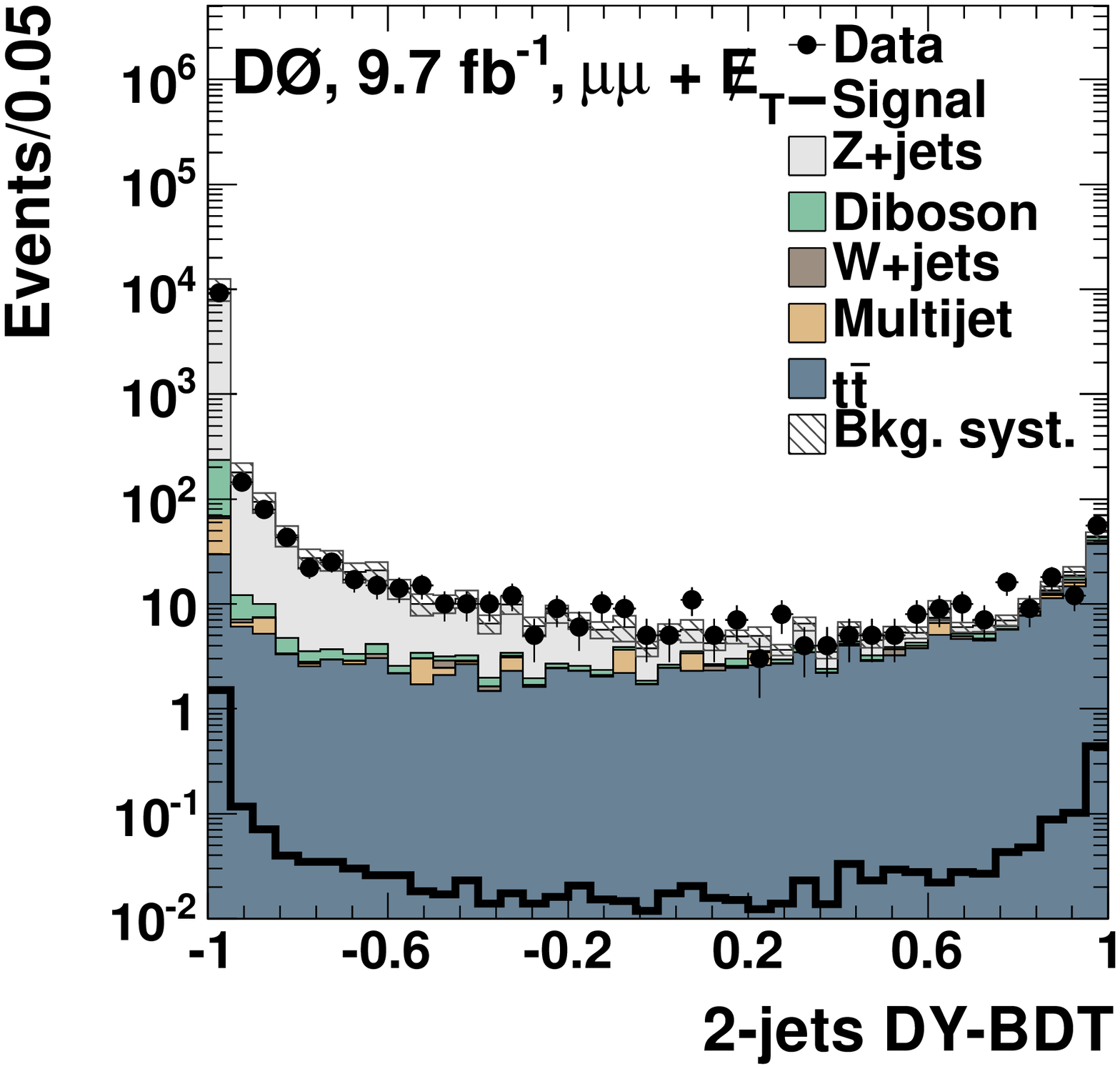}
\unitlength=1mm

\begin{picture}(00,00)
\if \mytwocolumn 1
\put(-77,50){\text{\bf (a)}} 
\put(-17,50){\text{\bf (b)}} 
\put(43,50){\text{\bf (c)}}
\else
\put(-47,106){\text{\bf (a)}} 
\put(-17,50){\text{\bf (c)}}
\put(13,106){\text{\bf (b)}}
\fi

\end{picture}

\caption{
  Distributions of the DY BDT discriminant for \mm\ channel in the (a) 0-jet bin, (b) 1-jet bin, and (c) $\geq$ 2-jets bin.
The BDTs are trained for a Higgs mass of 125 GeV. The signal distributions are those expected from a  Higgs boson of mass $M_H=125$~GeV.
}
  \label{fig:aux_DYDT_mm}
\end{figure*}

\begin{figure*}[!]

\includegraphics[height=0.237\textheight]{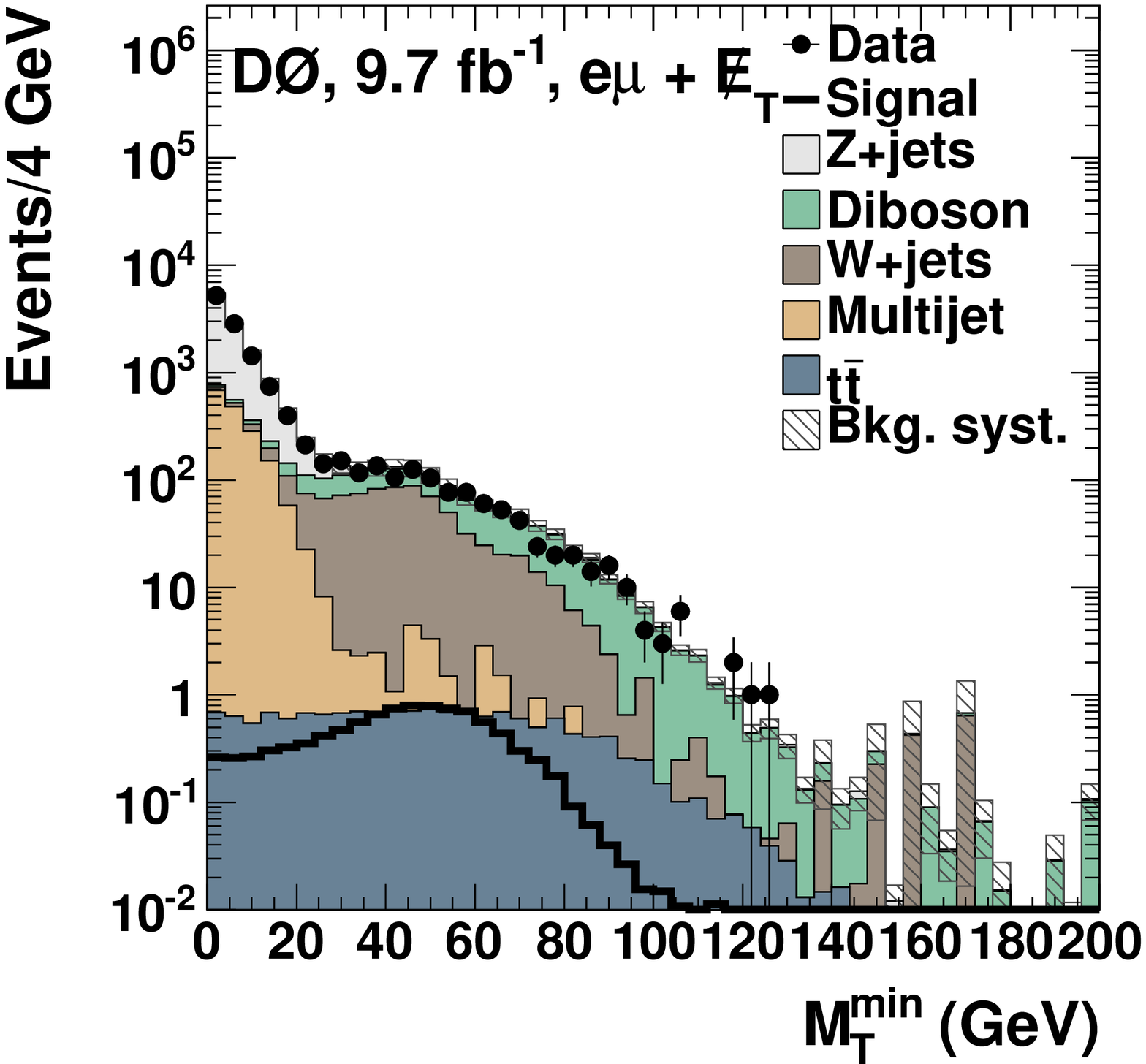}
\includegraphics[height=0.237\textheight]{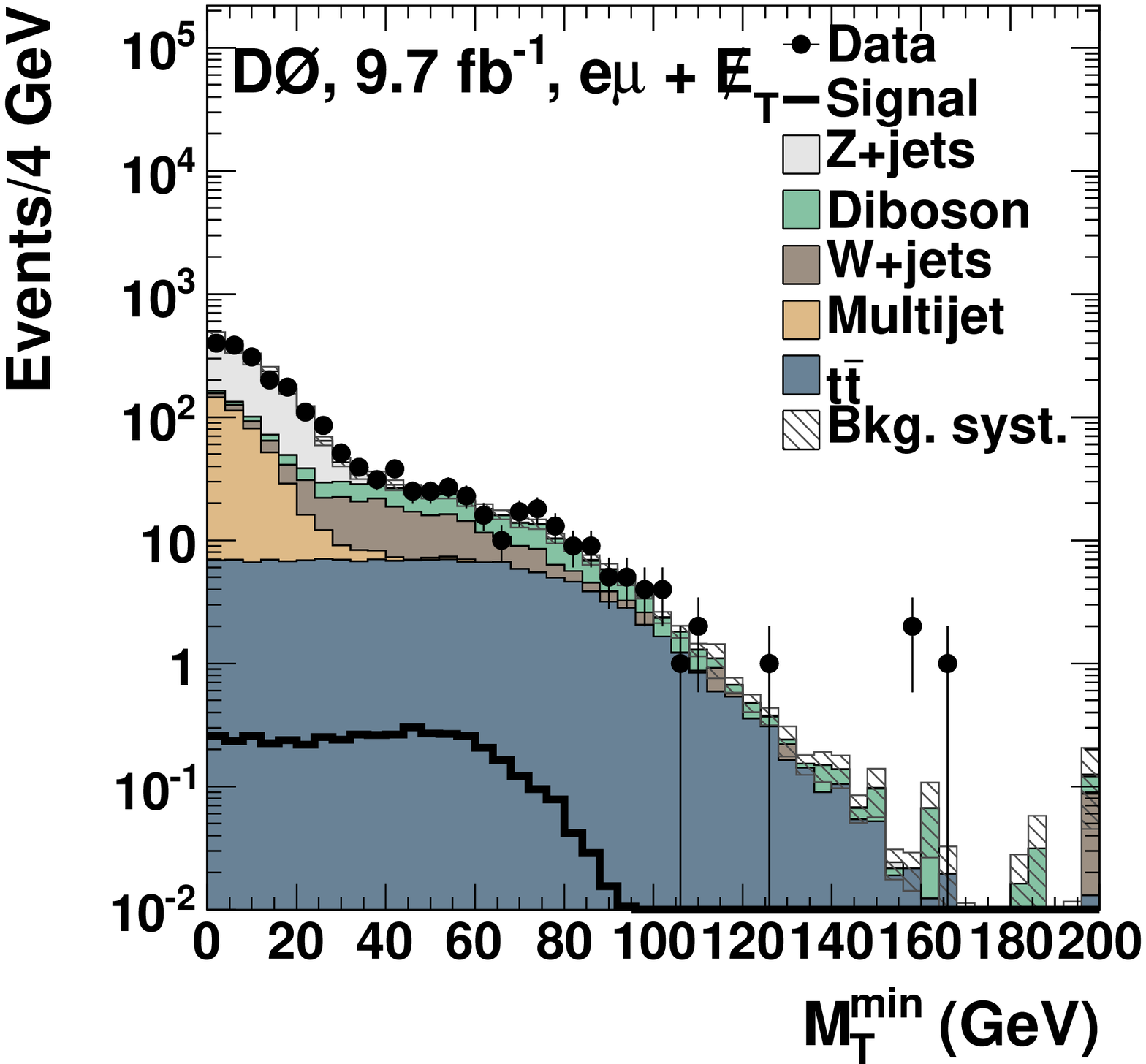}
\includegraphics[height=0.237\textheight]{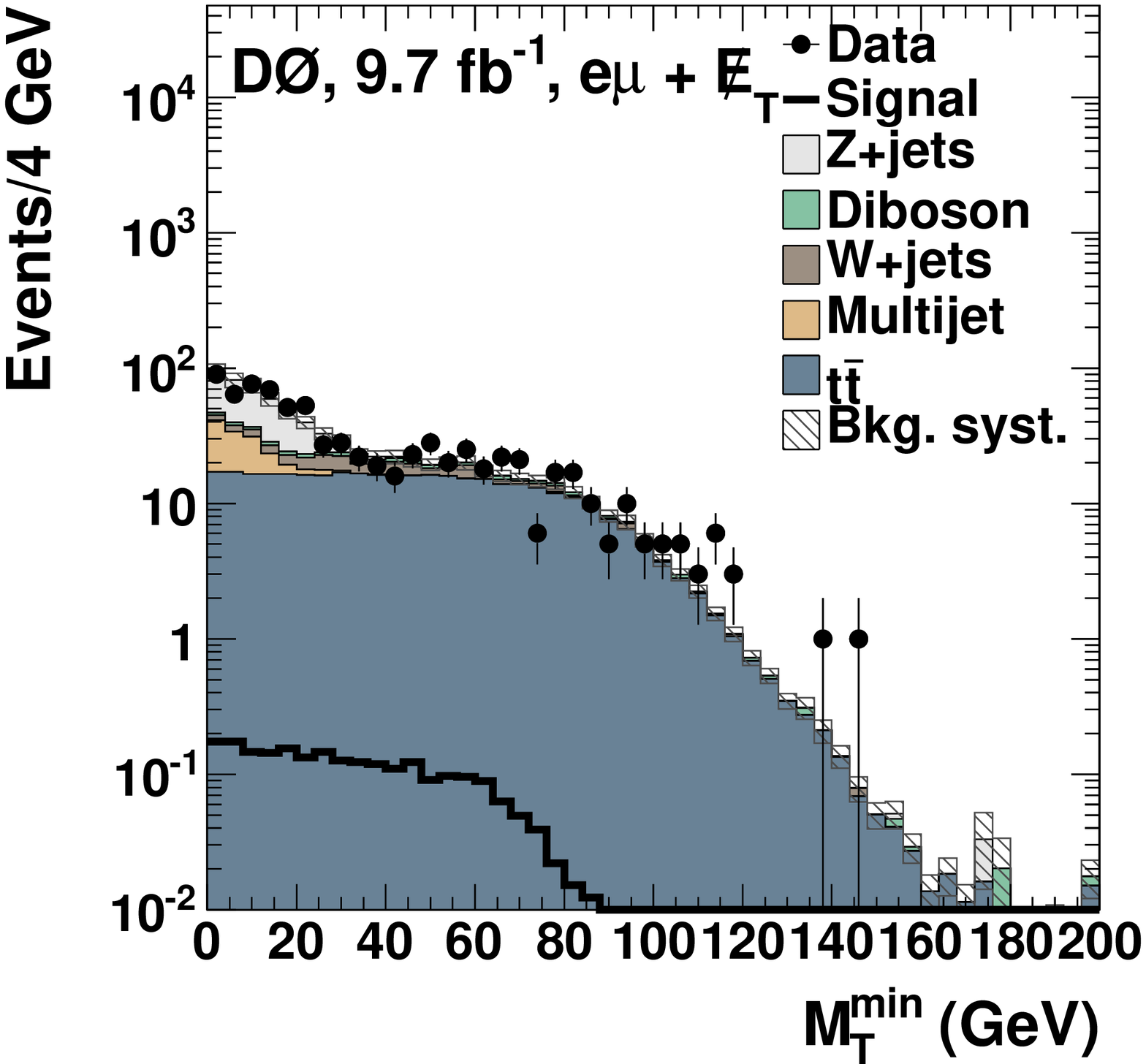}

\unitlength=1mm

\begin{picture}(00,00)
\if \mytwocolumn 1
\put(-77,50){\text{\bf (a)}} 
\put(-17,50){\text{\bf (b)}} 
\put(43,50){\text{\bf (c)}}
\else
\put(-47,106){\text{\bf (a)}} 
\put(-17,50){\text{\bf (c)}}
\put(13,106){\text{\bf (b)}}
\fi

\end{picture}

\includegraphics[height=0.237\textheight]{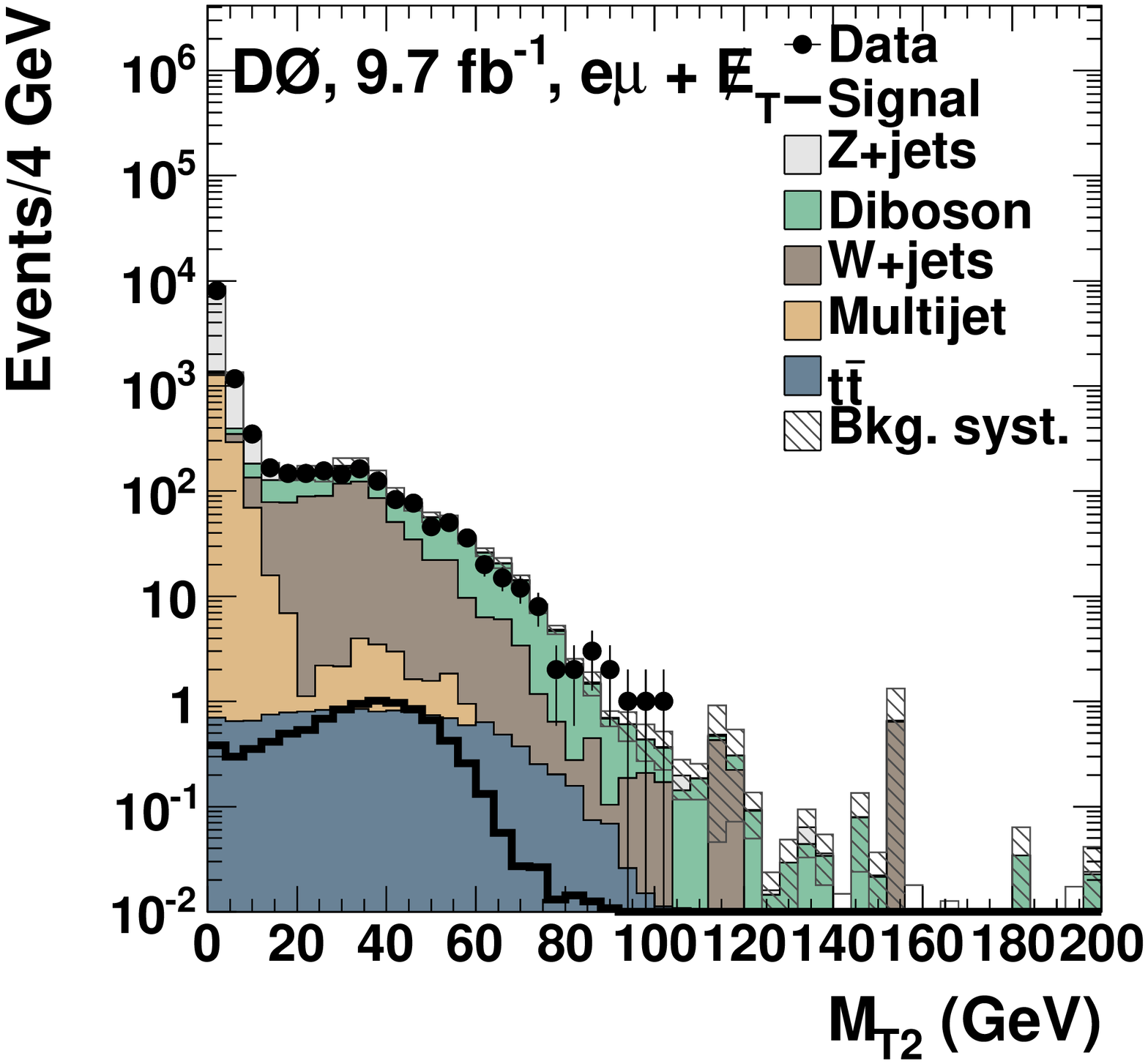}
\includegraphics[height=0.237\textheight]{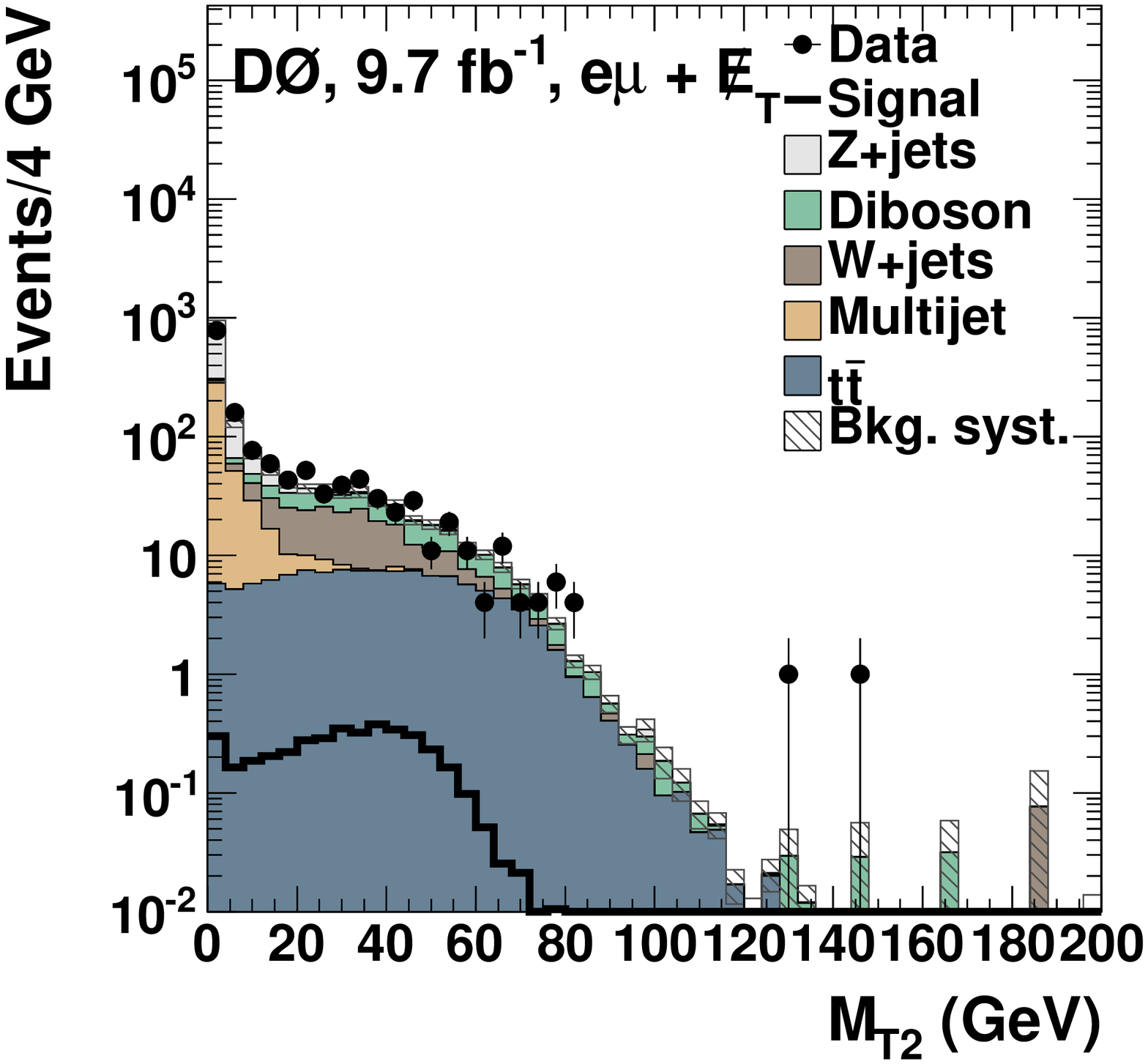}
\includegraphics[height=0.237\textheight]{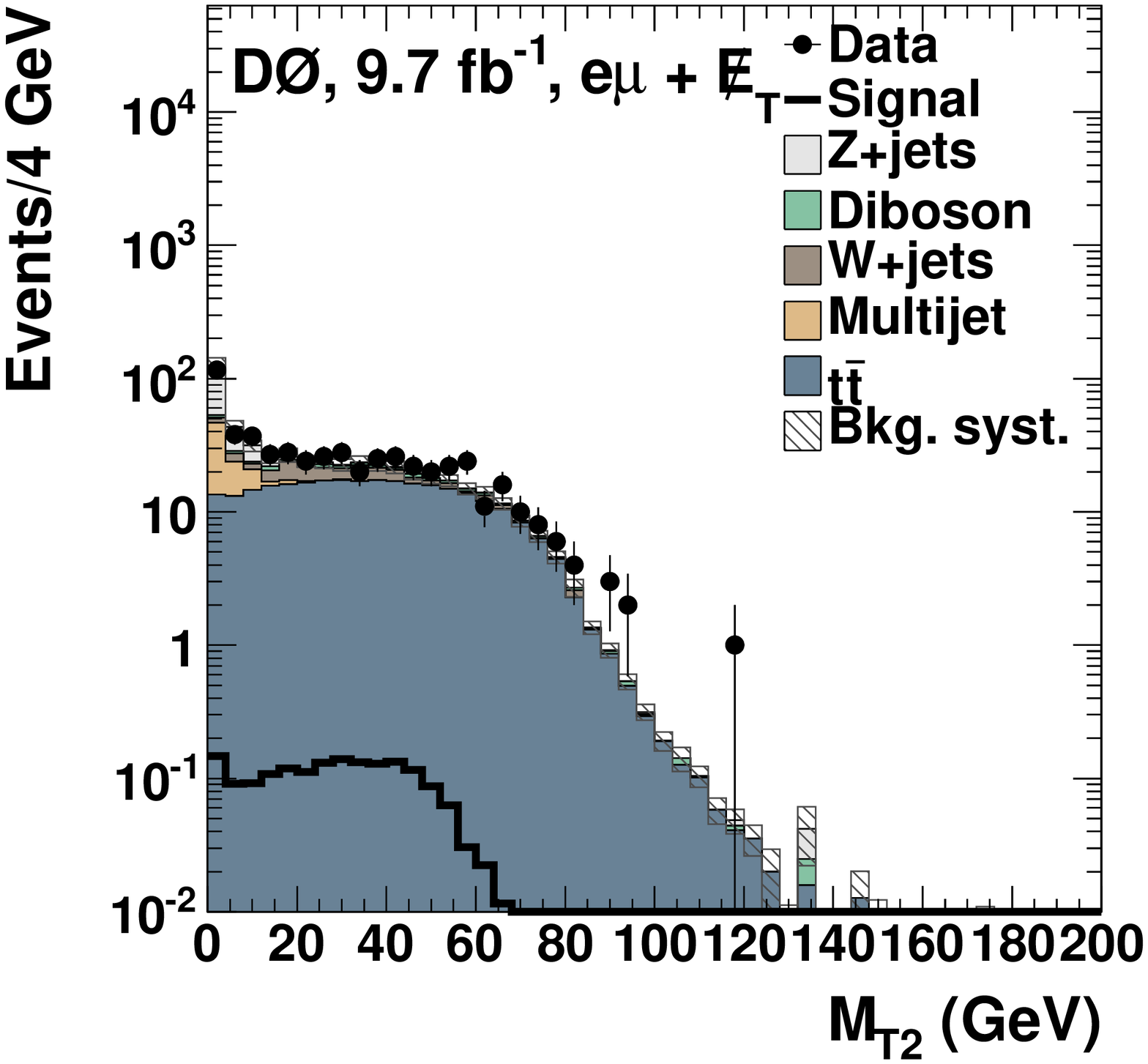}

\unitlength=1mm

\begin{picture}(00,00)
\if \mytwocolumn 1
\put(-77,50){\text{\bf (d)}} 
\put(-17,50){\text{\bf (e)}} 
\put(43,50){\text{\bf (f)}}
\else
\put(-47,106){\text{\bf (d)}} 
\put(-17,50){\text{\bf (e)}}
\put(13,106){\text{\bf (f)}}
\fi

\end{picture}

\caption{
\mtmin distribution for the \em\ channel in the (a) 0-jet bin, (b) 1-jet bin, and (c) $\geq$ 2-jets bin.
$M_{T2}$ distribution for the \em\ channel in the (d) 0-jet bin, (e) 1-jet bin, and (f) $\geq$ 2-jets bin.
For both distributions, the signal distributions are those expected from a  Higgs boson of mass $M_H=125$~GeV.
}
  \label{fig:aux_minMt_em}
\end{figure*}


\begin{figure*}[!] 
\section*{\Large Distributions at the final selection level}
 
\includegraphics[height=0.237\textheight]{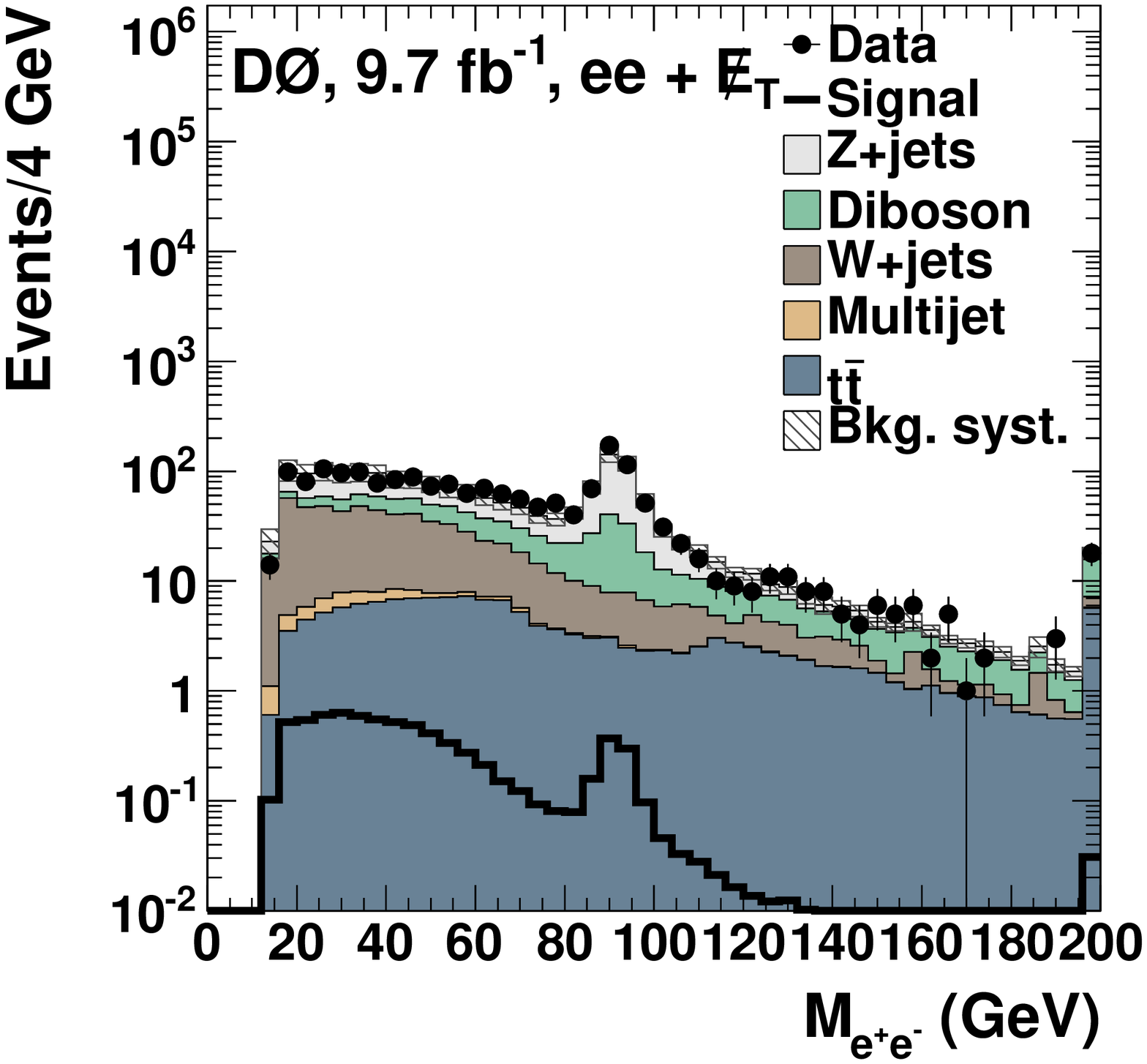}
\includegraphics[height=0.237\textheight]{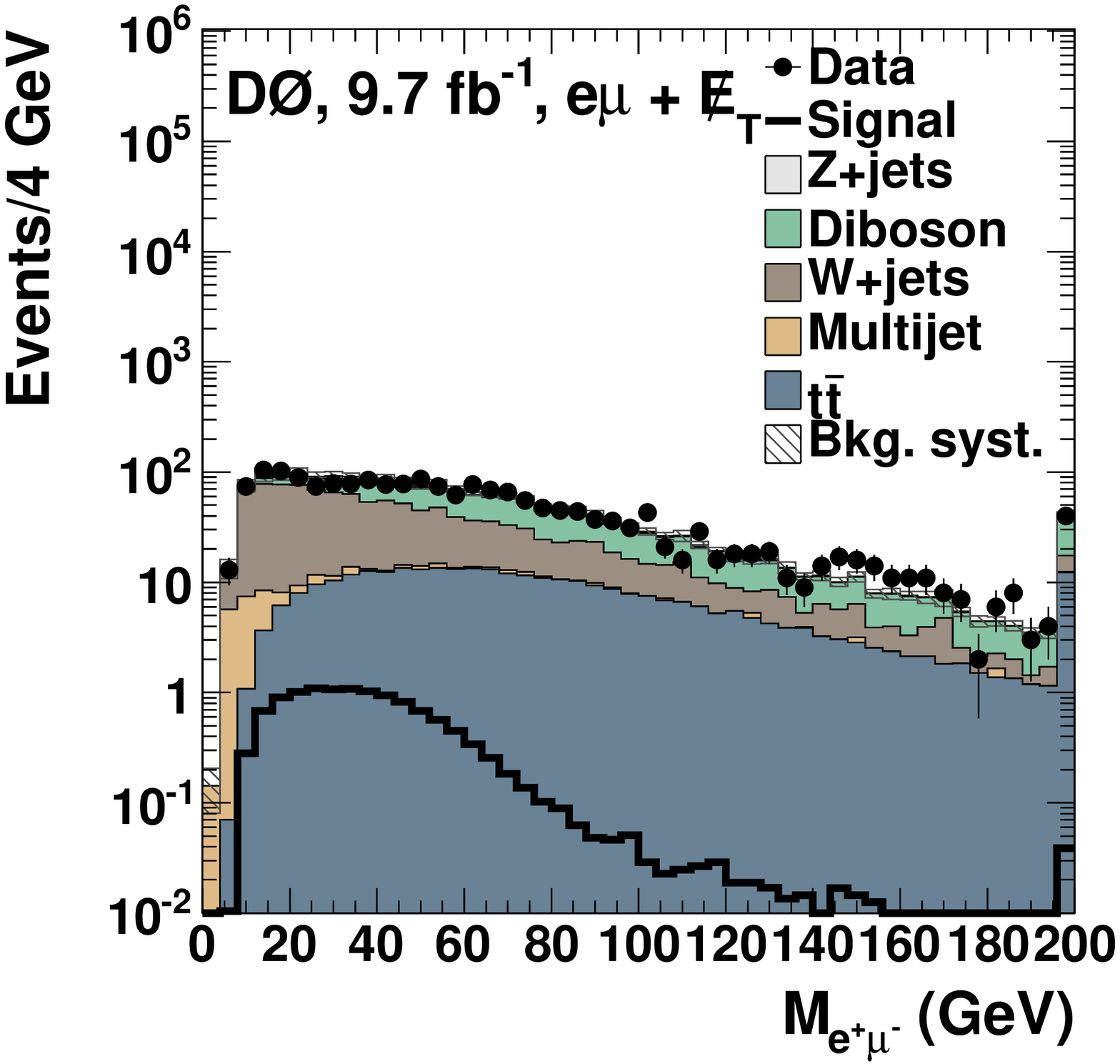}
\includegraphics[height=0.237\textheight]{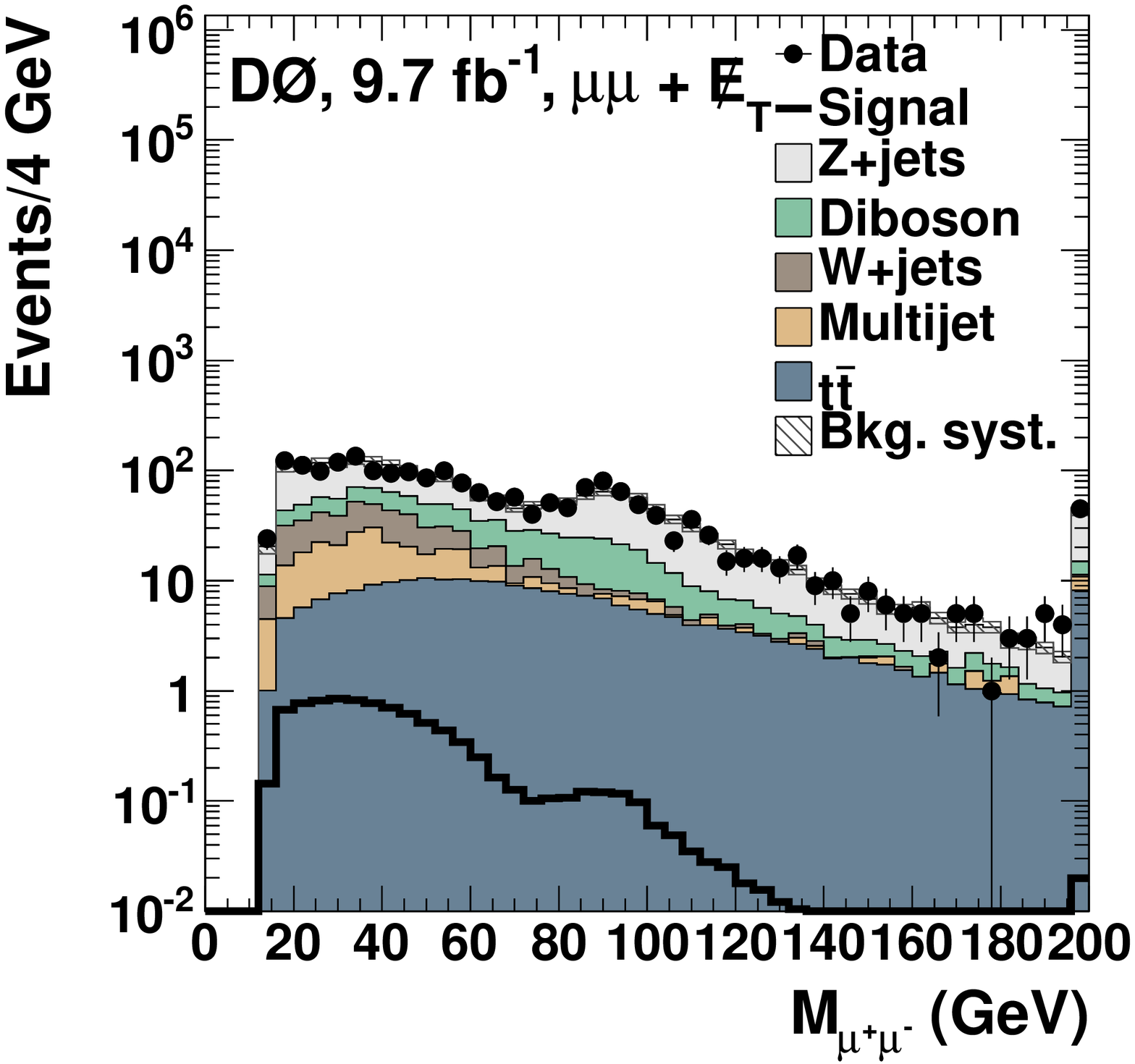}
\unitlength=1mm

\begin{picture}(00,00)
\if \mytwocolumn 1
\put(-77,50){\text{\bf (a)}} 
\put(-17,50){\text{\bf (b)}} 
\put(43,50){\text{\bf (c)}}
\else
\put(-47,106){\text{\bf (a)}} 
\put(-17,50){\text{\bf (c)}}
\put(13,106){\text{\bf (b)}}
\fi

\end{picture}

\caption{
  Distributions of the  dilepton invariant mass for the (a) \ee\ channel,  (b) \em\ channel,
  and (c) \mm\ channel  after the final selection.
  In (a), (b) and (c) the last bin includes all events above the upper bound of the histogram.
  In these plots, the hatched bands show the total  systematic uncertainty on the background predictions, and
  the signal distributions are those expected from a  Higgs boson of mass $M_H=125$ GeV.
}
  \label{fig:aux_finsel_M}
\end{figure*}

\begin{figure*}[!] 
 
\includegraphics[ height=0.237\textheight]{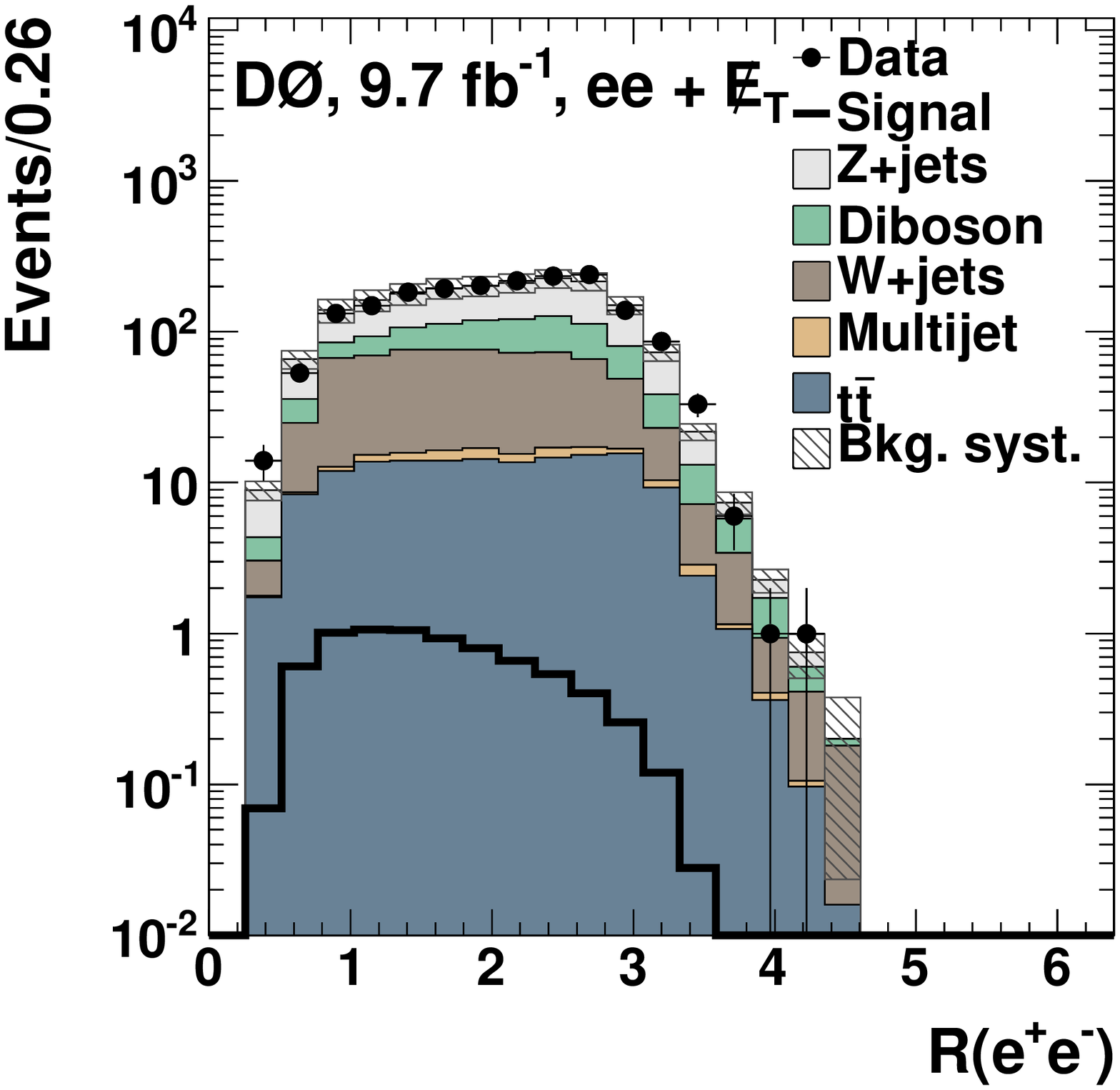}
\includegraphics[height=0.237\textheight]{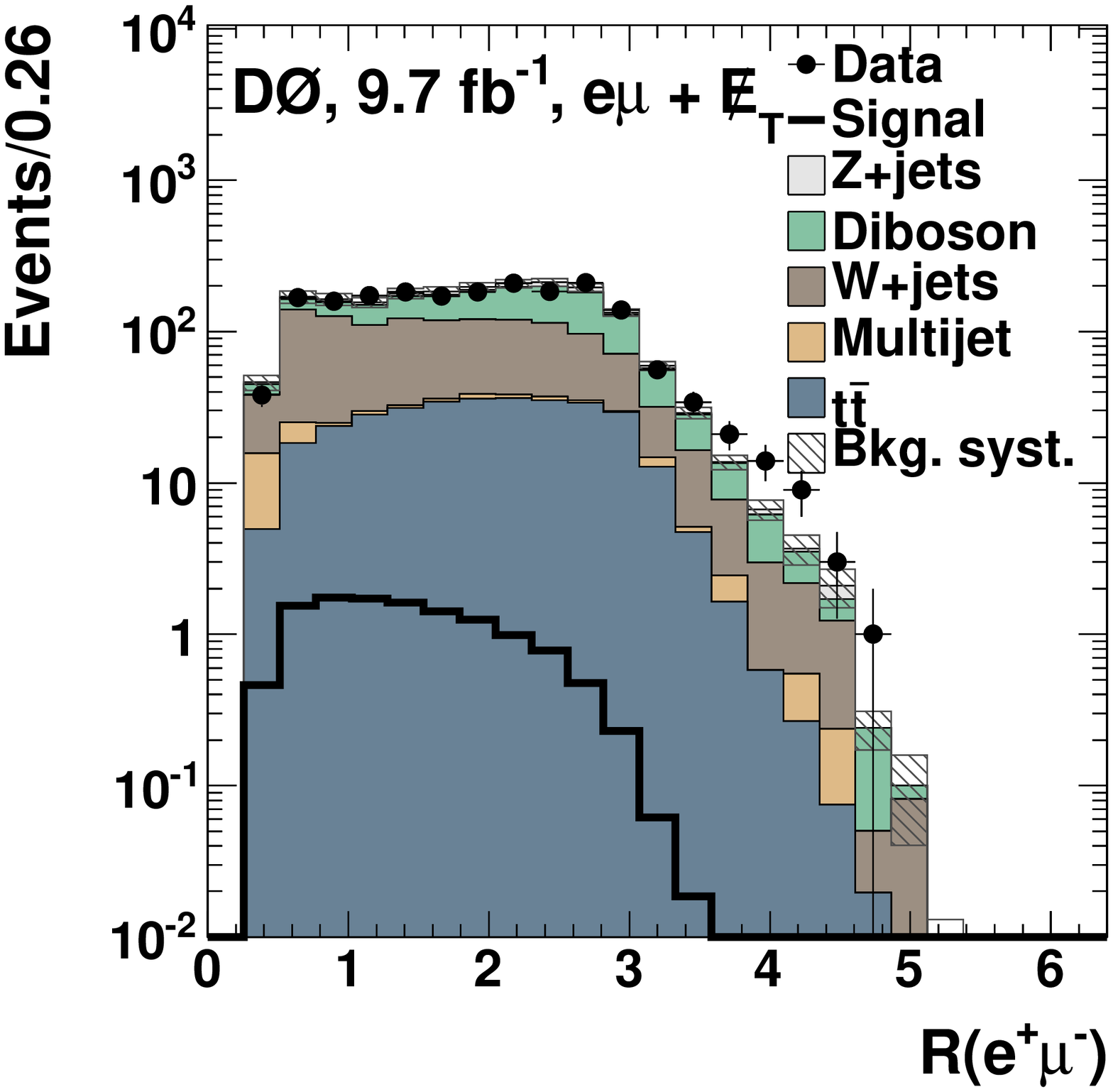}
\includegraphics[ height=0.237\textheight]{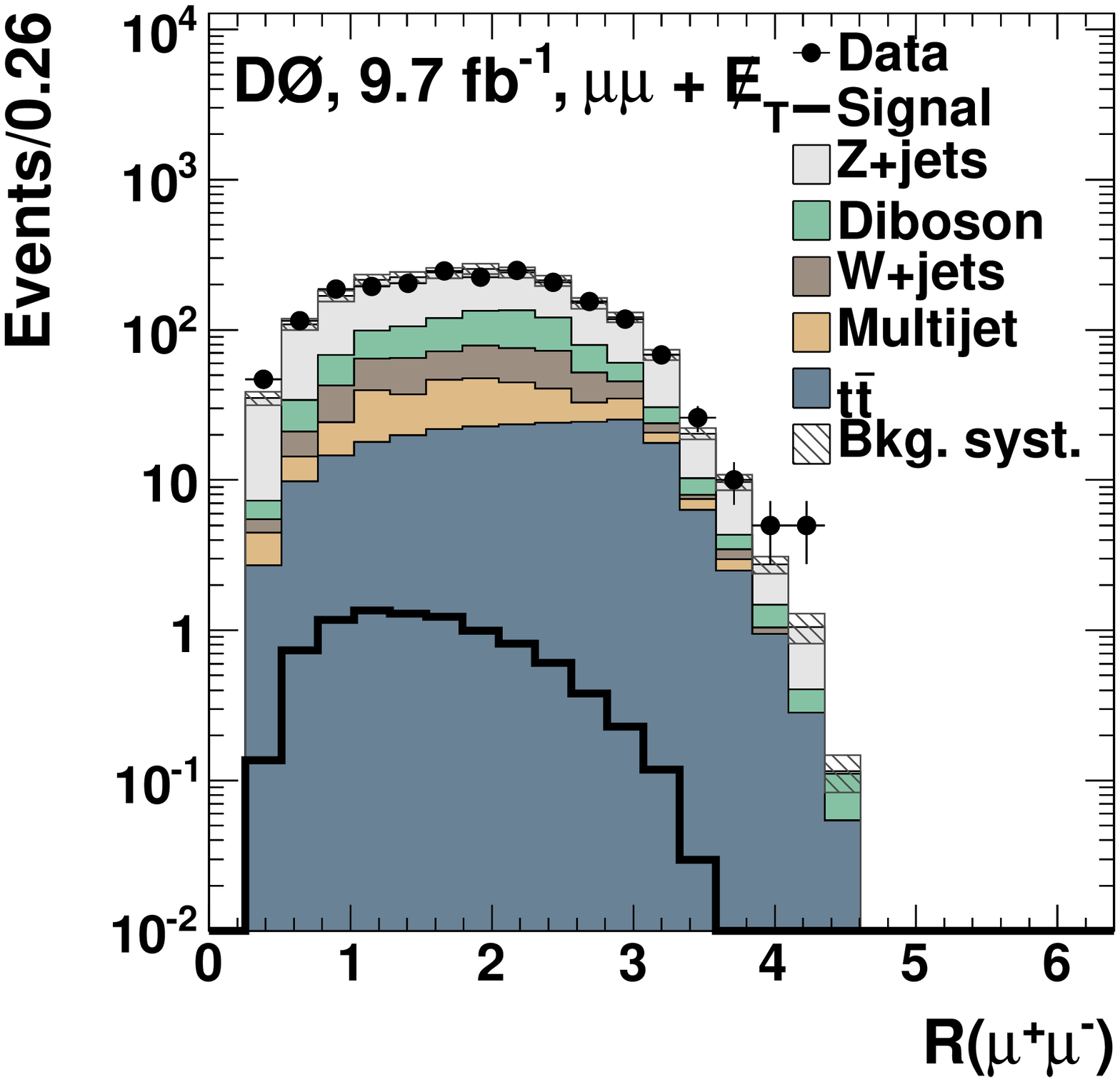}
\unitlength=1mm

\begin{picture}(00,00)
\if \mytwocolumn 1
\put(-77,50){\text{\bf (a)}} 
\put(-17,50){\text{\bf (b)}} 
\put(43,50){\text{\bf (c)}}
\else
\put(-47,106){\text{\bf (a)}} 
\put(-17,50){\text{\bf (c)}}
\put(13,106){\text{\bf (b)}}
\fi

\end{picture}

\caption{
  Distributions of the  angular separation $\cal R(\ell,\ell)$ between the leptons for the (a) \ee\ channel, 
   (b) \em\ channel,
  and (c) \mm\ channel  after the final selection.
  In these plots, the hatched bands show the total  systematic uncertainty on the background predictions, and
  the signal distributions are those expected from a  Higgs boson of mass $M_H=125$ GeV.
}
  \label{fig:aux_finsel_dr}
\end{figure*}

\begin{figure*}[!] 
 
\includegraphics[ height=0.237\textheight]{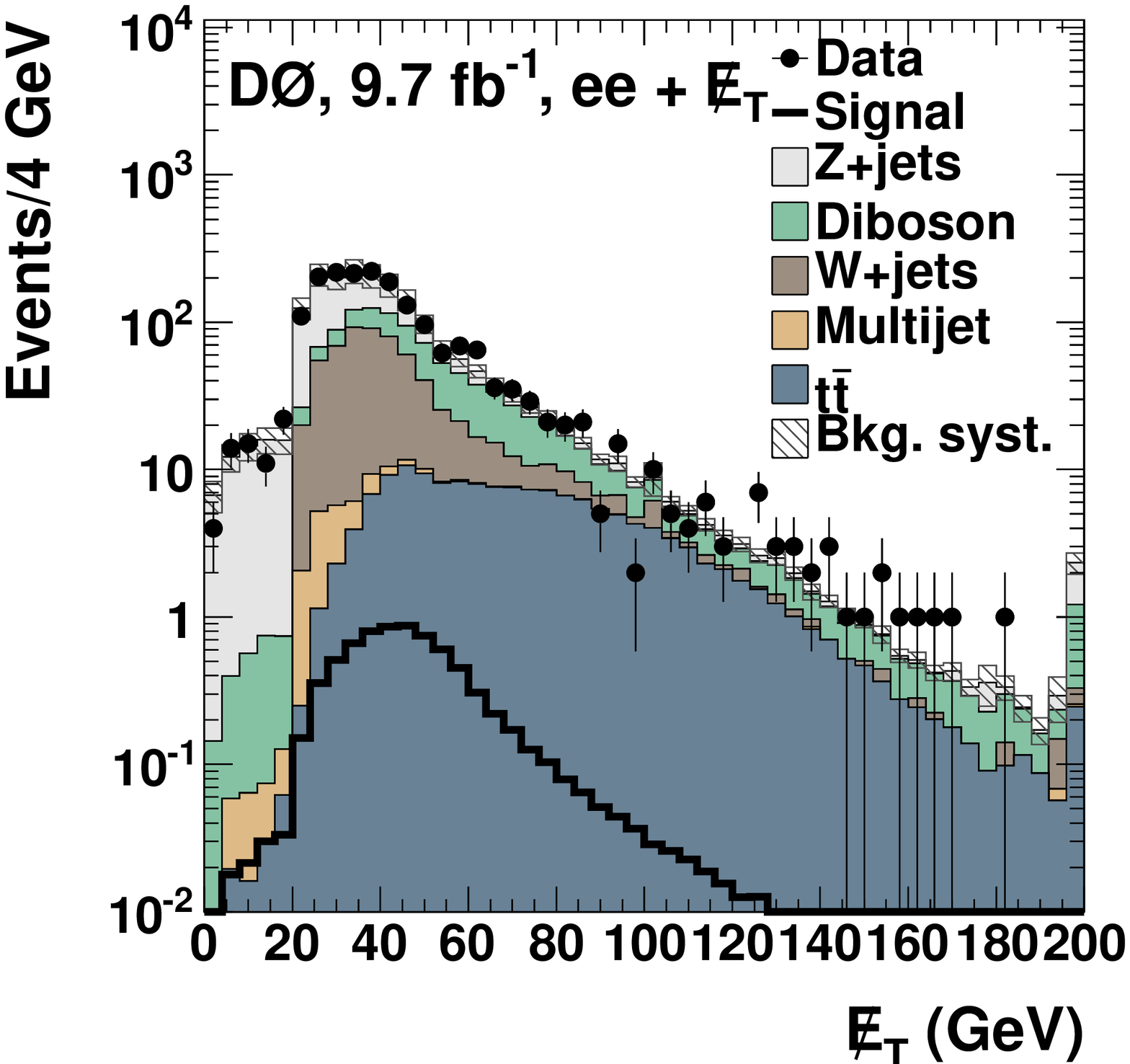}
\includegraphics[height=0.237\textheight]{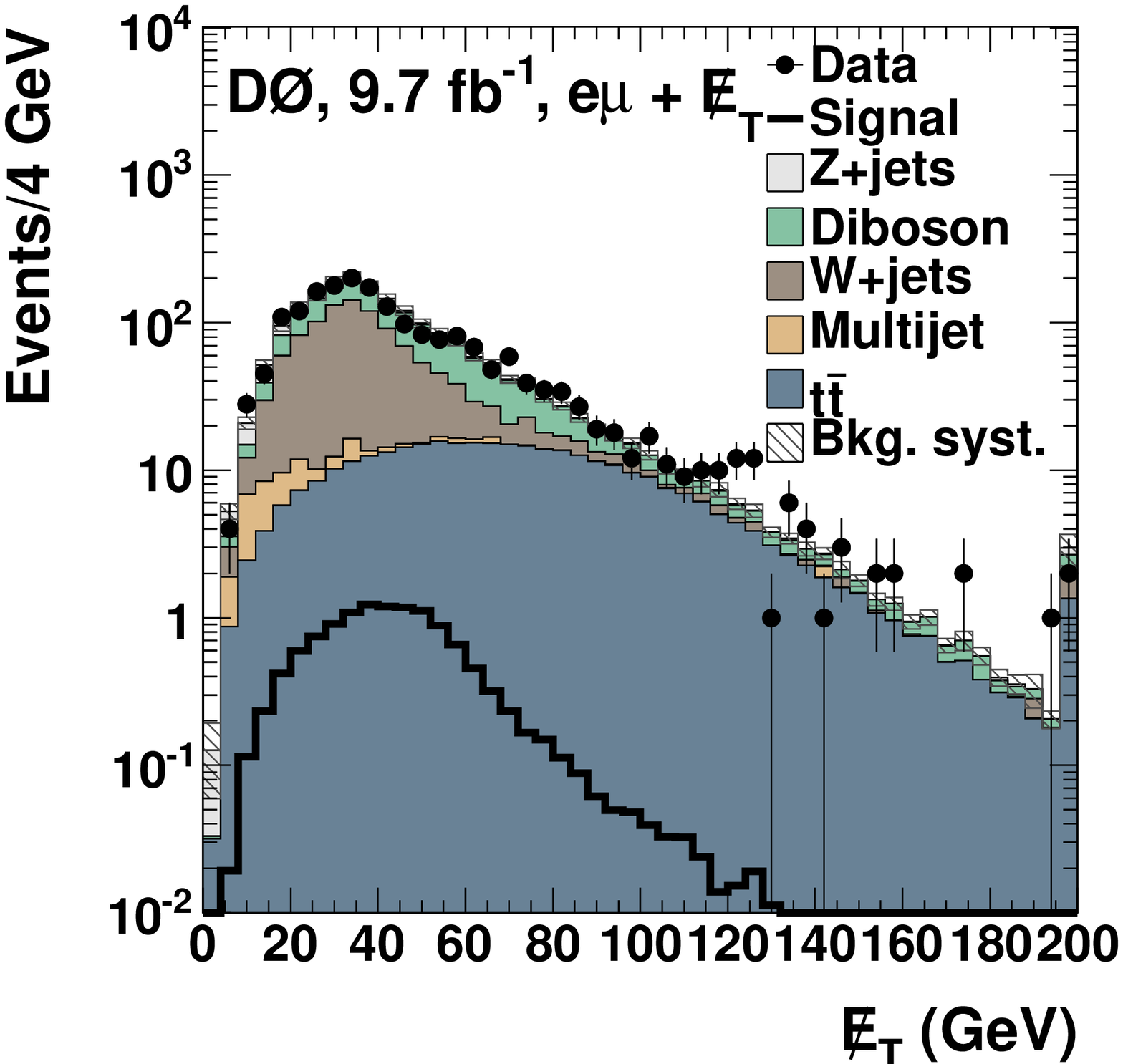}
\includegraphics[ height=0.237\textheight]{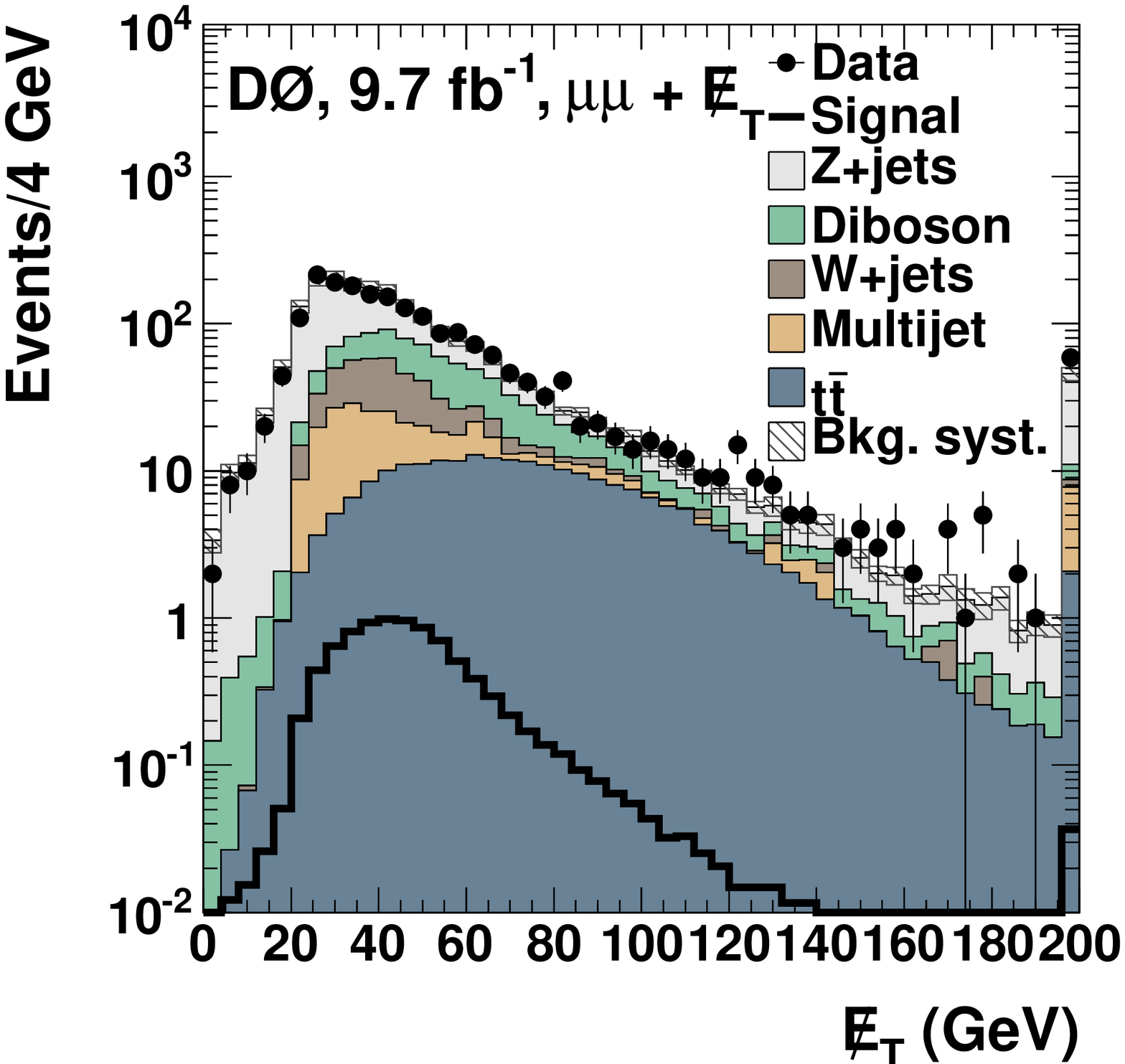}

\unitlength=1mm
\begin{picture}(00,00)
\if \mytwocolumn 1
\put(-77,50){\text{\bf (a)}} 
\put(-17,50){\text{\bf (b)}} 
\put(43,50){\text{\bf (c)}}
\else
\put(-47,106){\text{\bf (a)}} 
\put(-17,50){\text{\bf (c)}}
\put(13,106){\text{\bf (b)}}
\fi

\end{picture}

\caption{
  Distributions of the missing transverse energy for the (a) \ee\ channel, 
  (b) \em\ channel,
  and (c) \mm\ channel  after the final selection.
  In (a), (b) and (c) the last bin includes all events above the upper bound of the histogram.
  In these plots, the hatched bands show the total  systematic uncertainty on the background predictions, and
  the signal distributions are those expected from a  Higgs boson of mass $M_H=125$ GeV.
}
  \label{fig:aux_finsel_met}
\end{figure*}




\begin{figure*}[!]  

\includegraphics[height=0.237\textheight]{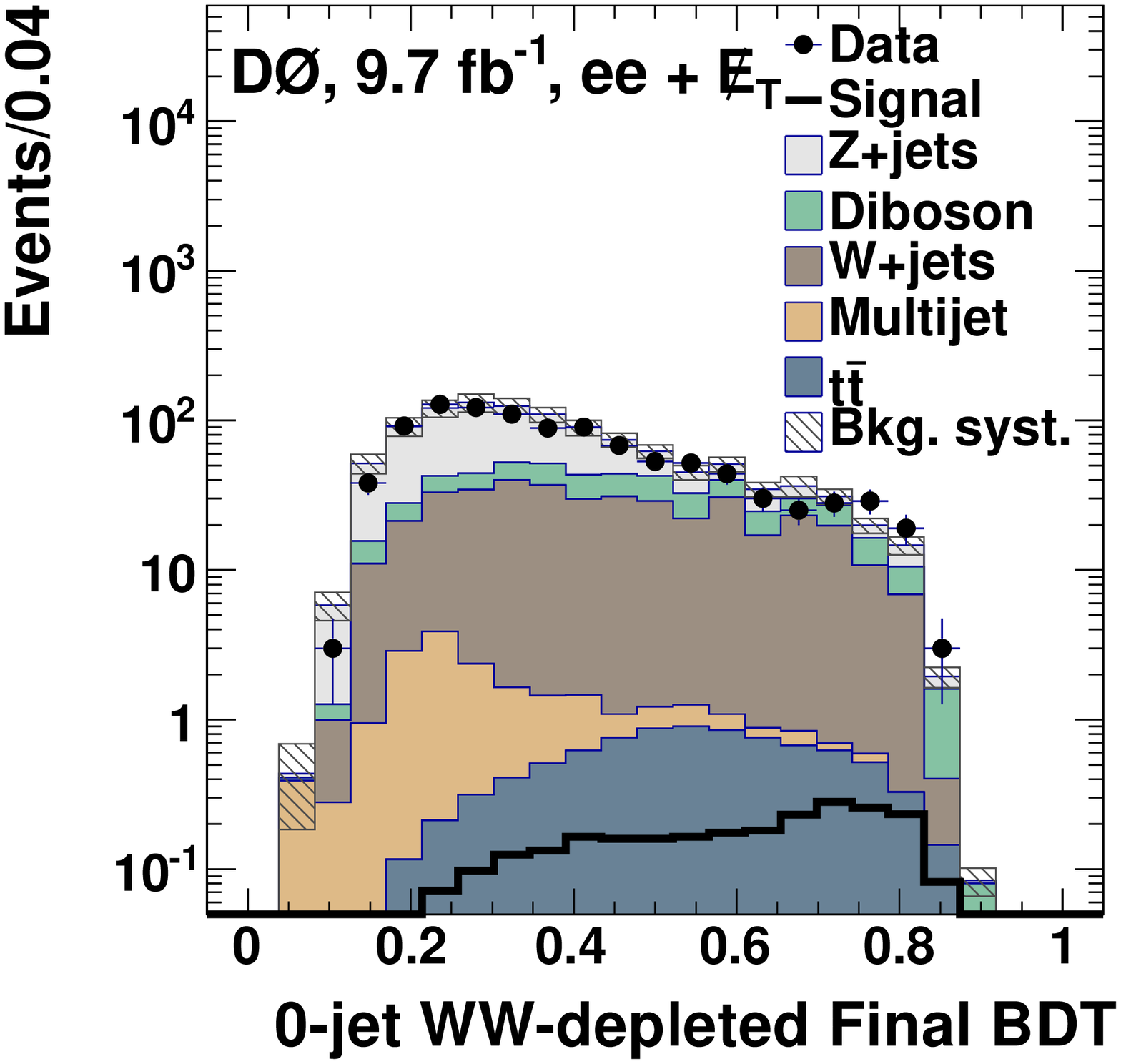}
\includegraphics[height=0.237\textheight]{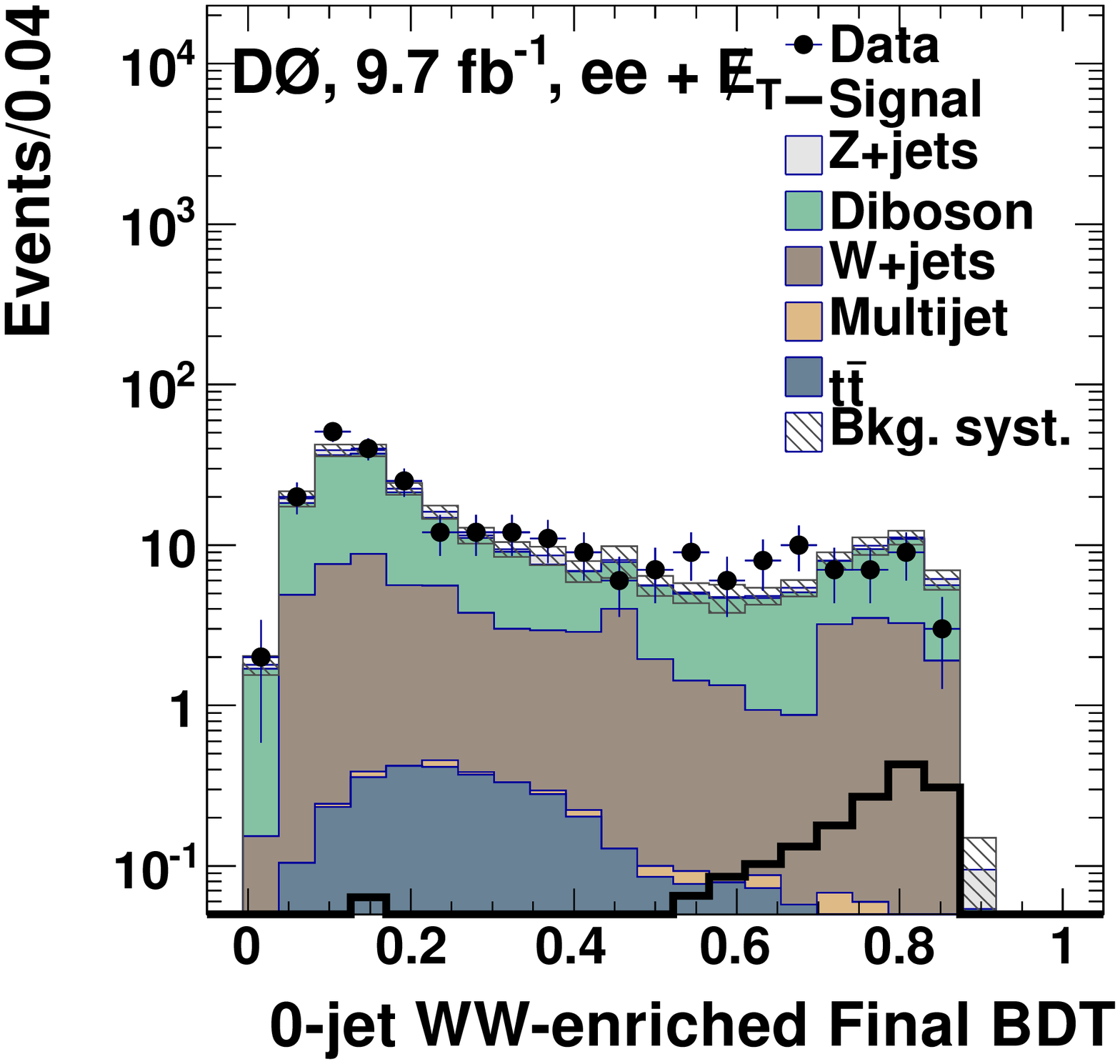}
\includegraphics[height=0.237\textheight]{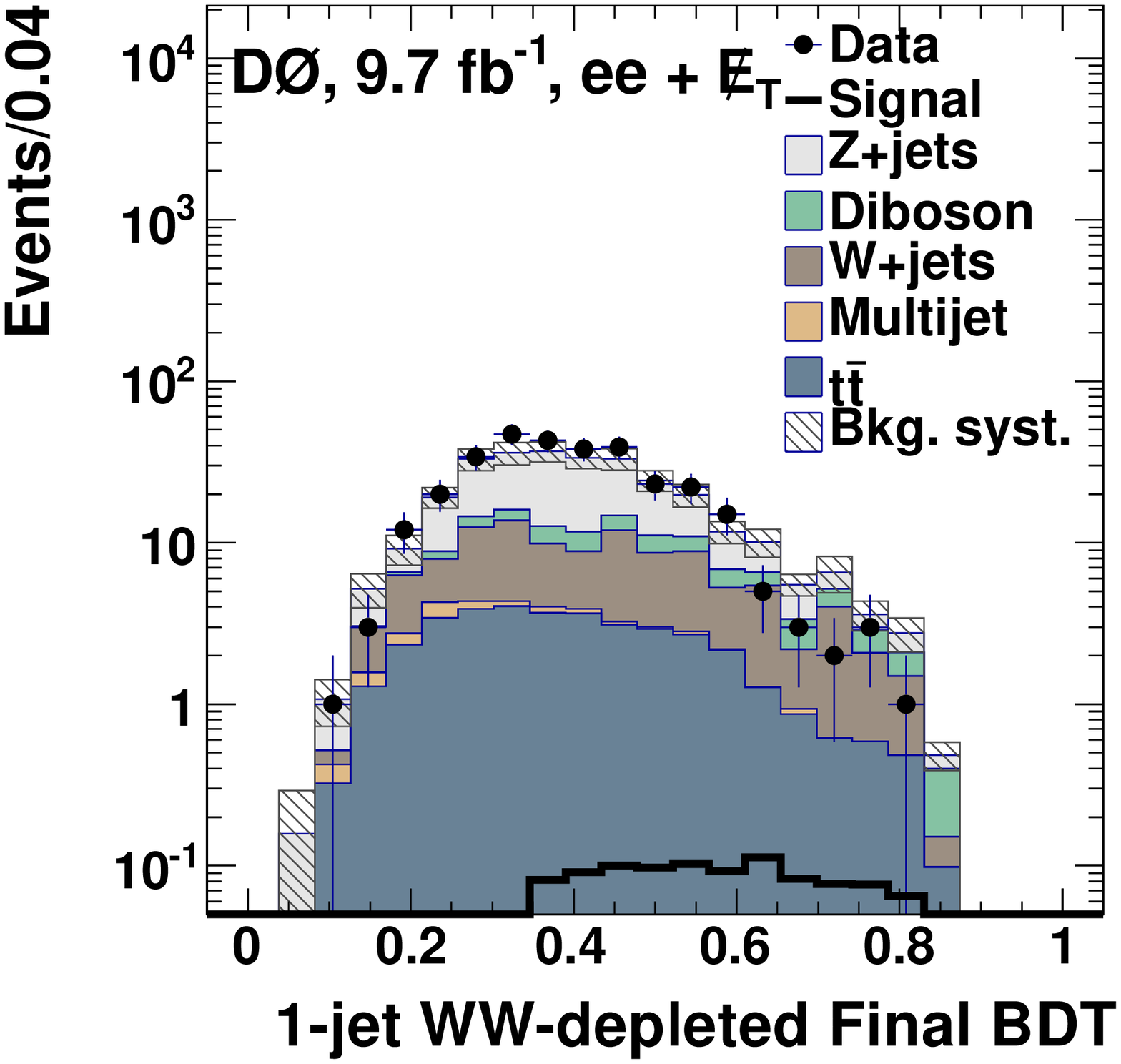}

\unitlength=1mm\begin{picture}(00,00)
\if \mytwocolumn 1
\put(-77,50){\text{\bf (a)}} 
\put(-17,50){\text{\bf (b)}} 
\put(43,50){\text{\bf (c)}}
\else
\put(-47,106){\text{\bf (a)}} 
\put(-17,50){\text{\bf (c)}}
\put(13,106){\text{\bf (b)}}
\fi

\end{picture}

\includegraphics[height=0.237\textheight]{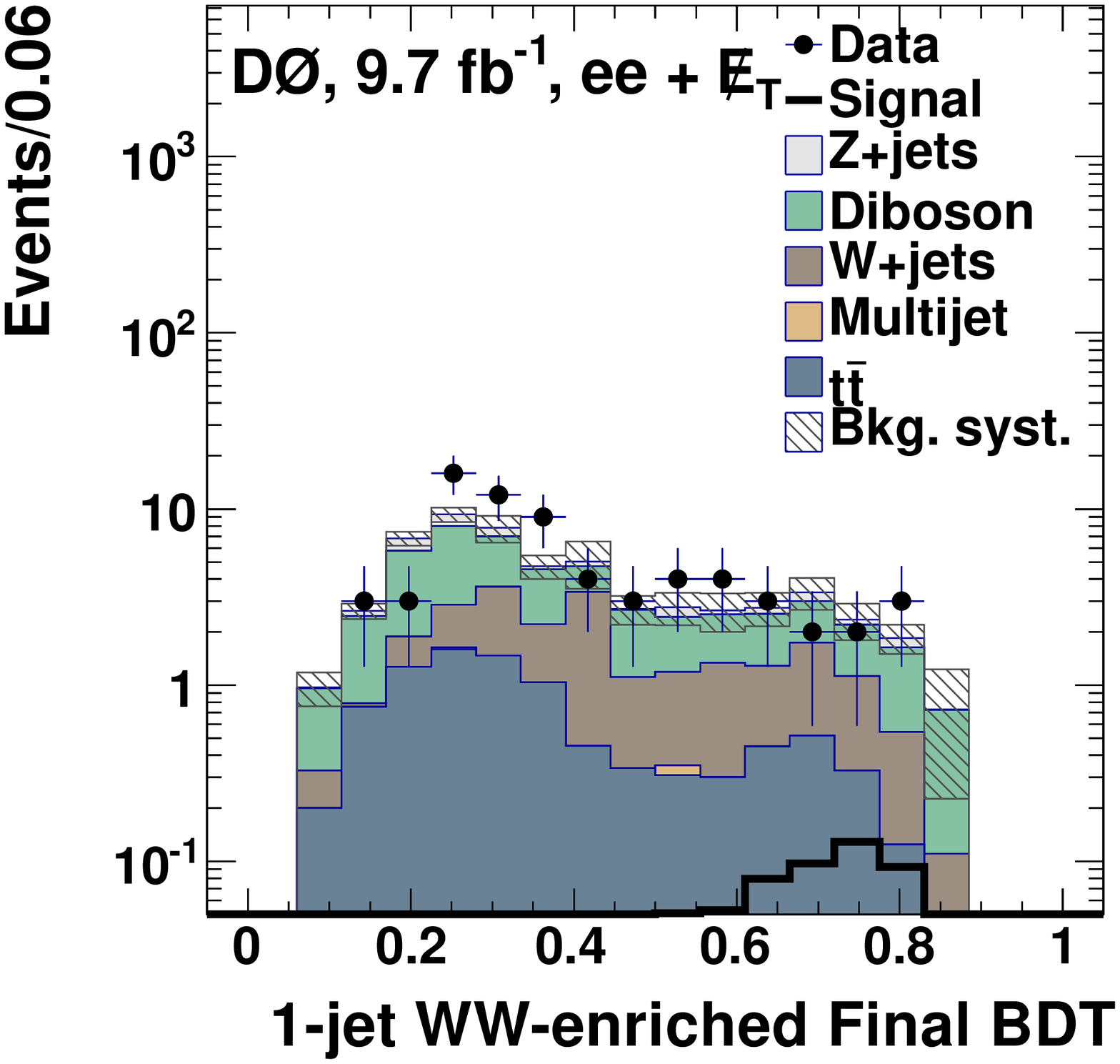}
\includegraphics[height=0.237\textheight]{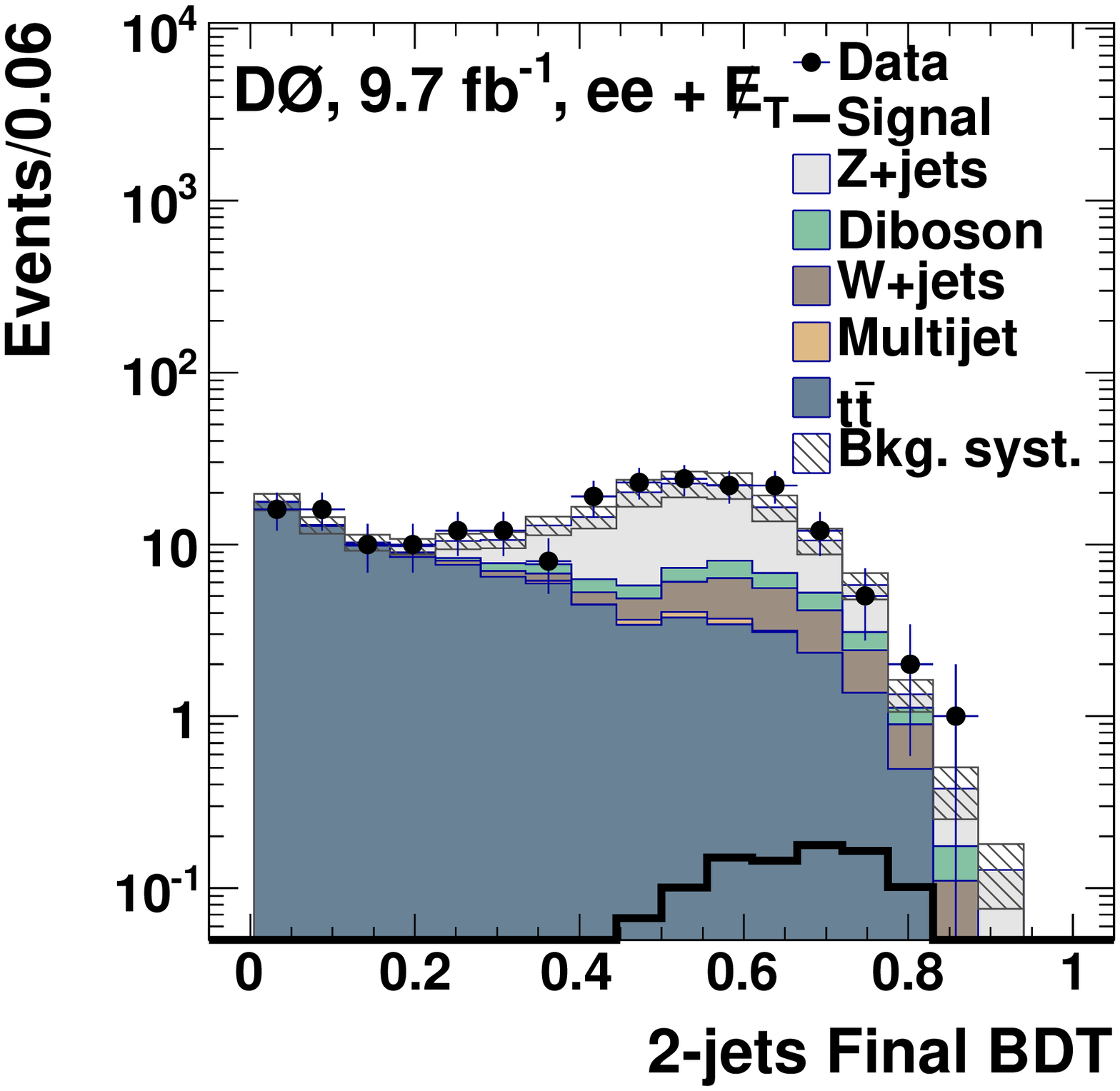}

\unitlength=1mm
\begin{picture}(00,00)
\put(-46,50){\text{\bf (d)}} 
\put(14,50){\text{\bf (e)}} 
\end{picture}

\caption{
Distributions of the final BDT discriminant for $M_{H}=125$ GeV for the \ee\ channel with  (a) 0-jet $WW$-depleted, (b) 0-jet $WW$-enriched, (c) 1-jet $WW$-depleted, 
(d) 1-jet $WW$-enriched, and (e) $\ge$  2-jets.
}
  \label{fig:aux_FDDT_ee}
\end{figure*}

\begin{figure*}[!]

\includegraphics[height=0.237\textheight]{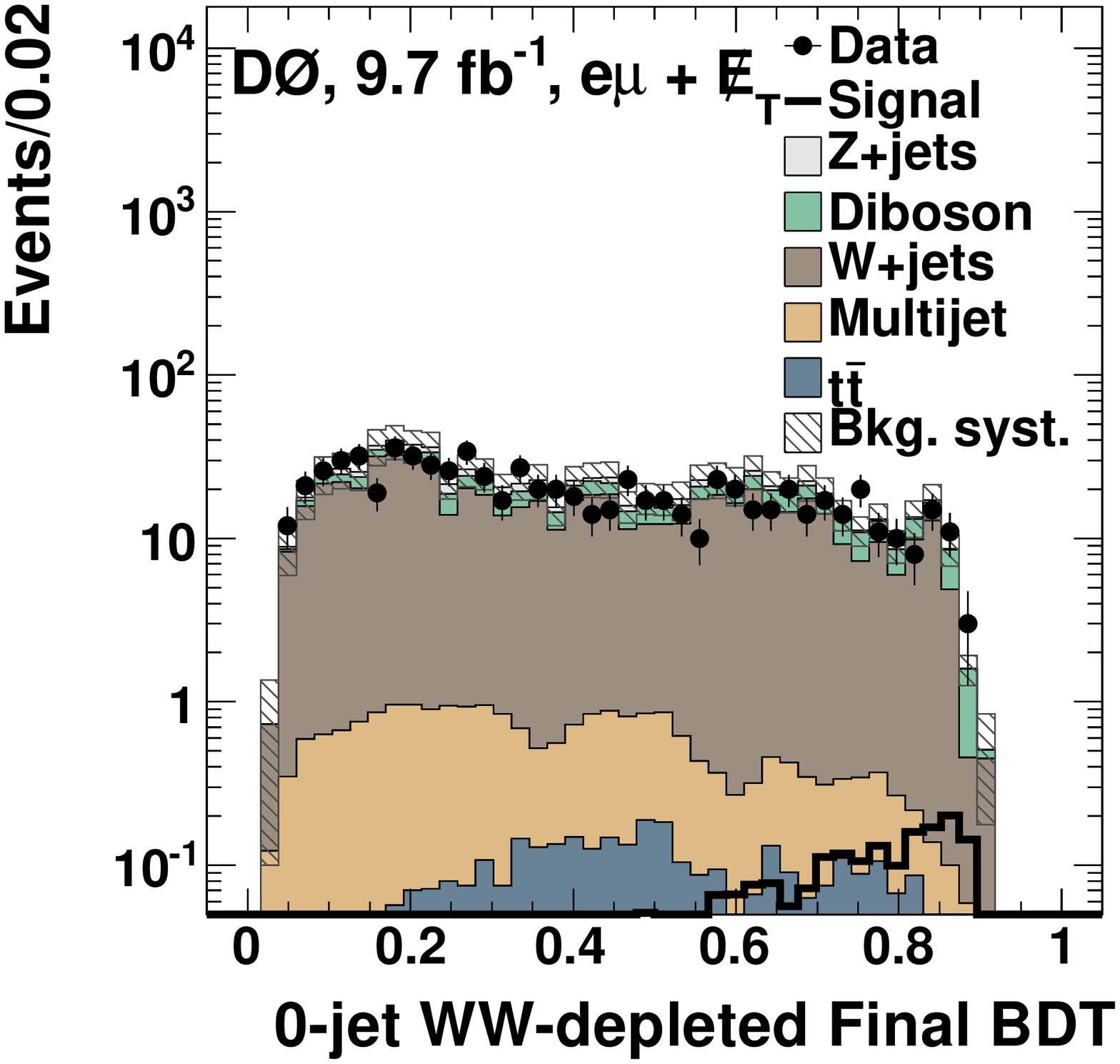}
\includegraphics[height=0.237\textheight]{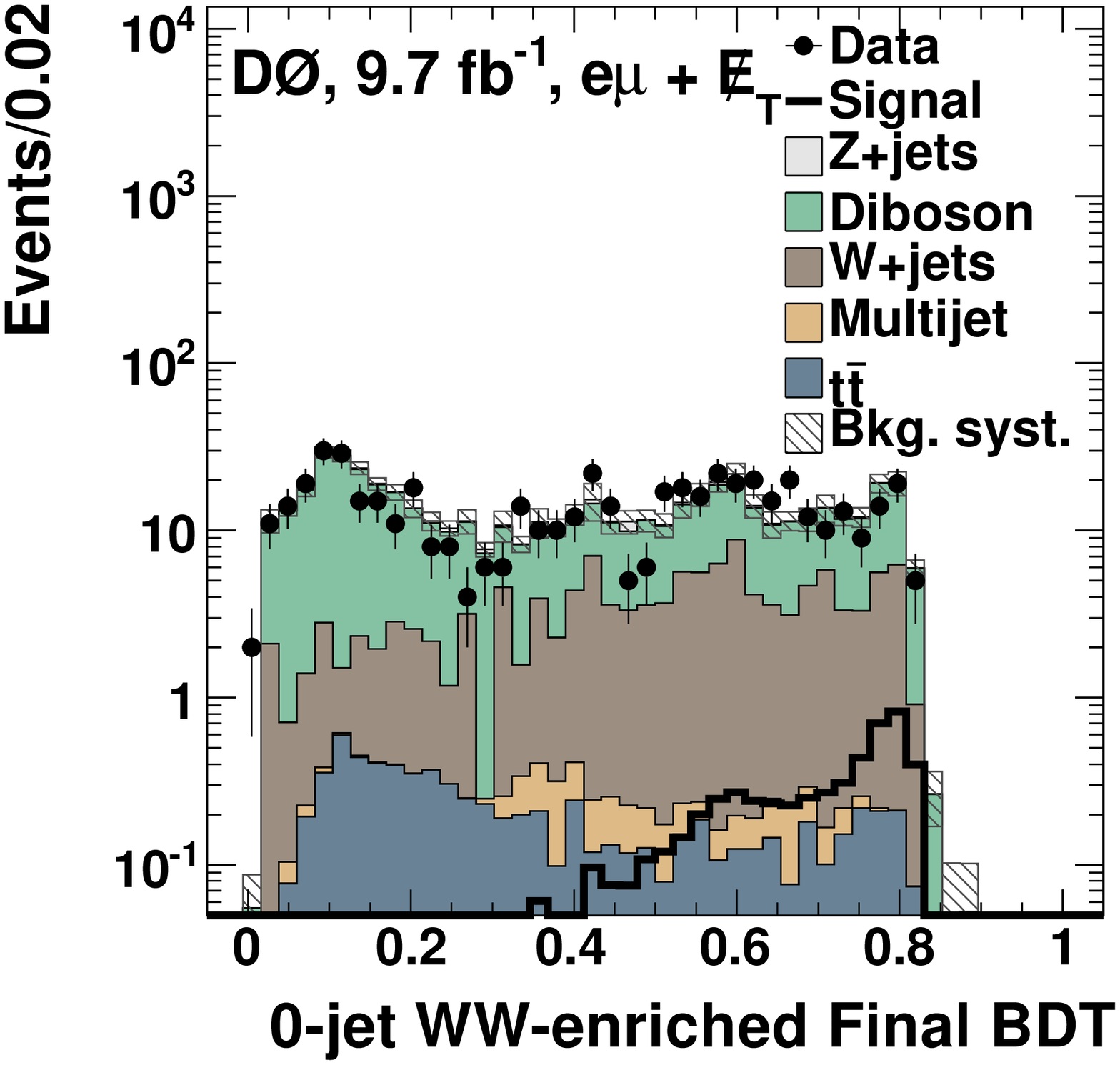}
\includegraphics[height=0.237\textheight]{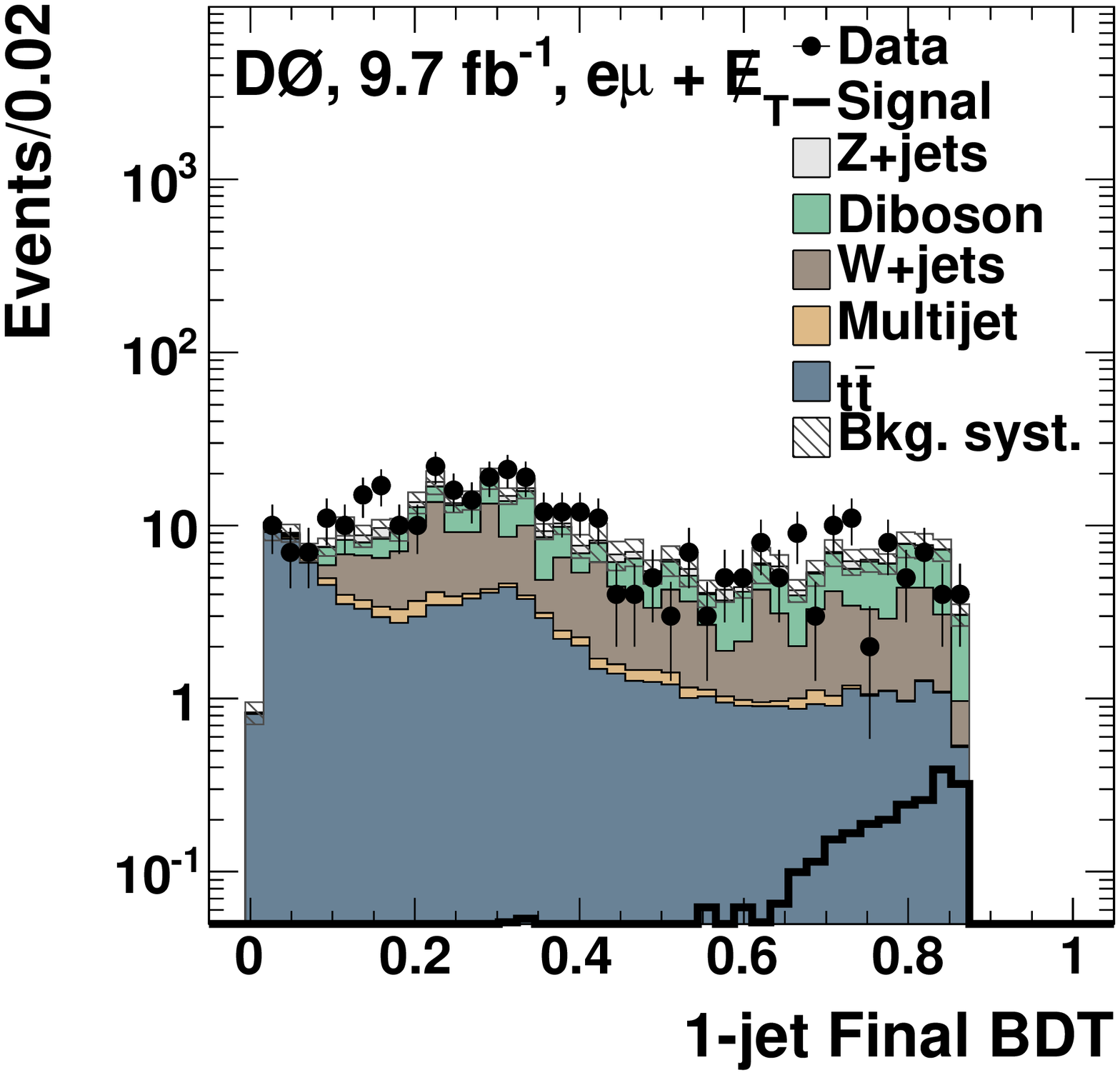}

\unitlength=1mm
\begin{picture}(00,00)
\if \mytwocolumn 1
\put(-77,50){\text{\bf (a)}} 
\put(-17,50){\text{\bf (b)}} 
\put(43,50){\text{\bf (c)}}
\else
\put(-47,106){\text{\bf (a)}} 
\put(-17,50){\text{\bf (c)}}
\put(13,106){\text{\bf (b)}}
\fi

\end{picture}

\includegraphics[height=0.237\textheight]{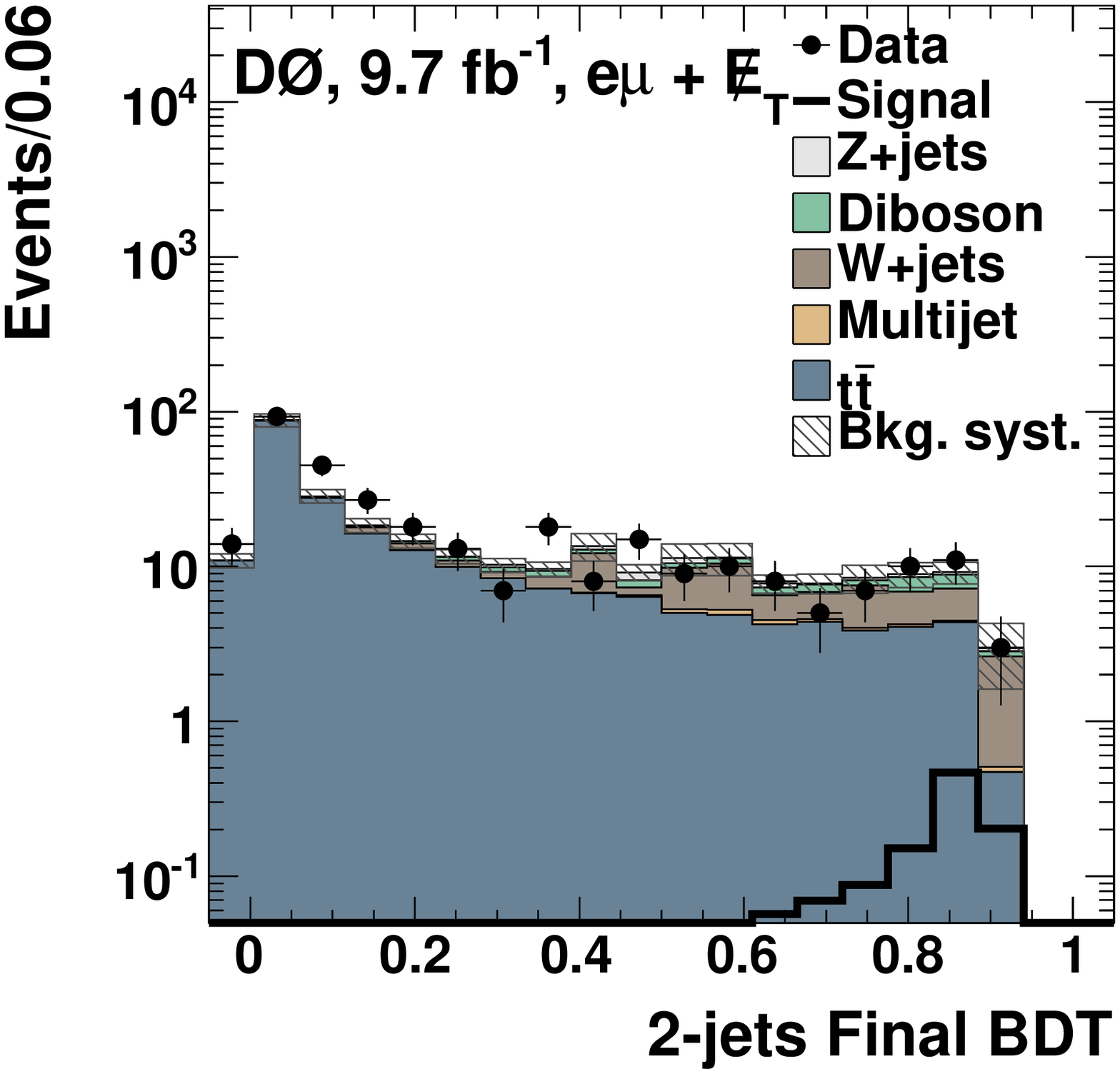}

\unitlength=1mm
\begin{picture}(00,00)
\put(-16,50){\text{\bf (d)}} 
\end{picture}

\caption{
Distributions of the final BDT discriminant for $M_{H}=125$ GeV for the \em\ channel with  (a) 0-jet $WW$-depleted, (b) 0-jet $WW$-enriched, (c) 1-jet , and (d)  $\ge$  2-jets.
}
  \label{fig:aux_FDDT_em}
\end{figure*}

\begin{figure*}[!]

\includegraphics[height=0.237\textheight]{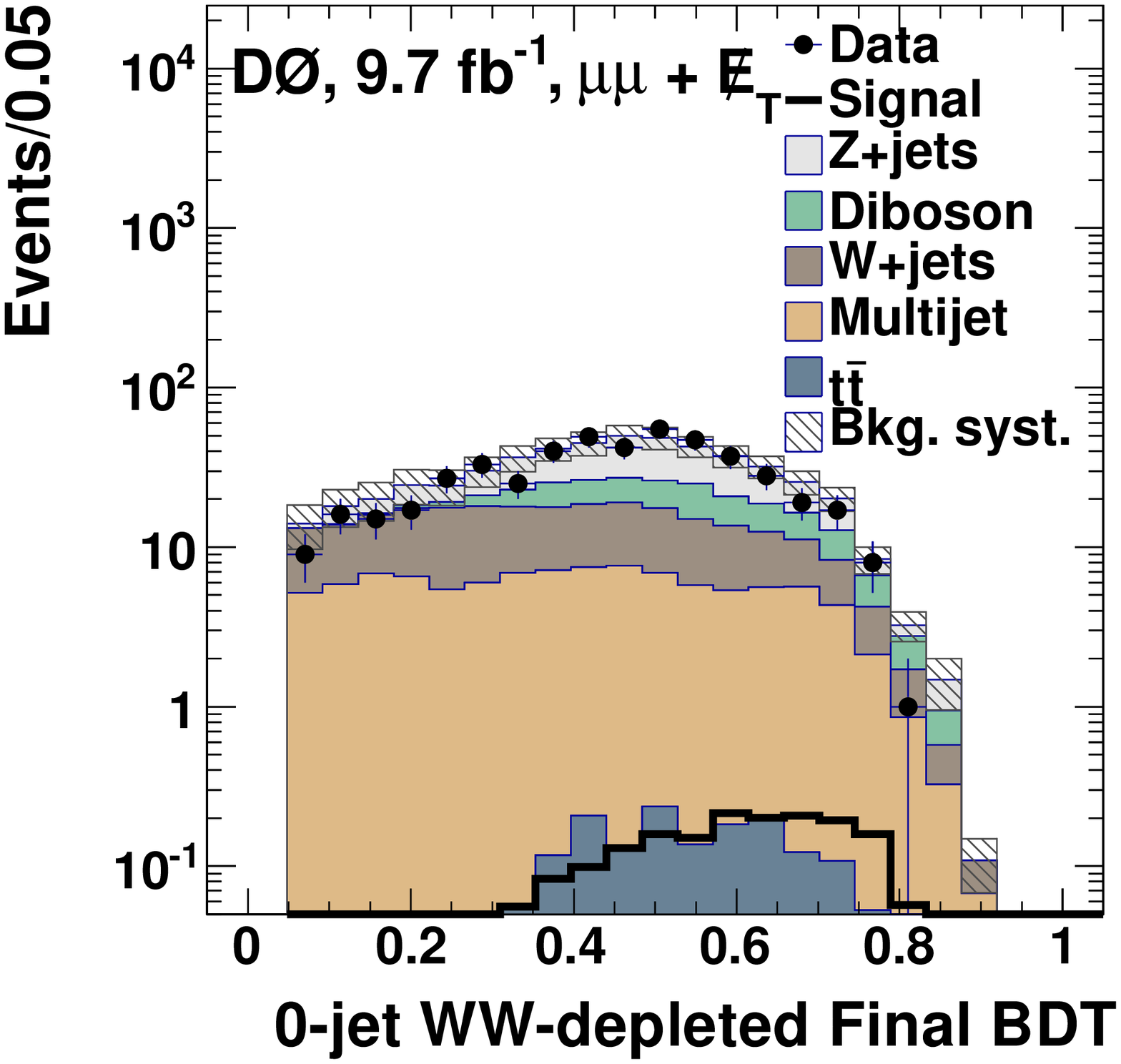}
\includegraphics[height=0.237\textheight]{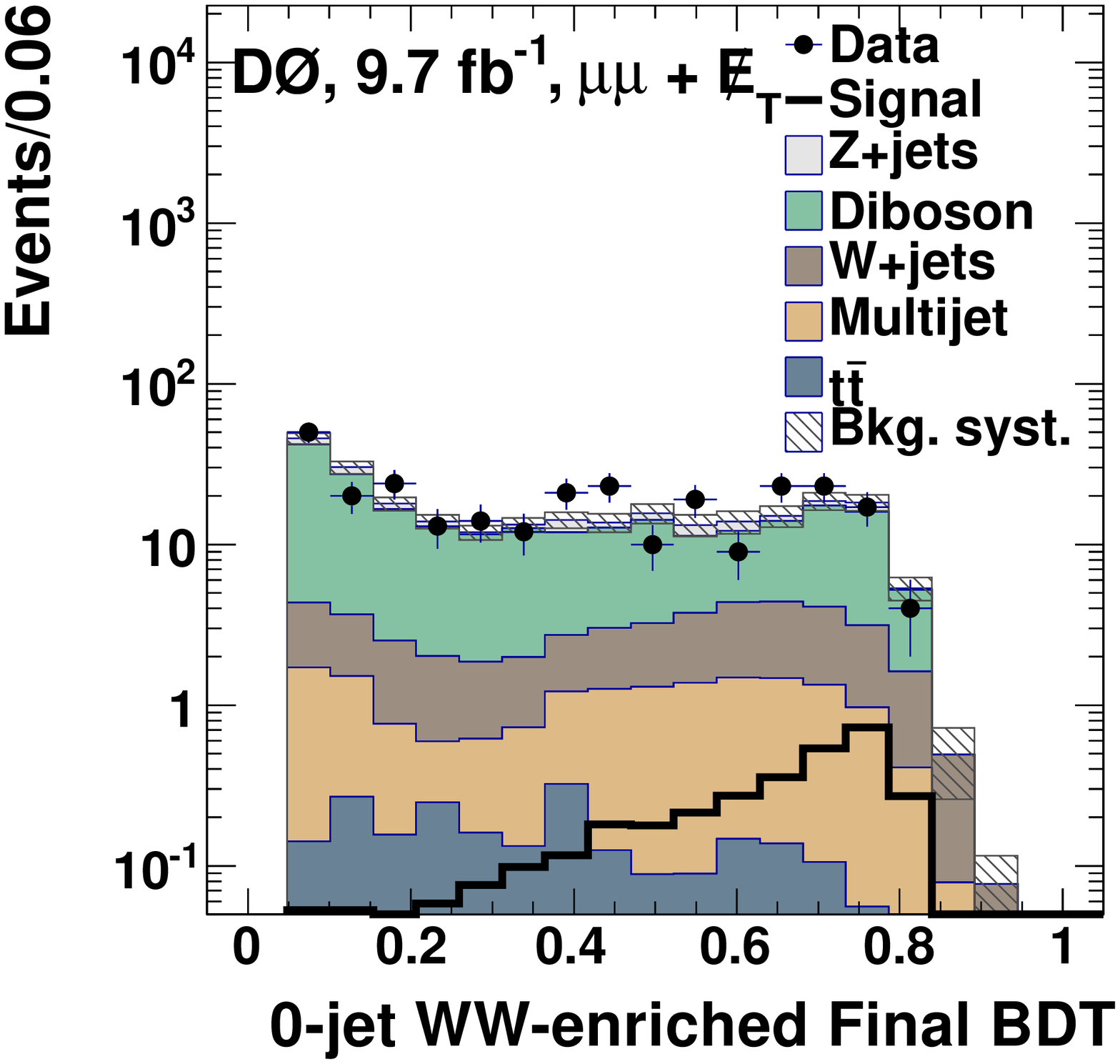}
\includegraphics[height=0.237\textheight]{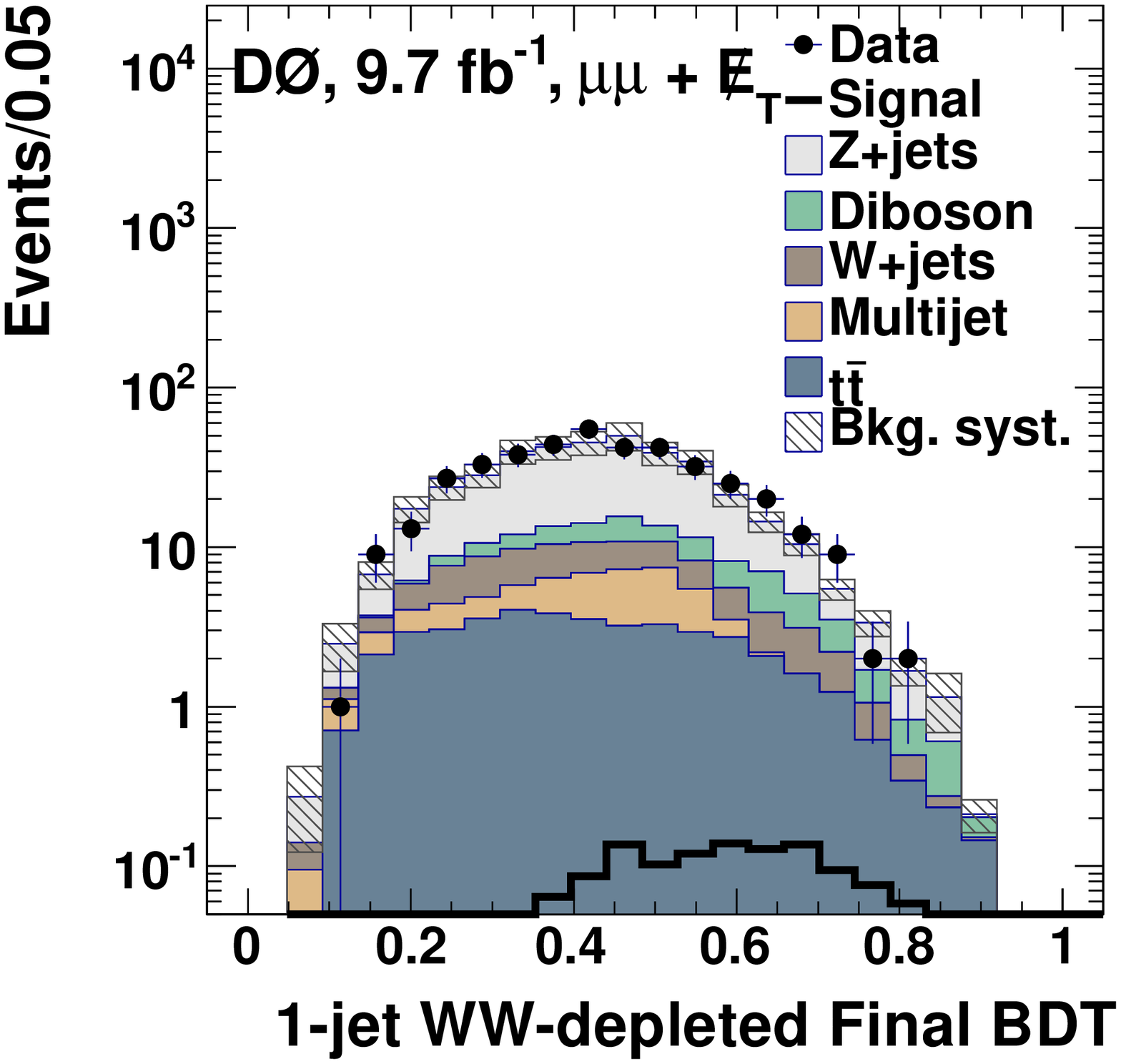}

\unitlength=1mm
\begin{picture}(00,00)
\if \mytwocolumn 1
\put(-77,50){\text{\bf (a)}} 
\put(-17,50){\text{\bf (b)}} 
\put(43,50){\text{\bf (c)}}
\else
\put(-47,106){\text{\bf (a)}} 
\put(-17,50){\text{\bf (c)}}
\put(13,106){\text{\bf (b)}}
\fi
\end{picture}

\includegraphics[height=0.237\textheight]{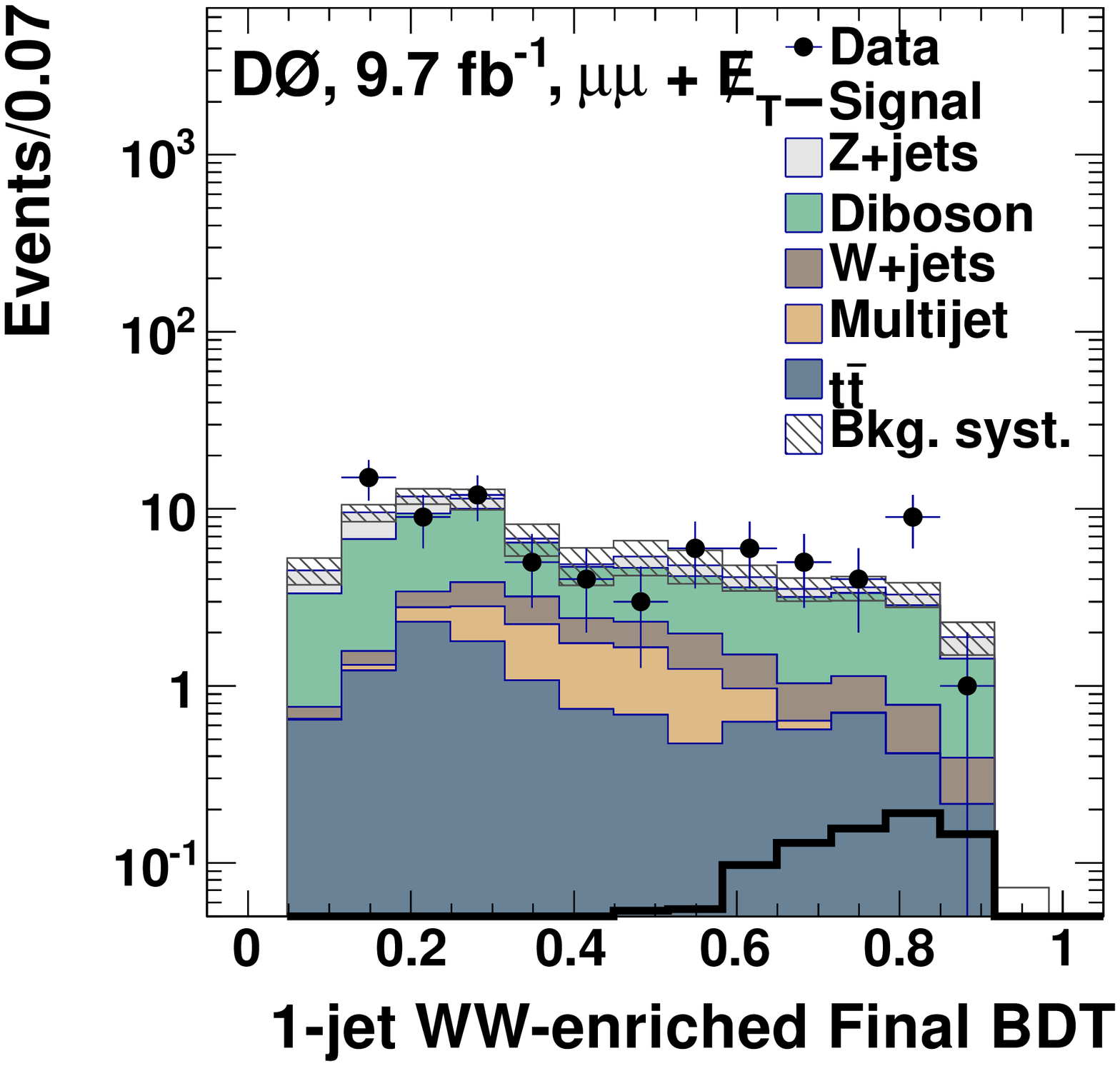}
\includegraphics[height=0.237\textheight]{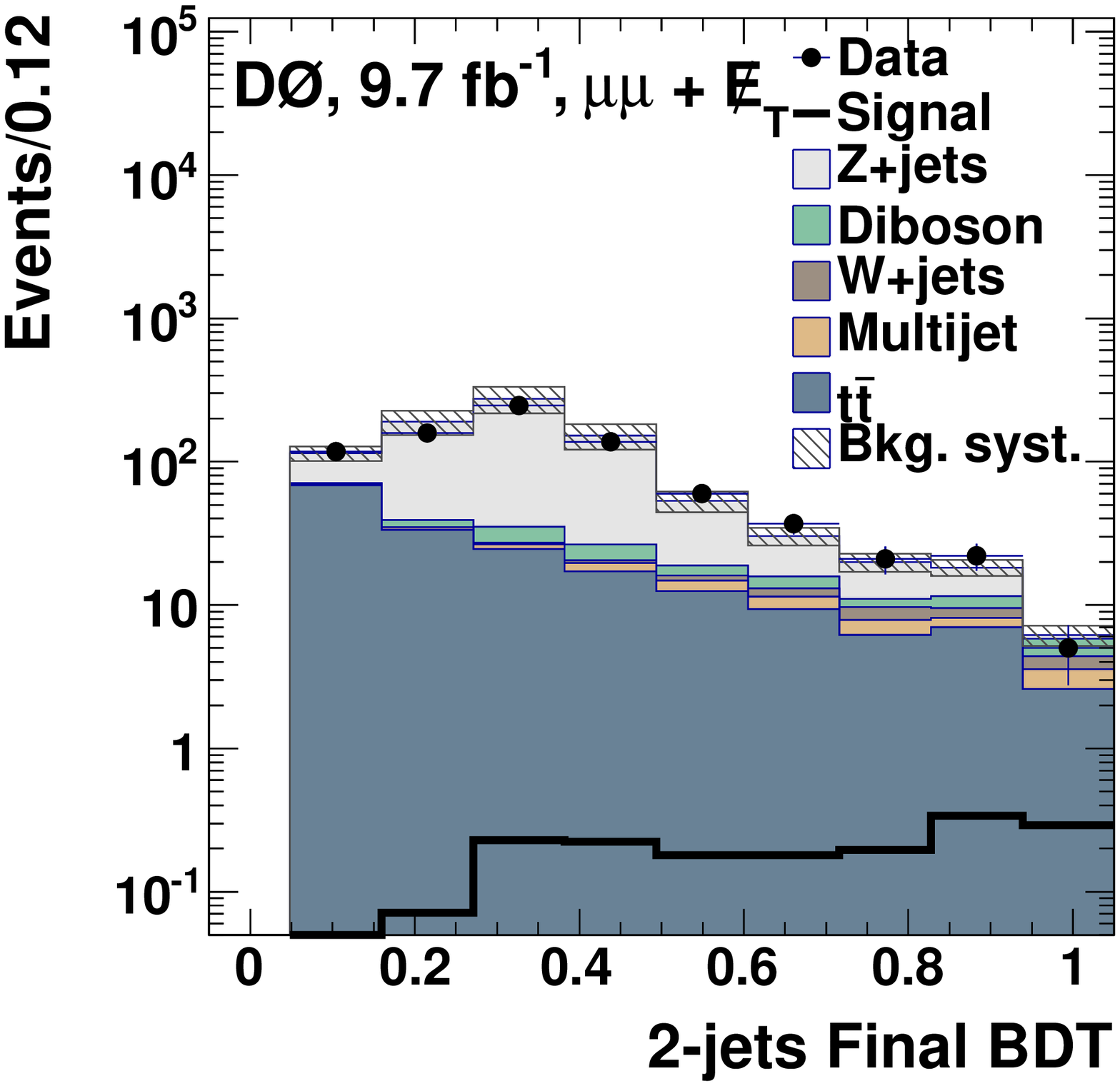}

\unitlength=1mm
\begin{picture}(00,00)
\put(-46,50){\text{\bf (d)}} 
\put(14,50){\text{\bf (e)}} 
\end{picture}

\caption{
Distributions of the final BDT discriminant for $M_{H}=125$ GeV for the \mm\ channel with  (a) 0-jet $WW$-depleted, (b) 0-jet $WW$-enriched, (c) 1-jet $WW$-depleted, 
(d) 1-jet $WW$-enriched, and (e) $\ge$  2-jets.
}
  \label{fig:aux_FDDT_mm}
\end{figure*}

\begin{figure*}[!]
    \includegraphics[height=0.25\textheight]{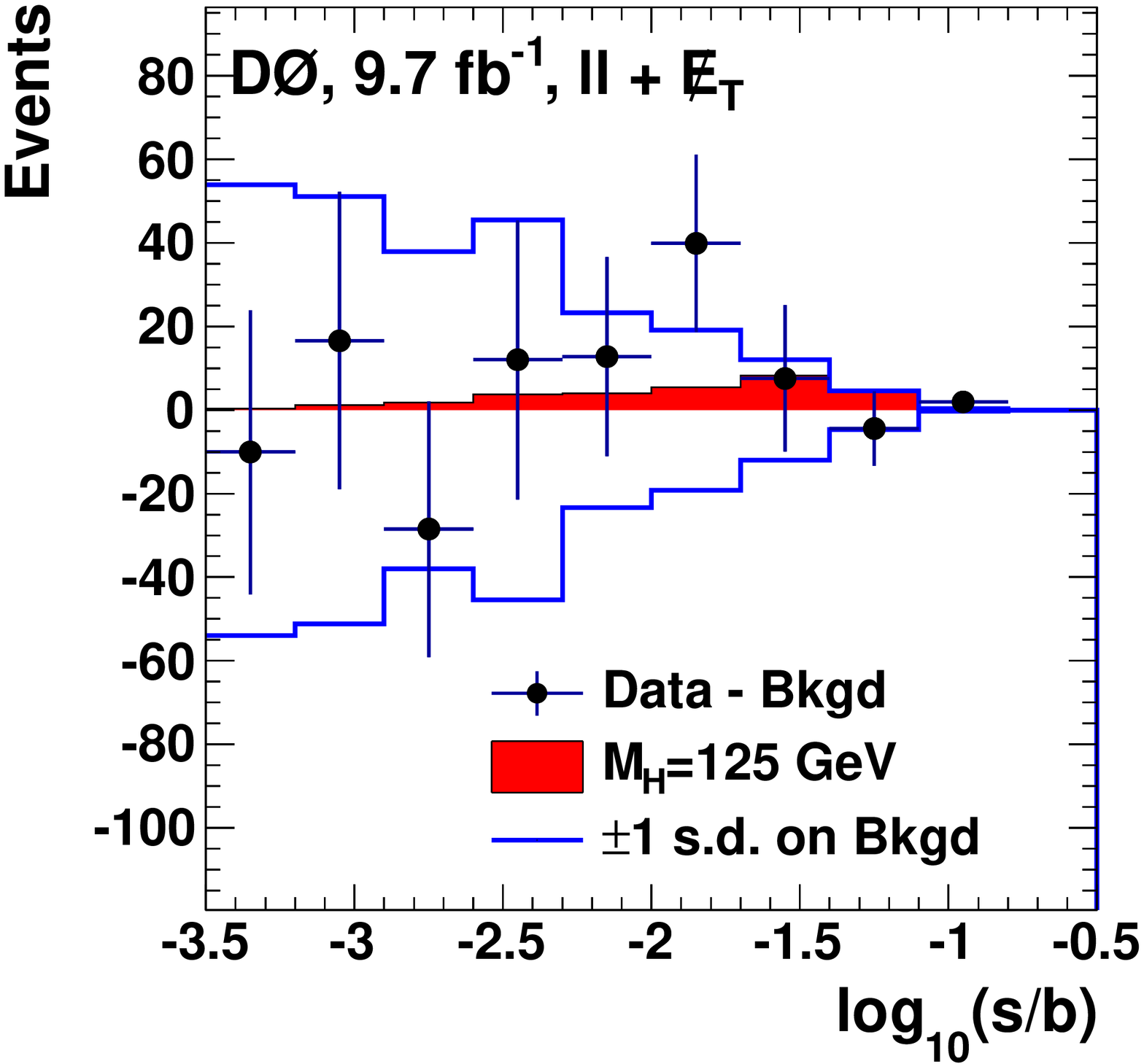}\hspace{2cm}
    \includegraphics[height=0.25\textheight]{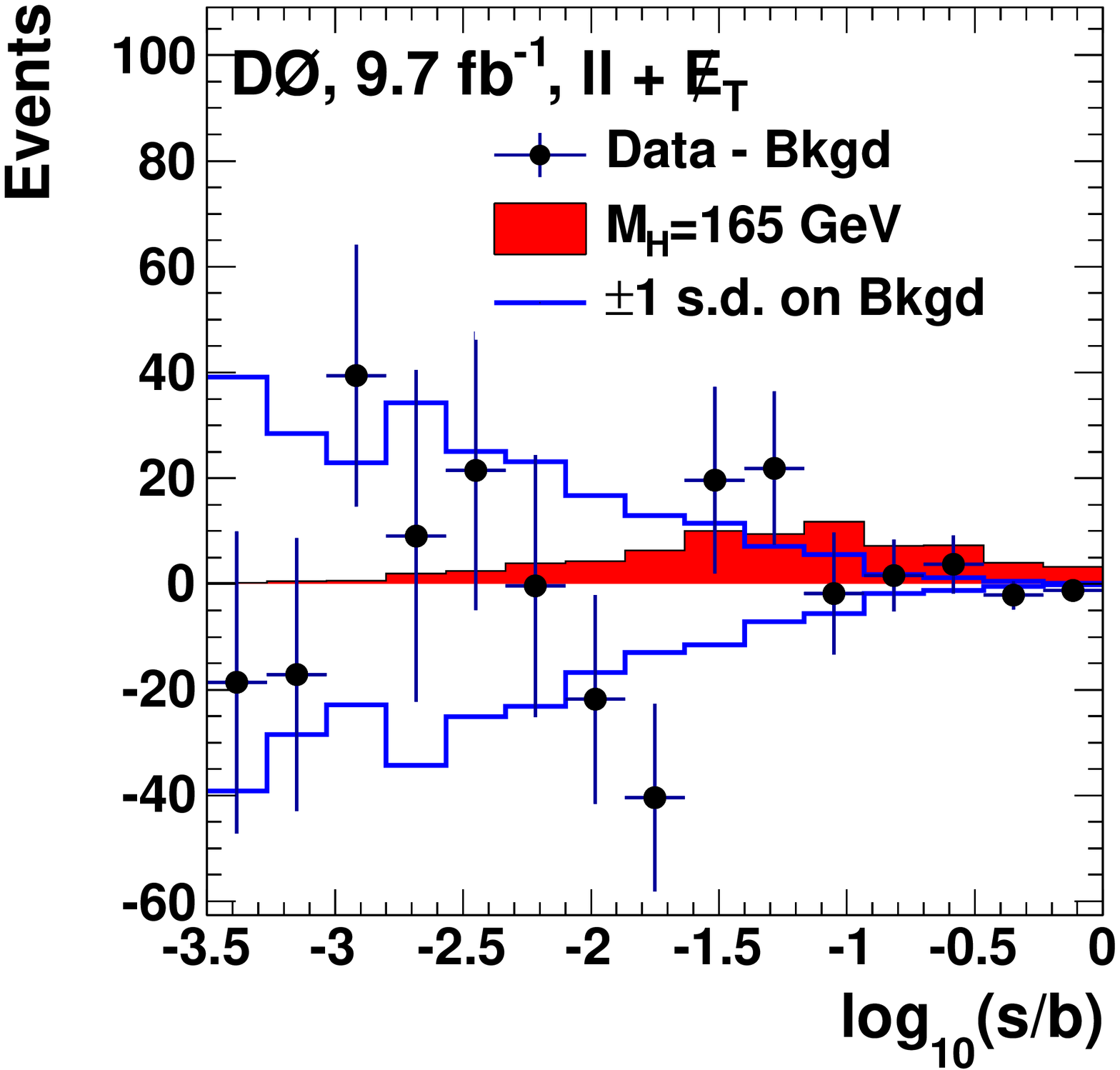}

\unitlength=1mm
\begin{picture}(00,00)
\put(-21,56){\text{\bf (a)}} 
\put(62,56){\text{\bf (b)}} 
\end{picture}

\caption{\label{fig:bk_sub_higgs} The
background-subtracted data distributions for the final discriminants,
summed in bins with similar signal to background ratios, for (a) $M_H
=125$~GeV, and (b) $M_H =165$~GeV. The uncertainties shown on the
background-subtracted data points are the square roots of the post-fit
background number of events predictions in each bin, representing the expected
statistical uncertainty on the data. Also shown is the $\pm$1 s.d.\ band
on the total background after fitting.
}
\end{figure*}

\clearpage

\begin{figure*}[!]
\section*{\Large Tables of limits for the different searches }
\end{figure*}

\begin{table*}[!]
\caption{\label{tab:alllimit} Expected and observed upper limits at the
95\% C.L. for $\sigma(p\bar{p}\rightarrow H+X)$ relative to the
SM for the total combination, and separately for the \ee, \em,
\mm\ channels and different Higgs boson masses.
}
\begin{ruledtabular}
\begin{tabular}{cccccccccccccccccccccc}
$M_H$ (GeV)&   100 &105 &110  &115   &120  &125  &130  &135  &140  &145  &150  &155  &160  &165  &170  &175  &180  &185  &190  &195  &200 \\
\hline


Exp. all:  & 11.8 &11.1 &8.84 &6.73 &4.70 &3.36 &2.64 &2.15 &1.88 &1.56 &1.32 &1.13 &0.82 &0.76 &0.94 &1.10 &1.34 &1.69 &2.11 &2.52 &2.91 \\
Obs. all:  & 17.2 &19.1 &13.9 &8.85 &5.58 &4.10 &2.88 &2.99 &2.50 &2.17 &1.73 &1.24 &0.96 &0.74 &0.84 &0.89 &1.20 &1.40 &2.20 &2.71 &2.48 \\

\hline
Exp. \ee. &  13.6 &14.2 &13.5 &13.0 &10.1 &7.21 &5.56 &4.32 &3.86 &3.22 &2.77 &2.30 &1.76 &1.64 &1.94 &2.24 &2.72 &3.44 &4.12 &4.90 &5.65 \\
Obs. \ee  &  20.6 &23.6 &16.9 &15.1 &8.41 &6.25 &4.67 &5.05 &3.80 &3.98 &3.13 &2.70 &2.11 &1.79 &2.02 &2.31 &2.68 &3.21 &5.53 &5.76 &5.76 \\
\hline

Exp. \em   &42.4 &27.9 &15.9 &10.0 &6.55 &4.65 &3.63 &2.97 &2.48 &2.02 &1.77 &1.50 &1.11 &1.06 &1.28 &1.51 &1.75 &2.27 &2.80 &3.30 &3.75\\
Obs. \em  &28.1 &20.0 &12.5 &7.95 &6.23 &4.75 &3.39 &2.88 &2.61 &2.14 &1.75 &1.28 &1.00 &0.86 &1.17 &1.33 &1.79 &2.16 &2.68 &3.31 &3.20 \\

\hline

Exp. $\mu\mu$ &   39.4 &29.0 &20.6 &13.9 &8.76 &6.25 &4.72 &3.95 &3.39 &3.09 &2.59 &2.31 &1.74 &1.61 &1.95 &2.30 &2.89 &3.49 &4.40 &5.05 &5.83 \\
Obs. $\mu\mu$ &   66.8 &68.2 &53.1 &27.4 &16.8 &10.1 &8.22 &7.65 &6.77 &5.53 &5.01 &3.95 &3.01 &2.31 &2.53 &2.79 &3.51 &4.14 &5.85 &7.45 &7.92 \\

\end{tabular}
\end{ruledtabular}
\end{table*}

\begin{table*}[bt]
\caption{\label{tab:alllimit_4thGen} Expected and observed upper limits at 95\% CL
for $\sigma(gg\to H)\times BR(H\to WW)$ in pb, for the combination of the $ee$, $e\mu$, $\mu\mu$ channels and different Higgs boson masses.}
\begin{ruledtabular}
\begin{tabular}{ccccccccccccccccc}
$M_H$  (GeV) &    100& 105 &110 &115  &120  &125  &130  &135  &140  &145  &150  &155  &160  &165  &170  &175  \\
\hline
Exp. all:  &    1.97 &1.37 &1.11 &0.97 &0.91 &0.85 &0.82 &0.80 &0.75 &0.70 &0.61 &0.51 &0.38 &0.32 &0.37 &0.40 \\
Obs. all:  &  2.55 &2.53 &1.60 &1.21 &1.15 &0.98 &1.00 &1.21 &0.98 &0.88 &0.69 &0.59 &0.42 &0.33 &0.34 &0.36\\ 
\hline\hline
$M_H$  (GeV)   & 180  &  185&  190&  195&  200&  210&  220&  230&  240&  250&  260& 270  & 280 & 290 & 300  \\
\hline
Exp. all:       &0.41 &0.45 &0.48 &0.50 &0.52 &0.56 &0.52 &0.51 &0.47 &0.44 &0.43 &0.43 &0.42 &0.42 &0.40 \\
Obs. all:     &0.34 &0.34 &0.48 &0.44 &0.47 &0.61 &0.64 &0.65 &0.71 &0.67 &0.63 &0.72 &0.67 &0.70 &0.86 \\
\end{tabular}
\end{ruledtabular}

\end{table*}

\begin{table*}[!]
\caption{\label{tab:FHlimit}  Assuming fermiophobic couplings,  expected and observed upper limits
for $\sigma(p\bar{p}\rightarrow H+X)$ relative to the fermiophobic Higgs expected yields
for the combination of the \ee, \em, \mm\ channels and different Higgs boson masses.
}
\begin{ruledtabular}
\begin{tabular}{cccccccccccccccccccccc}

$M_H$ (GeV) & 100 & 105 & 110 & 115& 120& 125& 130& 135& 140& 145& 150& 155& 160& 165& 170& 175& 180& 185& 190& 195& 200 \\
\hline                                                   
Exp. all: &1.53 &1.85 &1.85 &1.91 &2.03 &2.15 &2.29 &2.37 &2.53 &2.71 &2.81 &2.75 &2.70 &2.62 &3.01 &3.41 &3.78 &4.58 &5.40 &6.10 &6.60 \\
Obs. all: &1.97 &2.33 &2.85 &2.59 &2.23 &3.14 &2.96 &2.42 &3.05 &3.15 &2.61 &3.24 &3.16 &2.70 &3.23 &3.12 &3.89 &4.76 &5.23 &6.90 &8.70 \\
\end{tabular}
\end{ruledtabular}
\end{table*}

\clearpage
\begin{figure*}[!]
\section*{\Large $\boldsymbol{WW}$ cross section measurement}
\end{figure*}


\begin{figure*}[!]
\includegraphics[height=0.237\textheight]{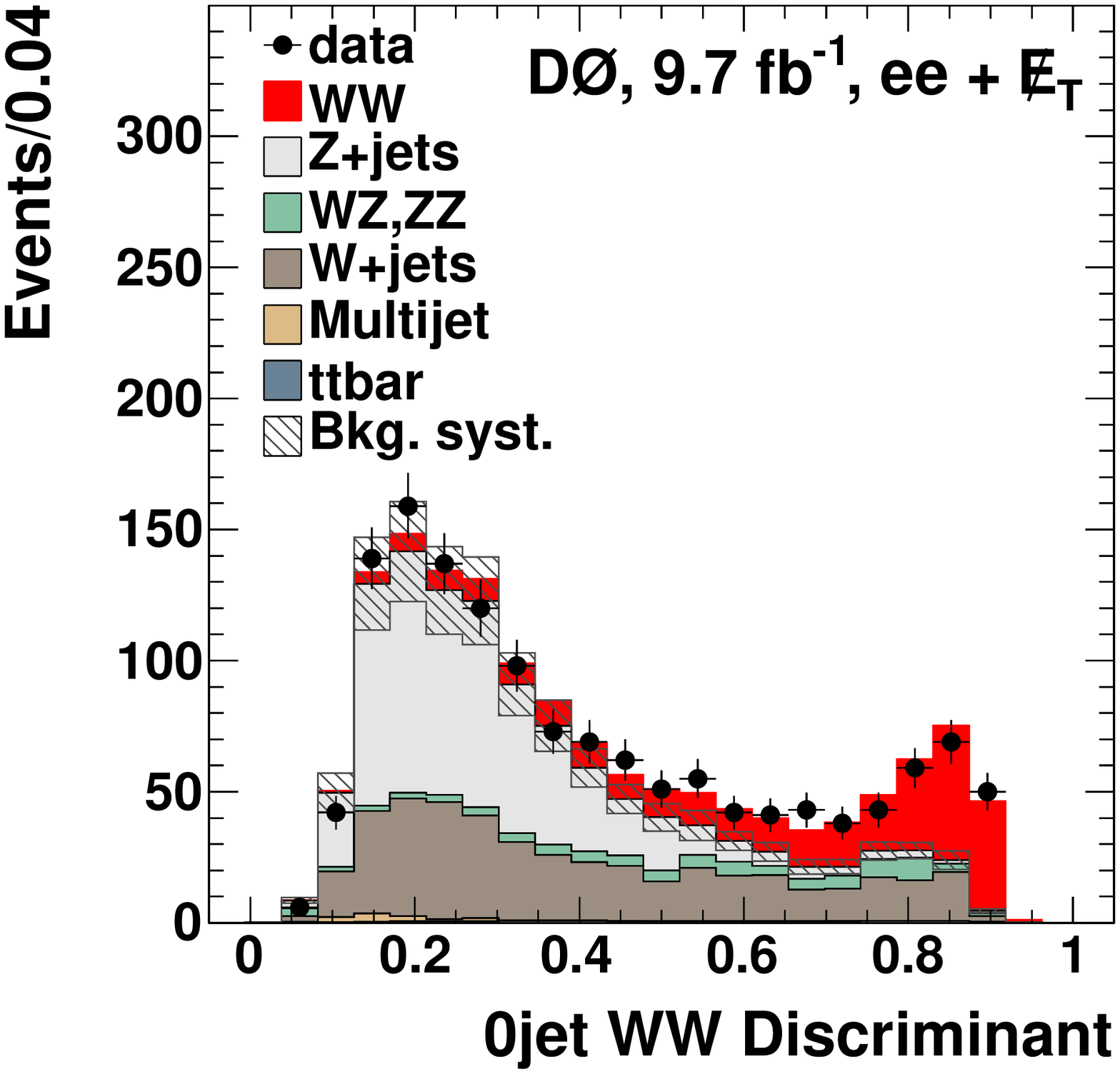}
\includegraphics[height=0.237\textheight]{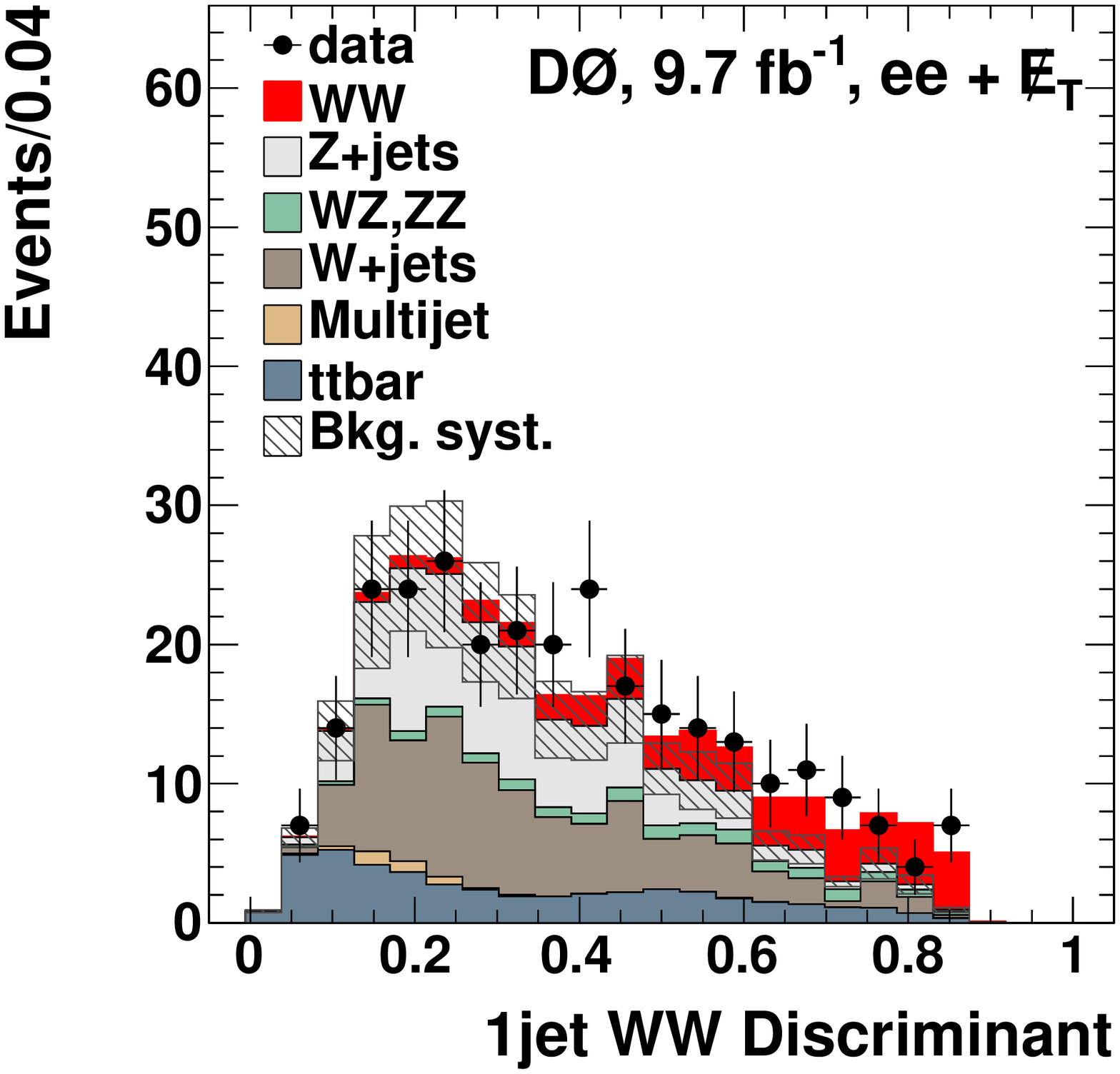}
\unitlength=1mm
\begin{picture}(00,00)
\put(-93,45){\text{\bf (a)}} 
\put(-33,45){\text{\bf (b)}} 
\end{picture}

\caption{
The $WW$ discriminant distributions in the \ee\ channel for (a) no jet and (b) one jet.
}
\label{fig:aux_WWDT_ee}
\end{figure*}

\begin{figure*}[!]
 \includegraphics[height=0.237\textheight]{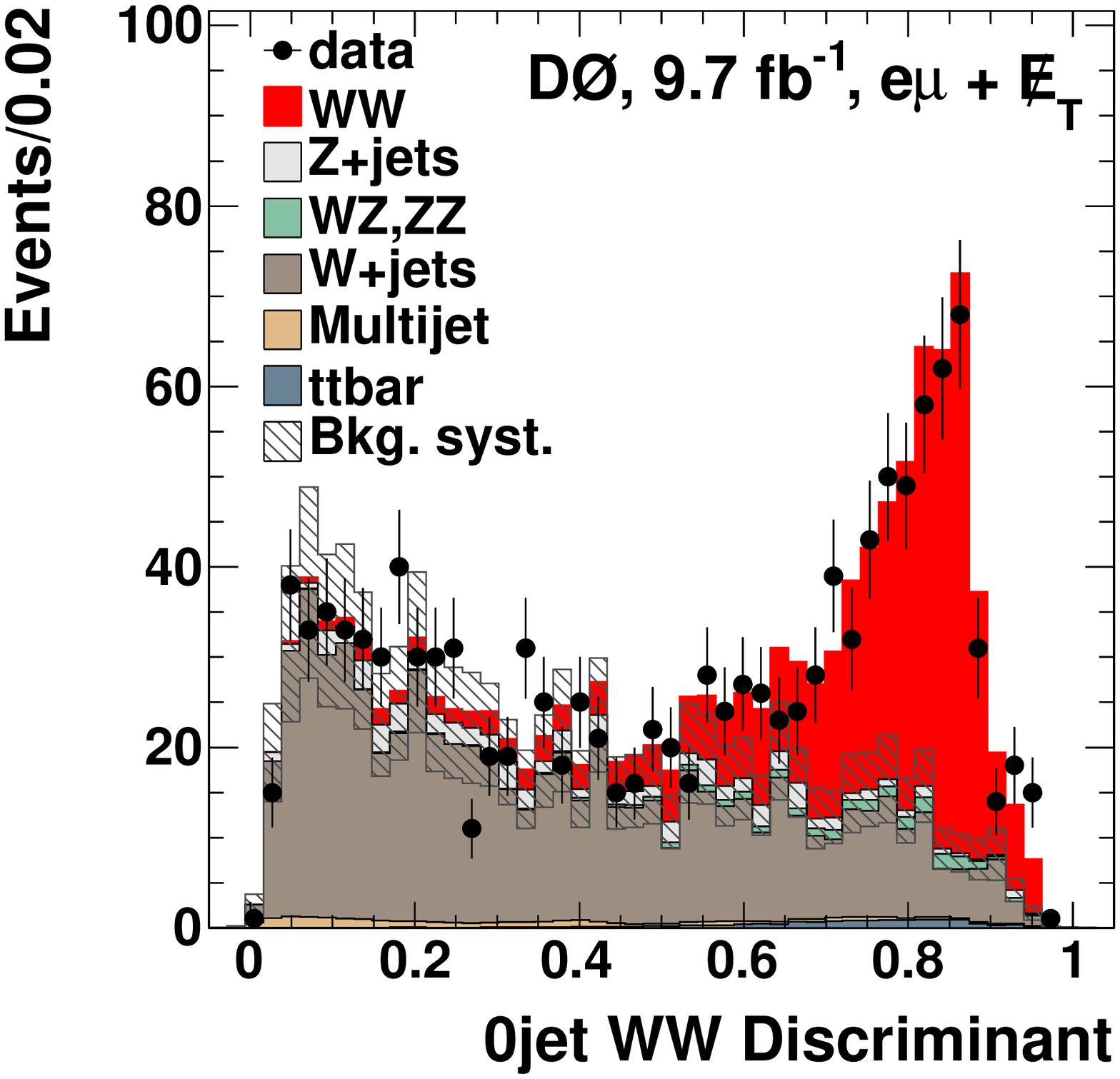}
\includegraphics[height=0.237\textheight]{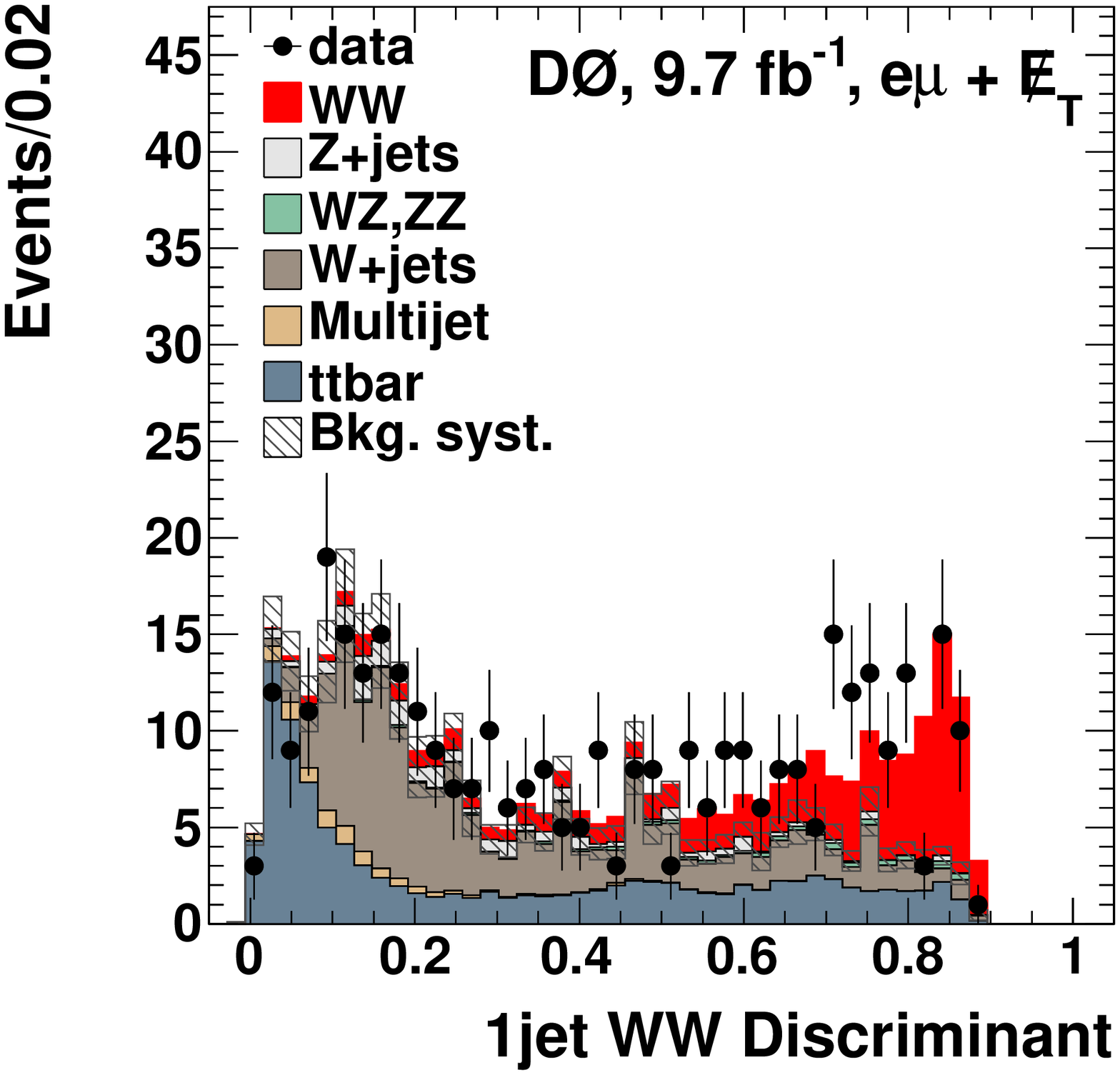}
\unitlength=1mm
\begin{picture}(00,00)
\put(-93,45){\text{\bf (a)}} 
\put(-33,45){\text{\bf (b)}} 
\end{picture}
\caption{
The $WW$ discriminant distributions in the \em\ channel for (a) no jet and (b) one jet.
}
\label{fig:aux_WWDT_em}
\end{figure*}

\begin{figure*}[!]
 \includegraphics[height=0.237\textheight]{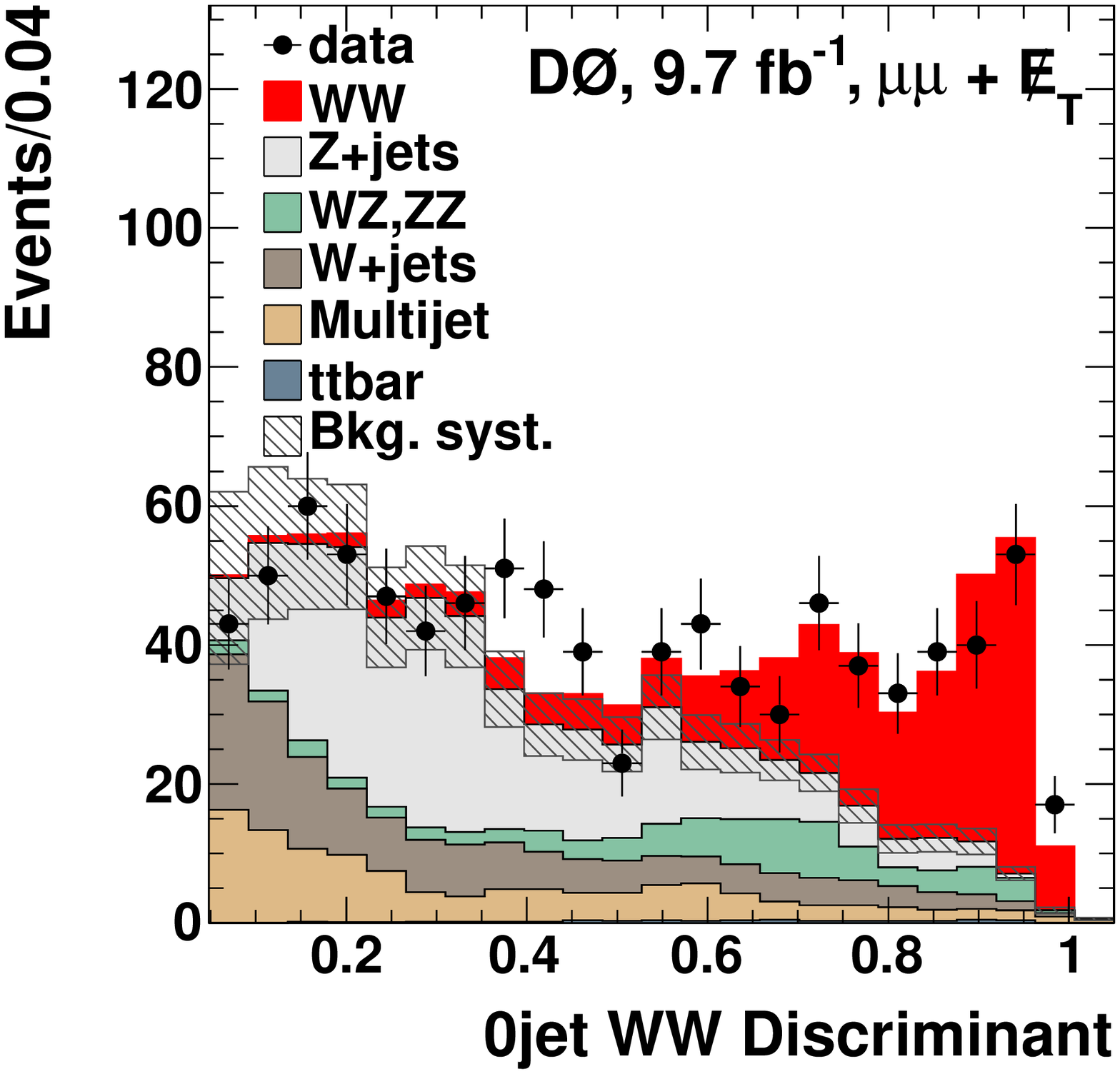}
\includegraphics[height=0.237\textheight]{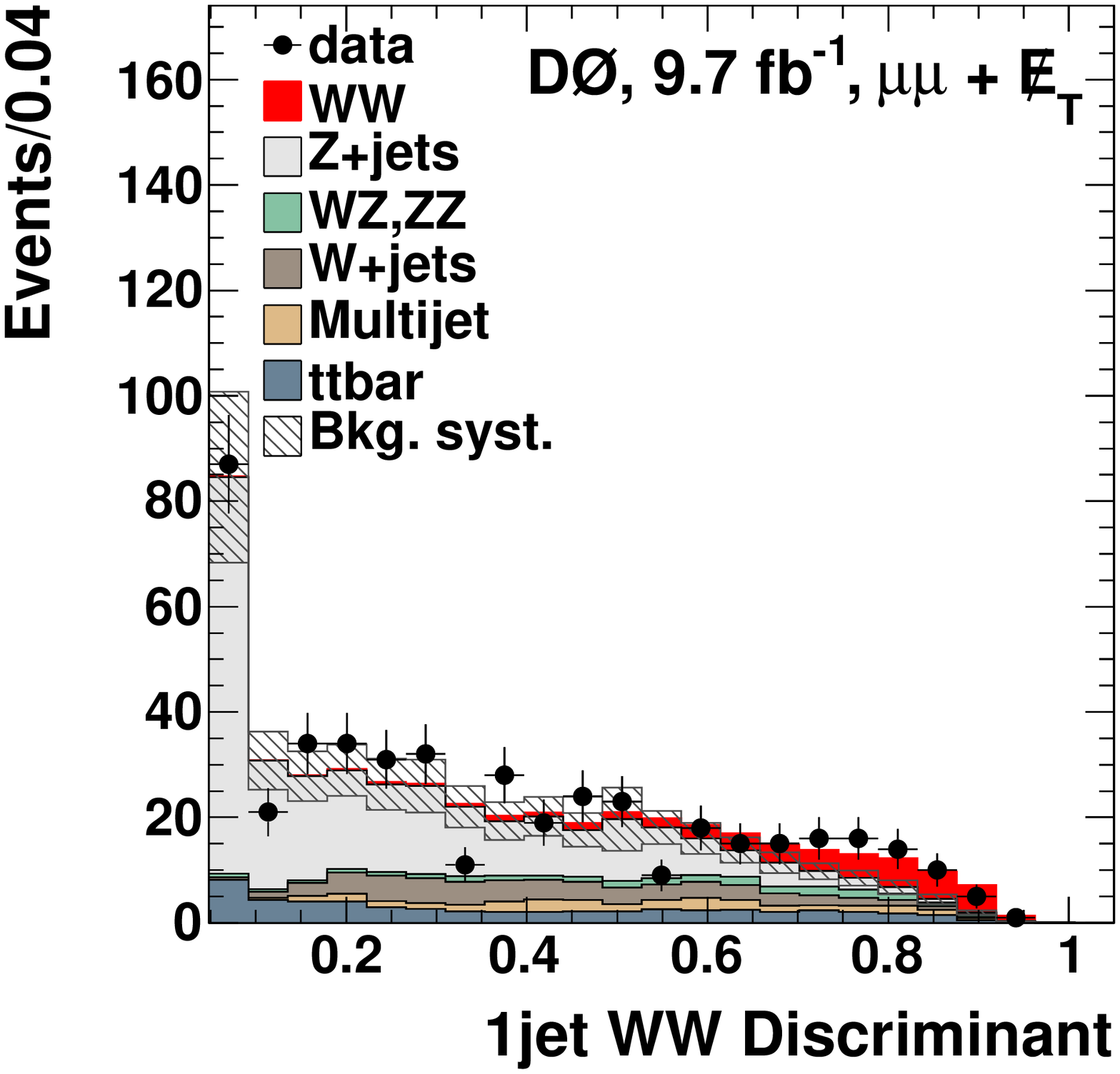}
\unitlength=1mm
\begin{picture}(00,00)
\put(-93,45){\text{\bf (a)}} 
\put(-33,45){\text{\bf (b)}} 
\end{picture}
\caption{
The $WW$ discriminant distributions in the \mm\ channel for (a) no jet and (b) one jet.
}
\label{fig:aux_WWDT_mm}
\end{figure*}

\begin{figure*}[!]
  \begin{center}
    \includegraphics[height=0.25\textheight]{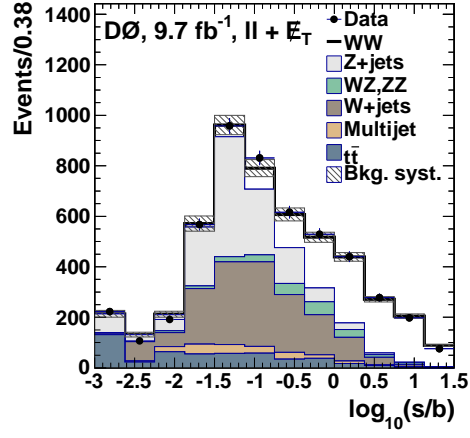}
  \end{center}
  \caption{\label{fig:WW_disc} 
Distribution of the combination of $WW$ discriminants from  \ee, \em, and \mm\ channels employed to measure the $WW$ cross section,
sorted as a function of signal over background ratios. The hatched bands show the total  systematic uncertainty on the background prediction.
}
\end{figure*}

\begin{figure*}[!]
  \includegraphics[width=0.48\textwidth]{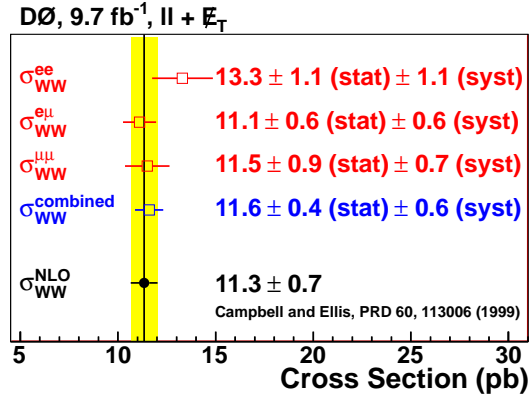}
\caption{Results  of the $p\bar{p}\rightarrow
    WW$ cross section measurements in each dilepton final state
    and their combination. 
 \label{fig:wwXsec} }
\end{figure*}



\begin{thebibliography}{99}
\bibitem{footnote:eta}{The pseudo-rapidity is defined as $\eta = -\ln{(\tan\theta/2)}$, where $\theta$ is the polar angle
relative to the proton beam direction.}
\bibitem{footnote:DR}{Here, ${\cal R} =
\sqrt{(\Delta\phi)^2+(\Delta\eta)^2}$, where $\phi$ is the azimuthal angle
around the proton beam direction.}
\end{thebibliography}
\end{document}
%